\documentclass[final]{ws-ijmpd}

\newcommand{\SN}{\operatorname{s}}
\newcommand{\CS}{\operatorname{c}}
\newcommand{\ddd}{\operatorname{d}}

\newcommand{\Ldot}{W}
\newcommand{\Lddot}{\dot \Ldot}
\newcommand{\eqdef}{:=}
\DeclareMathOperator{\sgn}{sgn}
\newcommand{\CHL}{\newline{}}

\begin{document}
	
	\markboth{Alain Riazuelo}{Seeing relativity -- III. }
	
	\title{Seeing Relativity -- III. Journeying within the Kerr metric toward the
		negative gravity region}
	
	\author{Alain Riazuelo} 
	
	\address{Sorbonne Universit\'e, CNRS, UMR~7095, Institut
		d’Astrophysique de Paris, 98 bis boulevard Arago, 75014~Paris,
		France}
	
	
	\maketitle
	
	\begin{abstract}
		In this paper we study some features of the Kerr metric both from an
		analytic and a visual point of view by performing accurate
		raytracing in various situations.  We focus on
		features that are unique to the maximal analytic extension of the Kerr
		metric as compared to that of the Schwarzschild or even the
		Reissner-Nordstr\"om one. A large number of new, yet underexplored
		phenomena appear, especially regarding the structure of bounded null geodesics
		and the aspect of the negative gravity regions whose visual characteristics are
		shown both from outside and inside it. 
	\end{abstract}

	\date{1st July 2020}
	
	\ccode{03.30.+p, 04.25.D-}
	
	
	
	\section{Introduction}
	
	The Kerr metric was discover by R.P.~Kerr in 1963~\cite{kerr65}. It
	was soon realized that it describes the gravitational field of a
	spinning black hole of mass $M$ and angular momentum per unit of mass
	$a$~\cite{boyer_price65}. Those are the sole parameters that describe
	the metric but despise this, the exterior part of this metric~(i.e., outside the
	horizon) corresponds to the exact description of real black holes, the number of
	which being several tens of millions in a Milky~Way-type galaxy. The existence
	of black hole has been a long-standing debate which was progressively settled by
	the discovery of a growing number of~(then) stellar ``black hole candidates''
	whose status slowly shifted toward real black holes, the black hole population
	being later expanded by that of supermassive black holes.
	
	The Kerr metric is therefore of utmost importance in astrophysics, and the aspect of a Kerr black hole has been the subjects of innumerable papers, e.g., 
	\cite{viergutz93,fanton97,falcke00,beckwith05,james15}. However most
	deal with the case of astrophysically realistic  black holes which were born at some epoch in the past.
	Before the black hole formation, space-time structure~(neglecting the expansion
	of the Universe) is that of a Minkowski space, and after black hole formation
	it becomes that of a spacetime with a future event horizon, separating the interior
	and the exterior of the black hole. But the Kerr metric is actually even
	richer if one considers its maximal analytic extension. Contrarily to
	astrophysical black holes, the maximal analytic extension of the Kerr metric
	describes eternal black holes, or more precisely eternal wormholes that connect
	several distinct universes and whose structure is much more complicated than that of the canonical Morris-Thorne wormholes~\cite{morris88}. Moreover there are several reasoons to doubt that such configurations are stable with respect to tiny ravitational perturbations~\cite{penrose68}. It is therefore clear that such extension is highly
	unlikely to exist in our Universe, however it possesses a number of fascinating
	properties whose study is barely sketched in a only very small number of papers, such as Ref.~\cite{beckwith05}.
The aim of this paper is to
	explore more deeply some of the mathematical features that arise in this metric.
	
	Indeed, the whole complexity of the Kerr metric does not appear immediately when one
	writes the metric in some coordinate system, since a metric is nothing more than
	a manifold which is locally isomorphic to ${\bf R}^4$, which is then locally
	isomorphic to Minkowski space. The Kerr metric is, on the other hand, a manifold
	with a much more complicated topological and causal structure than ${\bf R}^4$.
	Moreover, knowing the causal structure of a manifold does not suffice to
	describe the geodesic structures that exist in the manifold. We mean here that, for example,
	the fact that some part of spacetime are in an observer's past lightcone does
	not suffice for this region to be actually seen, that is, reached by null
	geodesics. But even when we know which regions can actually be seen, the
	knowledge of these geodesic structures is insufficient to give good insights on
	what this observer would actually see when travelling within this metric.
	Although this might seem provocative, we argue that one can hardly claim to
	understand what a metric is without having a good intuition of what it could
	look like from a visual point of view, and we think that the Kerr metric is a
	very good example of such claim. The aim of this paper is therefore to propose a
	numerical exploration of the maximal analytic extension of the Kerr metric from
	a visual point of view. 
	
	This paper is organized as follows. In \S\ref{sec_bkg}, we introduce the main mathematical quantities that are necessary in order to solve the geodesic equations. We use them in in
	\S\ref{views_1} to address a simple but unexplored case of the aspect of an astrophysical, realistic black seen from a very close distance. We then recall in \S\ref{sec_class} the main features of the
	maximal analytic extension of the metric as well as the coordinate systems that
	are needed to deal with horizon crossings. An important feature of the maximal
	analytic extension is the existence of bounded geodesics the properties of which
	are studied in \S\ref{sec_bound}. The visual consequences of them in some cases
	is shown in \S\ref{sec_dark}. Then, the most fascinating feature of the Kerr
	metric, that is an asymptotic region of negative gravity necessitates a new
	coordinate system and corresponding new geodesic equations, both of which are
	presented in \S\ref{sec_KScart}. With all this material we can now simulate in
	\S\ref{views_2} the journey of an observer all the way from a standard asymptotic
	region to a negative gravity region.
	
	\section{Background material}
	\label{sec_bkg}
	
	There exists a large number of review articles regarding the Kerr
	metric, e.g., \cite{visser07,teukolsky15,chandrasekhar83,oneill95}. Although at large amount of common materiel is present in each of these references~(and many others),
	each of them has its specificities. For example, Ref.~\cite{chandrasekhar83} addresses many mathematical aspects of the metric whereas Ref.~\cite{oneill95}
	focuses more on geodesics. 
	Still, these voluminous references do not exhaust such a vast subject,
	especially when one considers the maximal analytical extension of the metric,
	whose study is the aim of this paper. 
	
	Many coordinate systems can be used to study the Kerr metric,
	the most convenient one depending on the context. For example, the
	causal structure of the metric is more readily apparent when
	considering the Boyer-Lindquist coordinates, which are a
	generalization of the so-called Schwarzschild coordinates that are
	almost always used to introduce the Schwarzschild metric as well as
	the Reissner-Nordstr\"om one. However, if we need to handle horizon crossings, the Boyer-Lindquist coordinate system is
	inefficient. In the first part of this paper, we will not deal with horizon crossings, therefore we start by first using the Boyer-Lindquist coordinates.
	
	\subsection{Boyer Lindquist coordinate system}

Using the $(+---)$ metric signature
	convention, the non-zero metric coefficients in this coordinate system
	are written as
	\begin{eqnarray}
	g_{tt} & = & \frac{\Delta - a^2 \SN^2}{\Sigma} = 1 - \frac{2 M r}{\Sigma} , \\
	g_{t \varphi} & = & \frac{2 M r a \SN^2}{\Sigma} , \\
	g_{\varphi \varphi} & = & - \SN^2 \frac{(r^2 + a^2)^2 - \Delta a^2
		\SN^2}{\Sigma} = 
	- \SN^2 \left(r^2 + a^2 + \frac{2 M r a^2 \SN^2}{\Sigma} \right) , \\
	g_{rr} & = & - \frac{\Sigma}{\Delta} , \\
	g_{\theta \theta} & = & - \Sigma ,
	\end{eqnarray}
	where we have defined
	\begin{eqnarray}
	\SN & \equiv & \sin \theta , \\
	\CS & \equiv & \cos \theta , \\
	\Delta & \equiv & r^2 - 2 M r + a^2 , \\
	\Sigma & \equiv & r^2 + a^2 \cos^2 \theta .
	\end{eqnarray}
	This metric is asymptotically flat but the $r$ coordinate becomes
	timelike when $\Delta = 0$, which translates into the fact that there
	are two horizons situated at
	\begin{equation}
	r_\pm = M \pm \sqrt{M^2 - a^2} .
	\end{equation}
	The metric does not depend on the $t$ and $\varphi$ coordinates, which
	is a consequence of the fact that it is both stationary and
	axisymmetric. Consequently, for any geodesics there exists two obvious
	constants of motion, $E$ and $L_z$, defined as
	\begin{eqnarray}
	\label{def_E}
	E & \equiv & \pi_t = g_{t a} \dot x^a
	=   \left(1 - \frac{2 M r}{\Sigma} \right) \dot t
	+ \frac{2 M r a \SN^2}{\Sigma} \dot \varphi , \\
	\label{def_Lz}
	L_z & \equiv & - \pi_\varphi = - g_{\varphi a} \dot x^a
	=   \left(r^2 + a^2 + \frac{2 M r a^2 \SN^2}{\Sigma} \right) \SN^2 \dot \varphi
	- \frac{2 M r a \SN^2}{\Sigma} \dot t , 
	\end{eqnarray}
	where a dot corresponds to derivation of the corresponding coordinate
	with respect to one of the geodesic affine parameter. Furthermore, for
	any geodesic, the norm $\kappa$,
	\begin{equation}
	\kappa = g_{ab} \dot x^a \dot x^b , 
	\end{equation}
	is constant, and equal to either $1$ for timelike trajectories or $0$
	for null geodesics. Finally, there exists a far less obvious constant
	of motion, the Carter constant, $C$ which in this coordinate system is
	defined as
	\begin{equation}
	C = \pi_\theta^2 + a^2 \CS^2 \kappa 
	+ \left(a E \SN - \frac{L_z}{\SN} \right)^2 .
	\end{equation}
	
	\subsection{Equations of motion in Boyer-Lindquist coordinates}
	
	The constants of motions $E$ and $L_z$ allow to simplify the geodesic
	equations which, for coordinates $t$ and $\varphi$, can then be written
	as first order equations: one can invert the equations defining $E$
	and $L_z$ in order to obtain such closed forms for $\dot t$ and
	$\dot \varphi$. They are, respectively,
	\begin{eqnarray}
	\label{Delta_dot_t}
	\Delta \dot t & = &   \left( r^2 + a^2 + \frac{2 M r a^2 \SN^2}{\Sigma} \right)
	E
	- \frac{2 M r a}{\Sigma} L_z , \\
	\label{Delta_dot_varphi}
	\Delta \dot \varphi & = &   \left(1 - \frac{2 M r}{\Sigma} \right)
	\frac{L_z}{\SN^2}
	+ \frac{2 M r a}{\Sigma} E .
	\end{eqnarray}
	For the two other variables, the most compact form of the equations of
	motion reads
	\begin{eqnarray}
	\label{def_R}
	\Sigma^2 \dot r^2 & = & R(r)
	\equiv \left( (r^2 + a^2) E - a L_z \right)^2 - \Delta( \kappa r^2 + C) , \\
	\label{def_Th}
	\Sigma^2 \dot \theta^2 & = & \Theta(\theta)
	\equiv C - a^2 \CS^2 \kappa - \left(a E \SN - \frac{L_z}{\SN} \right)^2 .
	\end{eqnarray}
	These two equations are not usable in practice as they do not allow to
	notice a possible change of sign in $\dot r$ or $\dot \theta$. In
	order to do so, one needs to consider the time derivative of these,
	which give
	\begin{eqnarray}
	\Sigma^2 \ddot r & = &   \frac{R'}{2} - 2 r \Sigma \dot r^2
	+ 2 \Sigma a^2 \CS \SN \dot \theta \dot r , \\
	\Sigma^2 \ddot \theta & = &   \frac{\Theta'}{2}
	+ 2 a^2 \Sigma \CS \SN \dot \theta^2
	- 2 r \Sigma \dot r \dot \theta ,
	\end{eqnarray}
	where the primes denote a derivative with respect to $r$~(for $R$) and
	$\theta$~(for $\Theta$). These derivatives can be explicitly written
	as
	\begin{eqnarray}
	\label{def_Rp_2_v0}
	\frac{R'}{2}
	& = &   2 r E \left[ (r^2 + A^2) E - a L_z \right]
	- (r - M) (C + \kappa r^2) - \Delta \kappa r , \\
	\label{def_Thp_2}
	\frac{\SN}{\CS} \frac{\Theta'}{2}
	& = &   a^2 \SN^2 \kappa
	- \left(a E \SN - \frac{L_z}{\SN} \right)
	\left(a E \SN + \frac{L_z}{\SN} \right) .
	\end{eqnarray}
	Apart from the case of polar trajectory where the term $1 / \SN$ may
	diverge, this set of equations is regular everywhere outside the outer
	horizon at $r = r_+$.
	
	\section{Simulated views, 1st part}
	\label{views_1}
	
	The above equations are sufficient to address the issue of a Kerr
	black hole aspect for seen from the point of view of any observer
	outside the horizon. Almost all the literature dealing with raytracing
	in the Kerr metric focuses on the actual aspect of such black hole
	seen by a distant observer but almost none deals with the point of
	view of an observer much closer to the black hole. However, although
	this was not the initial aim of our work, we found that several
	interesting features arise when considering such observers, and we
	shall devote this Section to this issue. More specifically, whether or
	not the special case $a = M$ deserves attention in the study of the
	Kerr metric has been an open matter in the existing literature. For
	example, Ref.~\cite{chandrasekhar83} argues that it does not. Our work
	tentatively brings some elements to this debate as we have found that
	several strange phenomena can arise in this case and were so far overlooked in previous studies~\cite{james15}.
	
	Going from the geodesic equations to full-fledged views of a celestial
	sphere is essentially independent of the metric one is considering, at
	least as long that it is asymptotically flat so that we can consider a
	static celestial sphere lying at infinity. The technical details have
	already been explained in a previous paper~\cite{riazuelo19a}, so that
	we shall not give more details here.
	
	\subsection{Shape of the black hole silhouette for an equatorial observer}
	
	The case we have focused on is what we think to be the most extreme
	situation that can occur outside a rotating black hole horizon. It
	deals with an observer orbiting along a circular orbit around and
	close to an extremal Kerr black hole in the equatorial plane. The
	choice of an orbiting instead of a static observer is of course
	motivated by the fact that no static observer can lie within the
	ergosphere, i.e., close to the black hole. We therefore consider the
	case of an observer along a circular, equatorial geodesic and we
	impose the trajectory to be prograde as it allows smaller orbital
	radii. When the observer is at large distance, such a configuration is
	very close to that of a static observer since the Lorentz transform
	that allows to go from a moving to a static observer is close to
	identity, so that the resulting Doppler or aberration distortions are
	negligible. A well-known result regarding the shape of an extremal
	black hole silhouette it that from the point of view of a distant,
	static observer, it looks like a disk with a flattened side. Assuming
	that the black hole is spinning counterclockwise from the point of
	view of an equatorial observer, the flattened side is on the
	left. This asymmetry can more or less intuitively understood by the
	fact that there exist equatorial null geodesics, whose minimal
	coordinate radii are very different whether they and prograde or
	retrograde. These radii are given by $r_{\rm p}$ and $r_{\rm i}$ of
	Eq.~(\ref{def_r_eq}) and respectively tend to $M$ and $4 M$ when $a
	\to M$. Consequently, null equatorial geodesics performing a close
	flyby near the black hole can also reach values close to $M$ or $4 M$
	at their closest approach. Consequently the ``prograde radius'' of the
	black hole~(i.e., the side on which flyby geodesics reaching the
	observer travel in the same direction as prograde circular geodesics
	ones) is much smaller than the ``retrograde radius''. The net result
	of these features is that the black hole silhouette is asymmetric, it
	is flattened along the prograde side. However this silhouette is still
	convex but there is no obvious reason that it should be the case, nor
	that it should be the case for an observer standing at a finite
	distance. We investigate this question in the next paragraphs.
	
	In order to simulate images, we need to decide what the celestial
	sphere looks like. We~(rather arbitrarily) chose a celestial sphere
	corresponding to the sky seen from Earth we are familiar with. When
	computing a distorted celestial sphere, a different treatment must be
	applied of the celestial sphere itself and the pointlike images of the
	stars~\cite{riazuelo19a}. Since we are more interested in the
	distortions of the images, we decided to abruptly remove all the stars
	from the celestial sphere, which is taken as infrared sky as seen from
	the 2MASS survey~\cite{2mass06}.
	
	The images are now computed at a resolution of $3600\times 3600$
	pixels.  In order to reduce the number of views and to allow seeing
	every direction, pictures have been computed in fish-eye format that
	is adapted to digital planetarium, under the so-called DomeMaster
	format: the disk that fits inside the images correspond to exactly $2
	\pi$ steradians~(i.e., a half sphere). Aberration and Doppler effect
	are taken into account, although imperfectly for the latter, since
	only a bolometric correction is applied to pixel intensity~(intensity
	changes, but not hue). We shall show below and comment the views for
	decreasing values of the orbital radius $r$, which can be as small as
	$M$.
	
	\subsection{$r = 6 M$} 
	
	\begin{figure}[h]
		\centerline{\includegraphics*[width=4.8in]{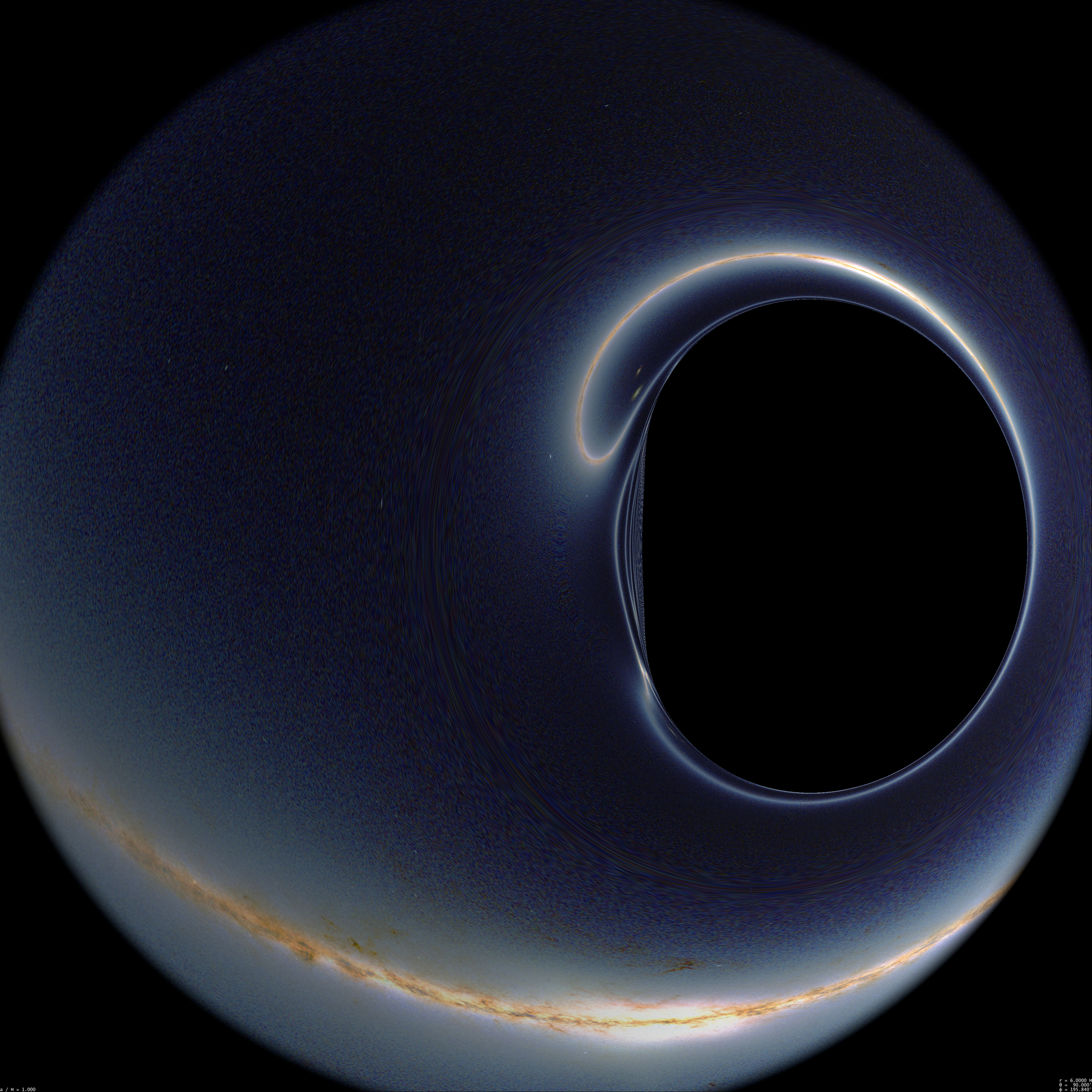}}
		\caption{$r = 6 M$, overall black hole silhouette.}
		\label{fig_r6m}
	\end{figure}
	
	We first show in Fig.~\ref{fig_r6m} the aspect of the black hole
	silhouette at moderate distance, i.e., coordinate distance $r = 6 M$,
	which is the innermost stable circular orbit in the Schwarzschild
	metric. The silhouette shape is qualitatively similar to that at
	infinite distance, although it is more clearly elongated along the
	polar direction than along the equatorial one. (This was already the
	case at large distance, but far less obvious to notice by eye.)
	Another interesting feature is that although we expect to have an
	infinite number of multiple images of the celestial sphere around the
	black hole silhouette, these images are easier to see on the prograde
	size in the sense that their angular separation is larger. In the
	rather featureless celestial sphere image we chose, the multiple
	images are those of the Milky Way disk.
	
	\subsection{$r = 1.5 M$}
	
\begin{figure}[ht]
		\centerline{
			\includegraphics*[width=3.2in]{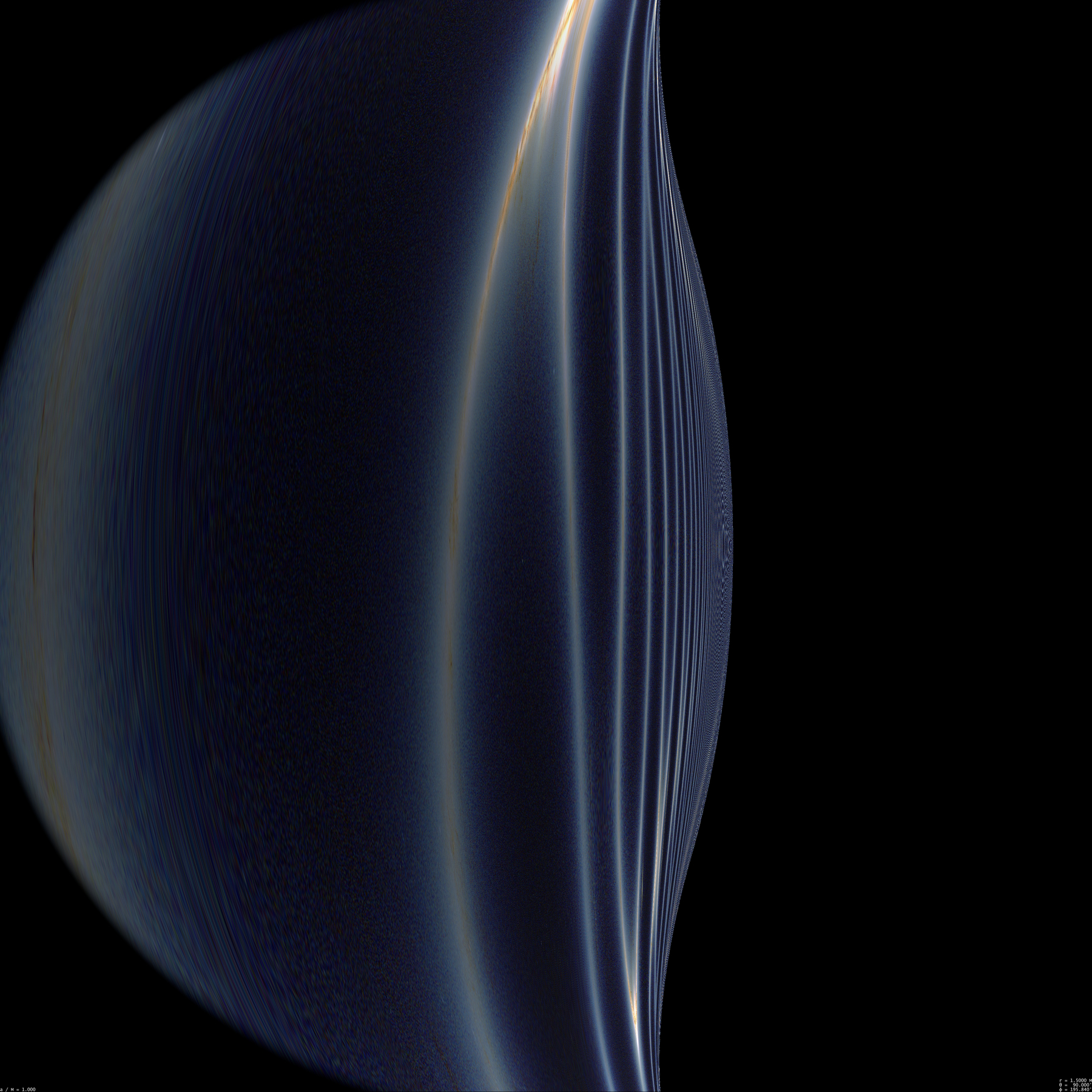}
			\includegraphics*[width=3.2in]{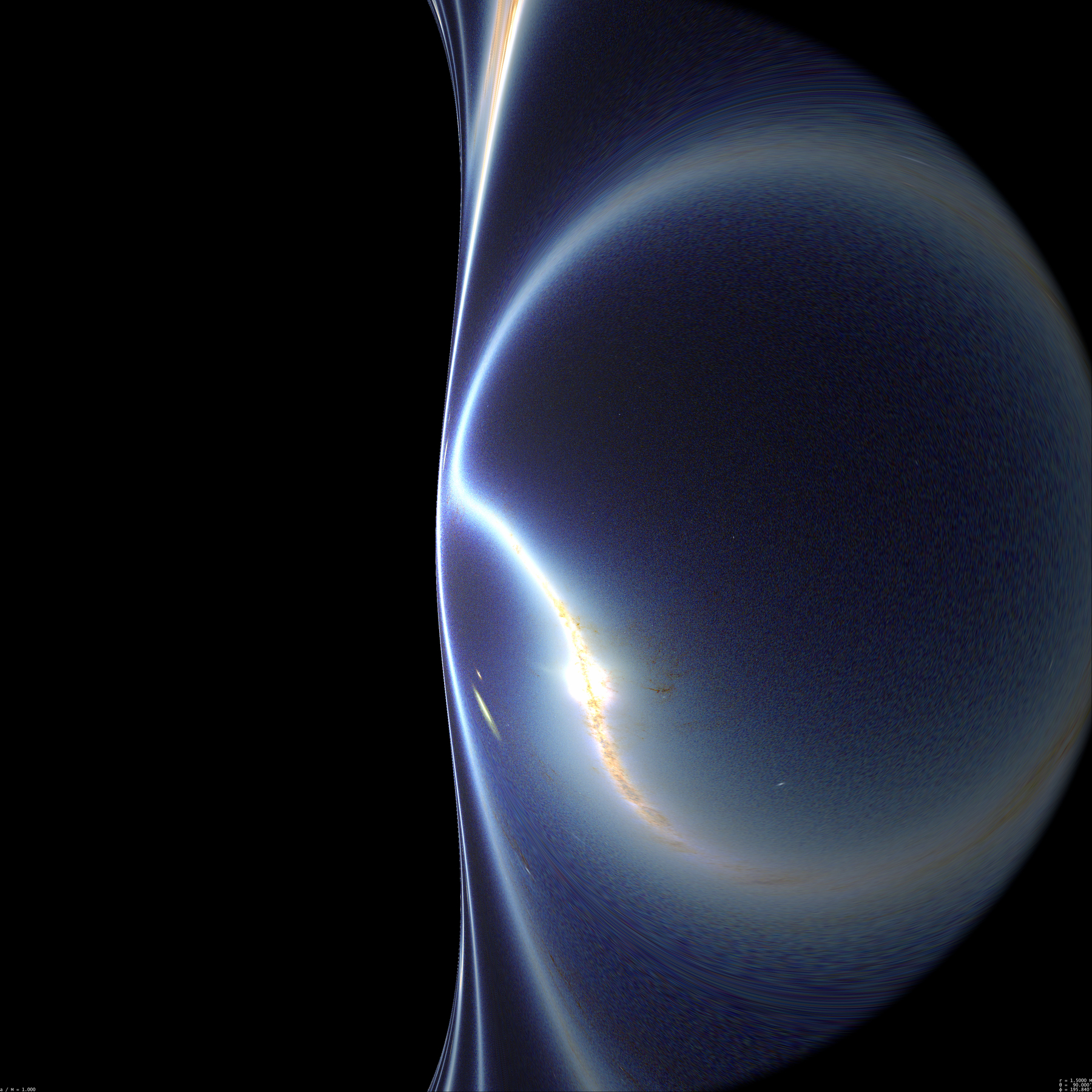}}
		\caption{$r = 1.5 M$, rear view~(left) and front view~(right).}
		\label{fig_r15m}
	\end{figure}

	Fig.~\ref{fig_r15m} shows what happens at $r = 1.5 M$. Below $r = 2
	M$, the observer lies in the ergoregion and the shape of the black
	hole silhouette significantly changes~(although in a regular and
	continuous manner). It is no longer convex, both along the prograde
	side and, more unexpectedly, along the retrograde side, taking overall
	a peanut-like shape. Another interesting feature is that, just as in
	the previous figure, the multiple images in the prograde side are
	separated by a much larger angle that on the retrograde size.
	
	Note also that determining the direction toward which the observer is
	heading is not easy from now on since we are beyond the static
	limit. The concept of direction makes sense in a static metric: as
	compared to a static observer of four-velocity $u_{\rm stat} =
	\partial / \partial t$, another observer's arbitrary four-velocity
	$v^a$ can always be written $v^a = \gamma (u_{\rm stat}^a + \beta n^a)$,
	where $n^a$ is a unit spacelike vector orthogonal to $u_{\rm stat}^a$
	and $\gamma = 1 / \sqrt{1 - \beta^2}$ is the Lorentz factor $v_a u^a_{\rm stat}$. In such a
	configuration, one can state that the second observer is heading
	toward direction $n^a$, but the same procedure can no longer be
	implemented within the ergoregion~(and, of course, within the
	horizon). From now on, the choice of the ``front'' and ``rear''
	direction is rather arbitrary. We shall adjust our images so that the
	center of the ``front'' image corresponds to the most blueshifted part
	of the celestial sphere, keeping in mind that even outside the
	ergoregion~(and even in the Schwarzschild case) this does not
	correspond to the direction of motion as we just have defined it.
	
	\subsection{Going deeper in the ergosphere}
	
	Going deeper in the ergoregion, several effects arise.
	
	Firstly the silhouette is still peanut-shaped but in an increasingly
	asymmetric way. The concave part along the prograde side is quite
	close to a circular arc, whereas on the retrograde side it becomes
	increasingly pinched.
	
	Secondly there are more multiple images that are visible on the
	prograde side than on the retrograde one but their relative positions
	are quite different. On the prograde side, their separation decreases
	as one gets closer to the black hole silhouette. This is indeed what
	one could expect given what happens in the Schwarzschild case, however
	the big difference here is that the separation between two multiple
	images decreases rather slowly~(as opposed to exponentially in the
	Schwarzschild case, see, e.g., Ref.~\cite{chandrasekhar83}).
	
	Thirdly, on the retrograde side, the multiple images get organized
	along a very regular pattern. For what we use as an image here~(the
	Milky Way disk), they appear as circles~(or water droplets) which are tangent at the the
	``pinch'' of the black hole silhouette.
	
\begin{figure}[ht]
		\centerline{
			\includegraphics*[width=3.2in]{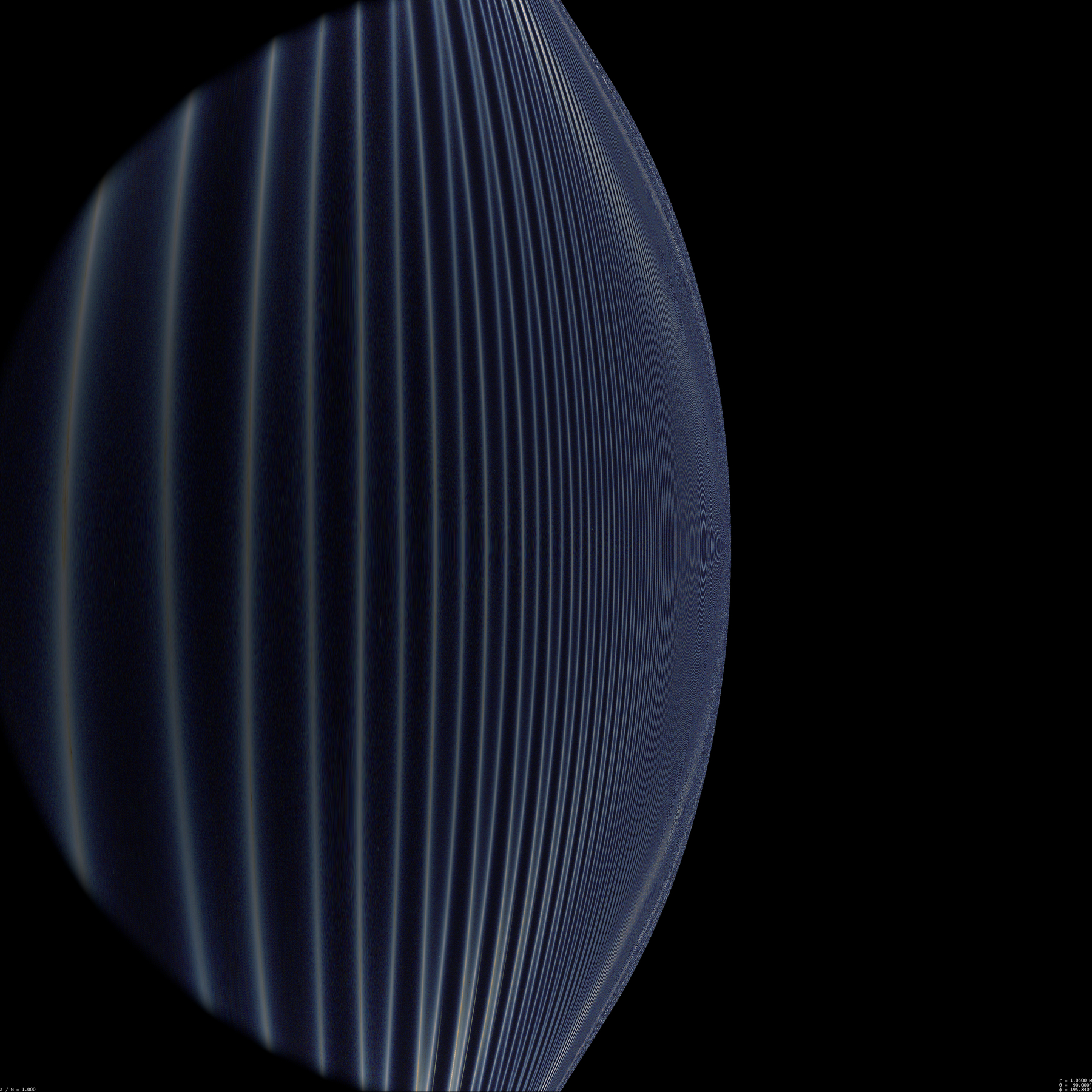}
			\includegraphics*[width=3.2in]{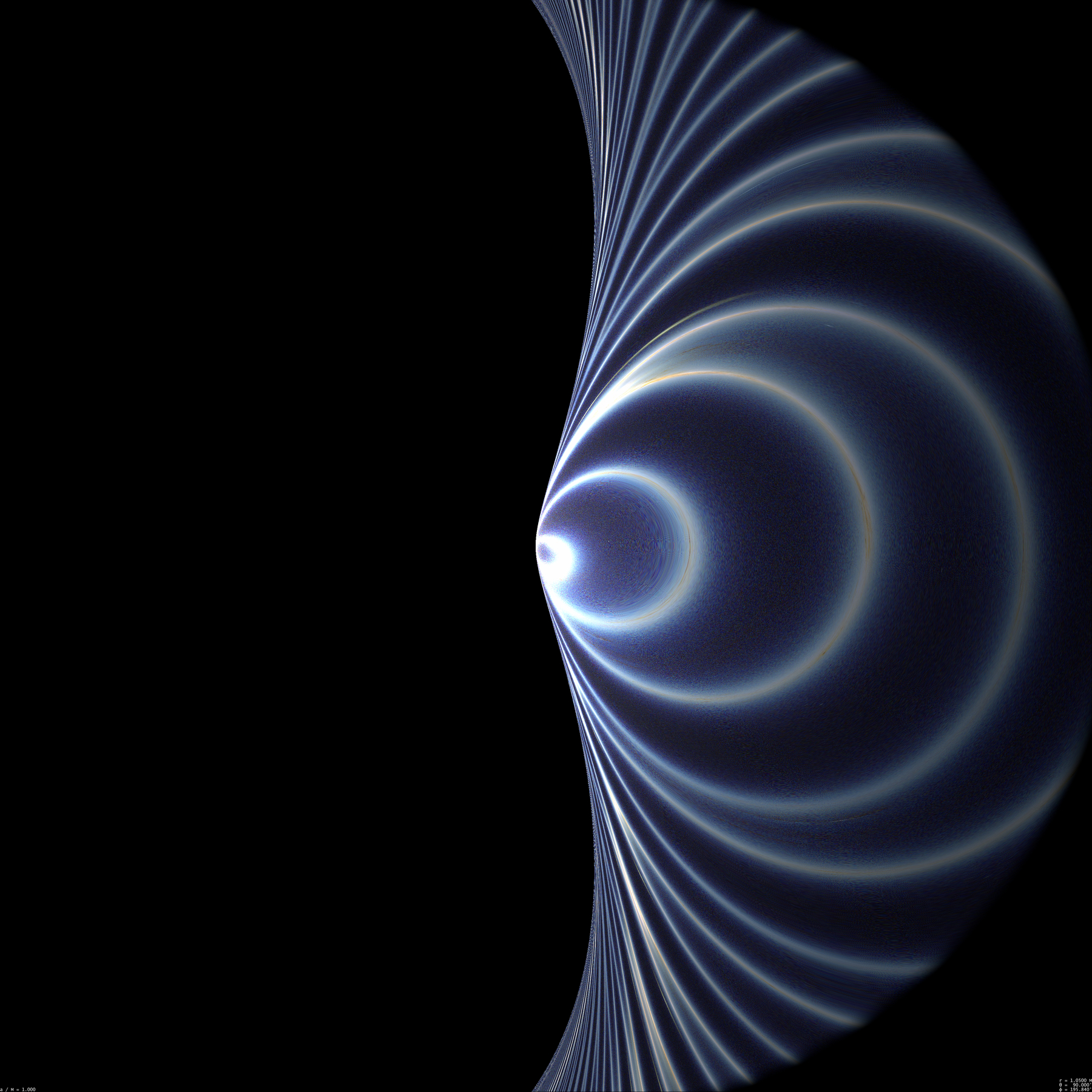}}
		\caption{$r = 1.05 M$, rear view~(left) and front view~(right).}
		\label{fig_r105m}
	\end{figure}

	Figure~\ref{fig_r105m} illustrates these three effects as they are see
	at $r = 1.05 M$
	
	\subsection{Skimming over the horizon}
	
\begin{figure}[ht]
		\centerline{
			\includegraphics*[width=3.2in]{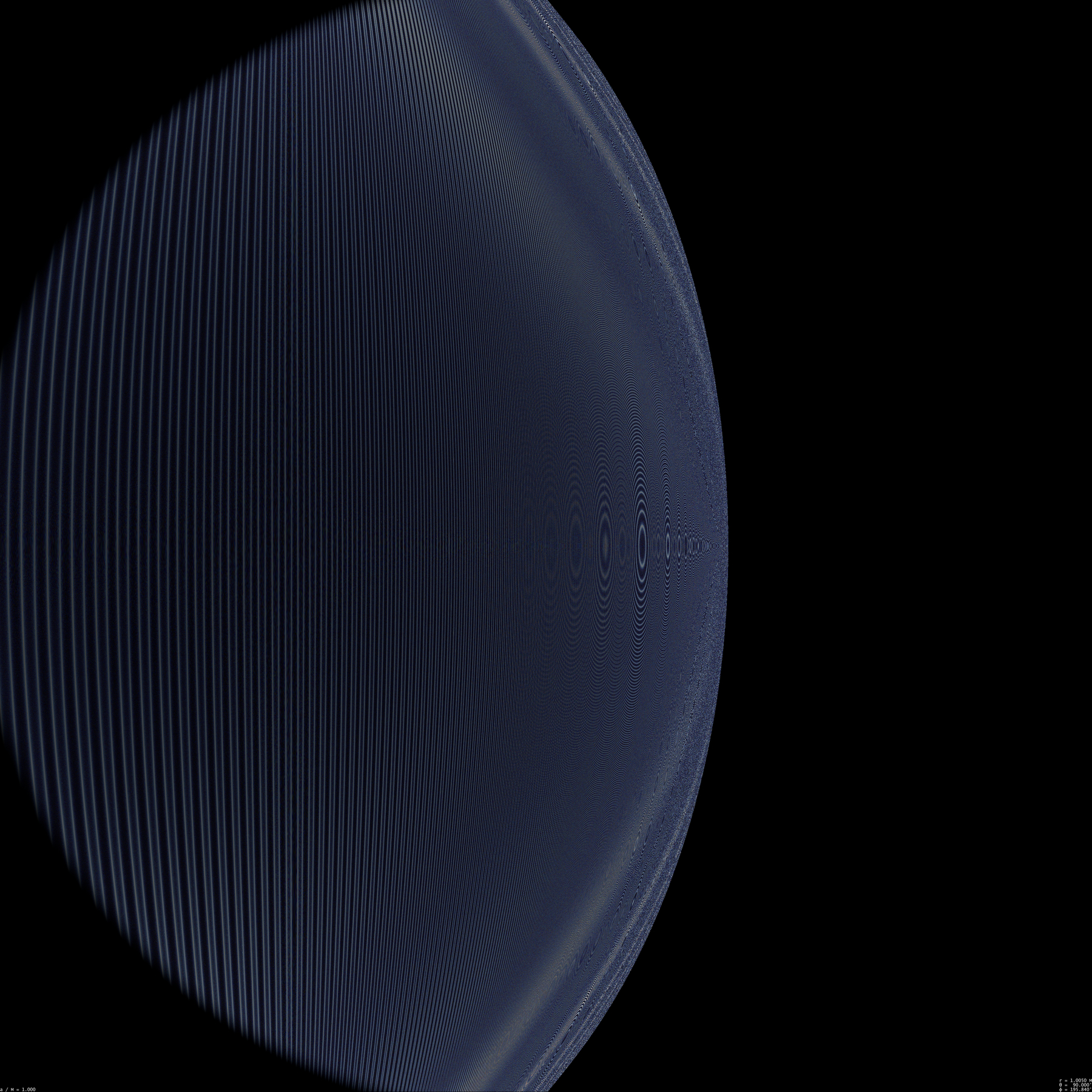}
			\includegraphics*[width=3.2in]{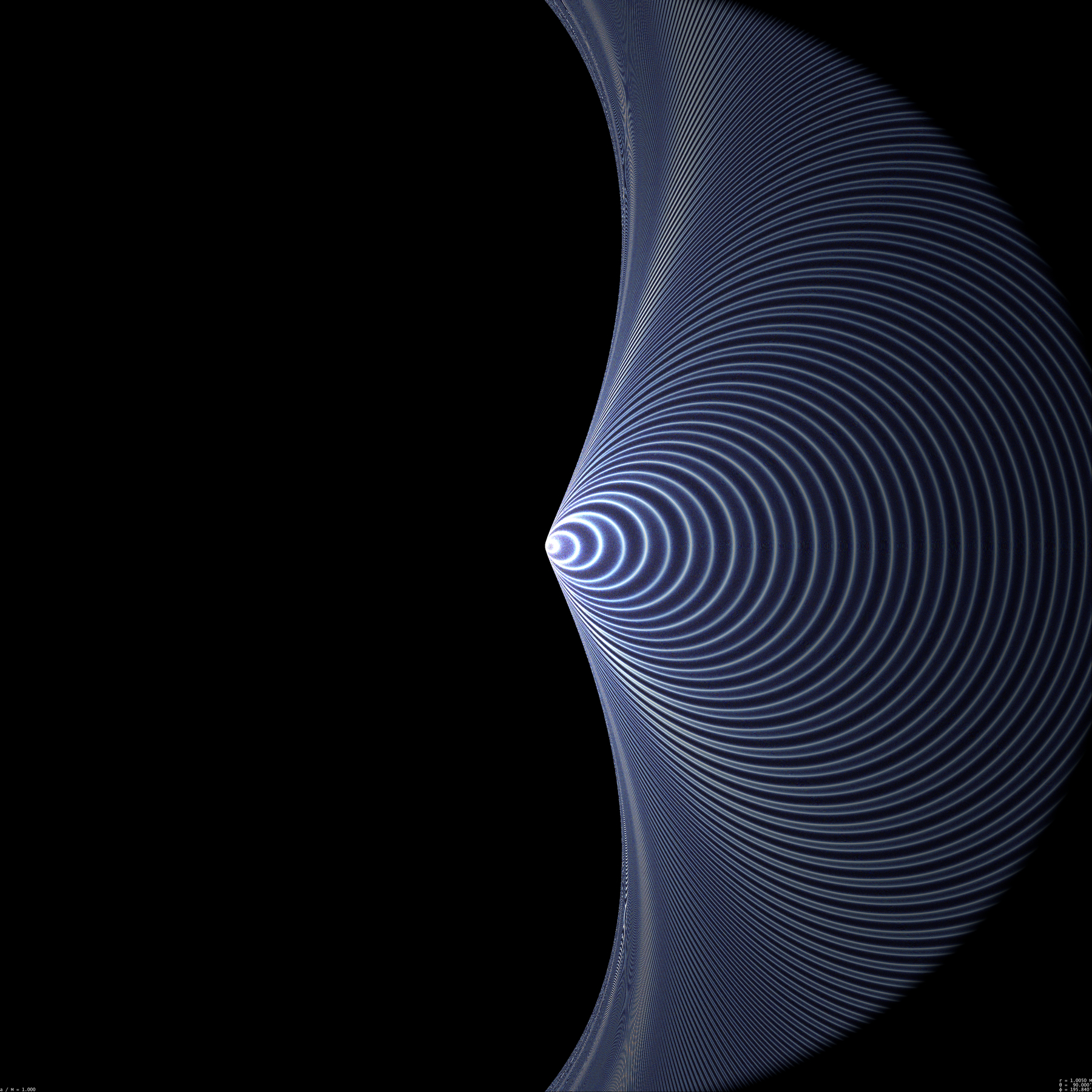}}
		\caption{$r = 1.005 M$, rear view~(left) and front view~(right). The
			Moir\'e-like effects on the left image are not physical features but
			instead limitation due to the insufficient resolution of the
			computer-generated view.}
		\label{fig_r1005m}
	\end{figure}

	In the Schwarzschild metric, the aspect of the black hole does not
	drastically change when a freely-falling observer approaches the
	horizon, but this is not what happens here in the Kerr metric,
	although in a slightly different context since we are considering a
	series of orbiting observer along similar circular orbits. The view at
	$r = 1.005 M$~(Fig.~\ref{fig_r1005m}) is qualitatively close to that
	of Fig.~\ref{fig_r105m} except that the patterns we mentioned on the
	retrograde side are even more regular, with something like ten times
	more images on the retrograde side. This may suggest that the spacing between the images scales as $M / (r - M)$,
	but this is merely a conjecture. The number of images on the
	prograde side is not so high that one is limited by the screen
	resolution.
	
\begin{figure}[ht]
		\centerline{
			\includegraphics*[width=3.2in]{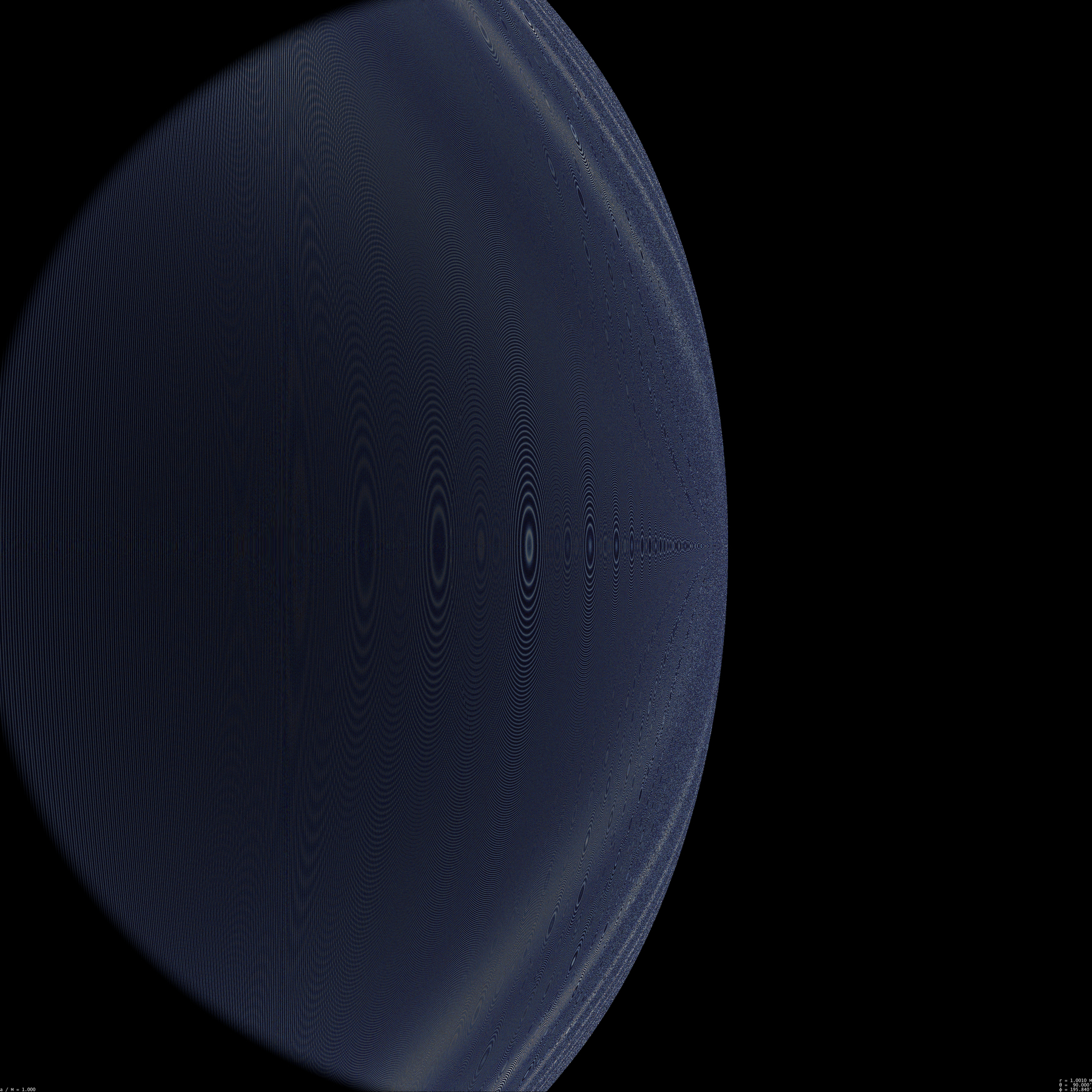}
			\includegraphics*[width=3.2in]{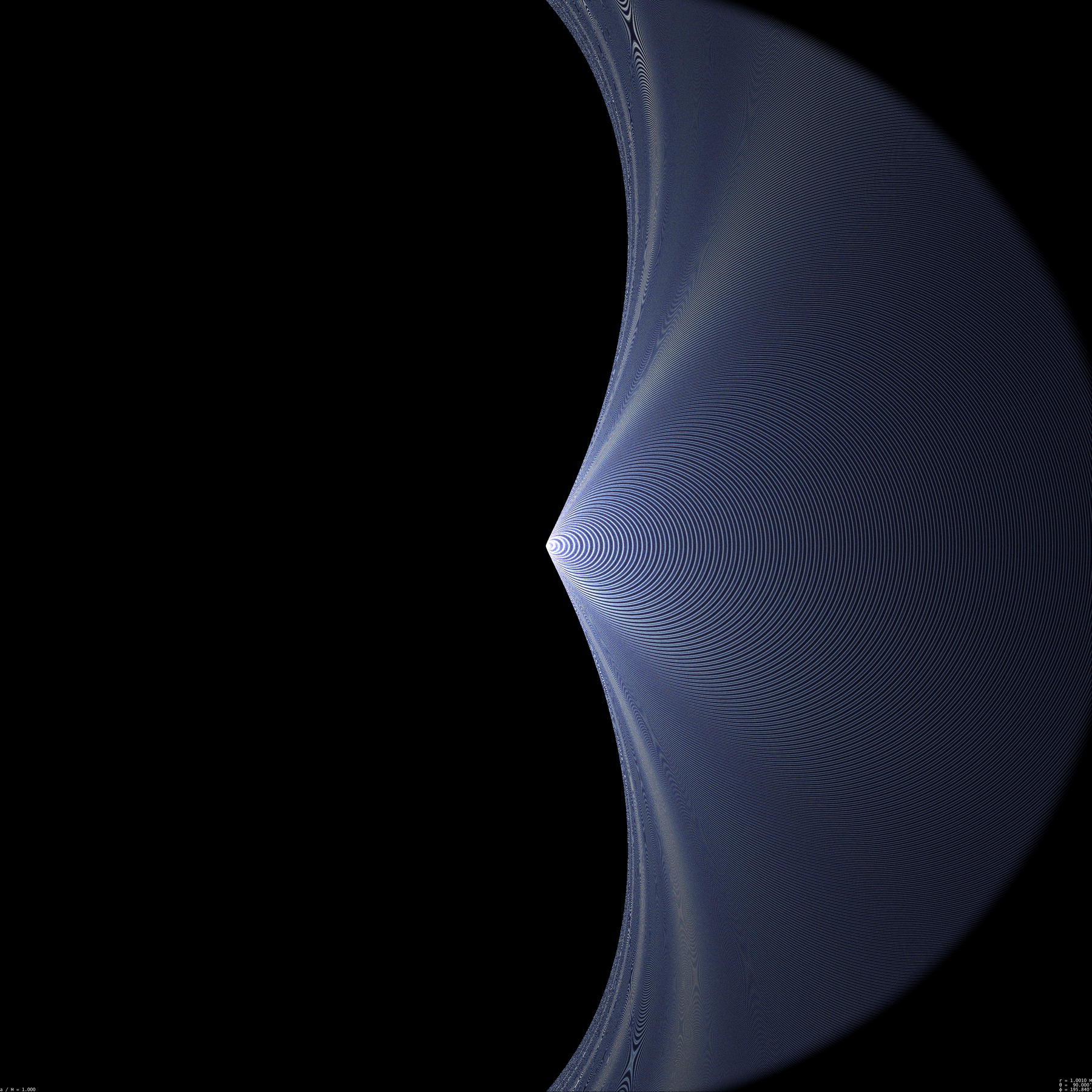}}
		\caption{$r = 1.001 M$, rear view~(left) and front view~(right). At
			this resolution, only the front view remains~(barely) accurate.}
		\label{fig_r1001m}
	\end{figure}

	The visual limit of what can be obtained with $1800 \times 1000$
	hemispheric view is reached around $r = 1.001 M$ on the retrograde
	side(see Fig.~\ref{fig_r1001m}). For the prograde side, the limit
	depends on where on is looking since the spacing between multiple
	image is less regular, but the limit is obtained for larger $r$~(see
	Fig.~\ref{fig_r1005m}).
	
\begin{figure}[ht]
		\centerline{
			\includegraphics*[width=4.8in]{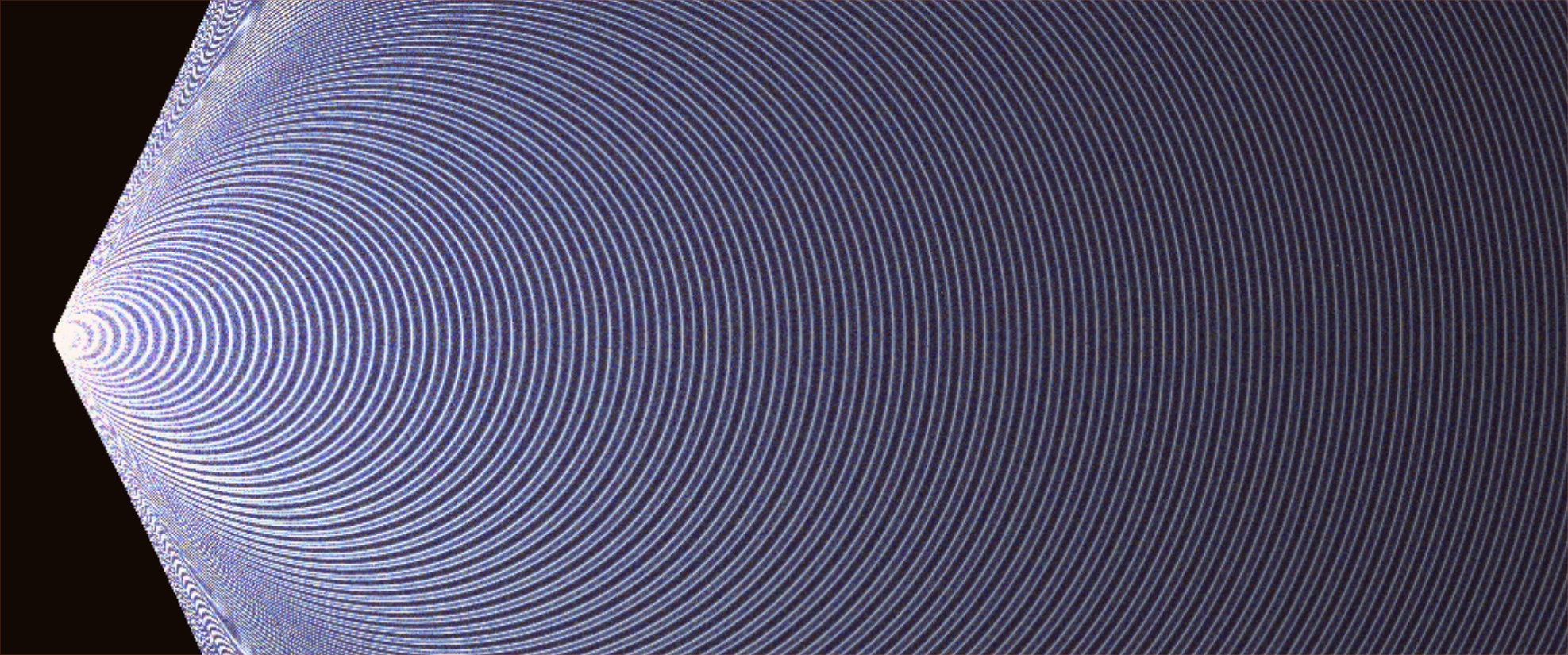}}
		\caption{Snapshot of a 3600 pixel wide hemispheric view at $r =
			1.0005 M$. This view covers a $50^\circ$ wide field, i.e. with an
			effective screen resolution of $3$~arcmin, not much worse than
			human eye resolution.}
		\label{fig_r10005mz}
	\end{figure}

	If we improve the picture resolution, so does the limit at which one
	can see details~(Fig.~\ref{fig_r10005mz}). However a larger number of
	technical difficulties arise at some point which all come from the
	combination of the observer's position and velocity. When $r$ is very
	close to the horizon, there exists some risk that the Boyer Lindquist
	coordinate system becomes inappropriate. This is true, unfortunately,
	this is also the case for the Kerr-Schild coordinates. Indeed, one
	might think that since the observer is looking in all
	directions~(since we compute two hemispheric views), many of the
	geodesics make a significant angle with respect to the
	horizon. However, the observer we are considering is very specific
	here. As compared with a freely falling observer with zero velocity
	and angular momentum, it is necessary to perform a huge Lorentz boost
	to pass from this first fiducial observer's reference frame to that of
	the observer that actually sees the celestial sphere. Because of this
	most of the directions of the circularly orbiting observer sees
	correspond to a very thin bundle of directions from the point of view
	of the freely-falling observer, and this bundle of direction
	correspond to photons that reach the observer with a very low $\dot
	r$. For this reason, the coordinate change from Boyer-Lindquist to
	Kerr-Schild coordinate does {\em not} affect the $t, \varphi$
	coordinates: we have by construction $\dot T \sim \dot t$ and $\dot
	{\tilde \varphi} \sim \dot \varphi$, which are both very large. Such
	situation is numerically very unstable and has a very bad impact on
	the computational time for each geodesic of the screen~(but a very
	small number of these). Consequently, the CPU time increases very
	rapidly when the circularly observer's orbital radius gets closer and
	closer to the horizon, $r_{\rm hor} = M$.
	
	We stop here our numerical investigation of the exterior aspect of the
	Kerr metric. Several of the features we discovered here, most notably
	the regular image spacing on the retrograde side, as well as the
	pinched aspect of the black hole silhouette from this side as well,
	obviously deserve a rigorous derivation but we leave this for a future
	work.
	
	\section{Classifying geodesics}
	\label{sec_class}
	
	The ultimate aim of this paper is to visualize the maximal analytic
	extension of the Kerr metric. Wherever the observer lies, it will
	intersect geodesics coming from some past horizons and traveling from
	different asymptotic regions than that it originates from. We
	therefore first need to address which regions can be seen as a
	function of the region the observer lies in.
	
	\subsection{Reminder on the maximal analytic extension of the Kerr metric}
	\label{sec_carter}
	
	The Carter-Penrose diagram of the Kerr metric has been described
	decades ago by several authors, see, e.g., \cite{hawking_ellis73}
	and~\cite{boyer_lindquist67} for the historical reference. We shall
	first consider an observer lying in some asymptotic region, i.e., a
	region of positive $r$ and outside the outer horizon~(that is, $r >
	r_+$). This region is bounded by the black hole outer horizon at $r =
	r_+$. In a causal diagram, it is represented by a diamond-shaped
	region whose edges are inclined at $\pm 45^\circ$ with respect to
	horizontal/vertical directions. The null ingoing and outgoing
	directions~($E^a$ and $S^a$, respectively) are defined by the null
	vectors such that $\kappa = C = 0, L_z = a \SN^2 E$ whose components
	are therefore, up to some arbitrary constant,
	\begin{eqnarray}
	\label{def_E_a_S_a}
	S^a = \left(\begin{array}{c} \displaystyle \frac{r^2 + a^2}{\Delta} \\  1 \\ 0
	\\  \displaystyle \frac{a}{\Delta} \end{array}
	\right) & , &   E^a = \left(\begin{array}{c}  \displaystyle \frac{r^2 +
		a^2}{\Delta} \\ - 1 \\ 0 \\ \displaystyle \frac{a}{\Delta} \end{array} \right)
	\end{eqnarray}
	Then, switching from $r$ to the 	tortoise coordinate $r^*$ such that $\ddd r^*
	= \frac{r^2 + a^2}{\Delta} \ddd r$,
	the hypersurface spanned by $E^a$ and $S^a$ is causally equivalent to that of a
	two-dimensional Minkowski spacetime in each intervals of $r$ where $r^*$ is
	regular, that is in the three intervals $] r_+, \infty [$, $]r_-, r_+ [$, $]-
	\infty, r_- [$. Consequently, in a Carter-Penrose diagram, all these regions are
	diamond-shaped. Considering the region $r > r_+$, the lower right edge of its
	patch corresponds to
	past null infinity ($t - r \to -\infty$, or, equivalently, $t - r^* \to
	-\infty$), the upper right edge to future
	null infinity ($t + r \to \infty$). The upper left edge corresponds to
	the future outer horizon ($r = r_+, t \to \infty$, or $t - r^* \to \infty$),
	and the lower left
	edge to the past outer horizon ($r = r_+, t \to -\infty$). In this
	region, $\Delta$ is always positive, so that $r$ is a spacelike
	coordinate. Conversely, $t$ is almost everywhere a future oriented
	timelike coordinate, except near the black hole when $g_{tt} \propto
	\Delta - a^2 \SN^2$ can become negative for sufficiently small
	$r$~(unless $\theta = 0, \pi$), a configuration which forms the outer
	ergoregion. The fact that $t$ is both timelike and future-oriented
	means that null geodesics coming from or heading toward null infinity
	have a positive $E = \pi_t$. Geodesics with negative $E$ can exist in
	this region, however they cannot go further~(in term of $r$
	coordinate) than the limit of the outer ergoregion. We shall label
	this region, including the ergoregion,~$1$.
	
	The next patch of the Carter-Penrose diagram is the inter horizon
	region, which we call region~$2$. There, $\Delta < 0$, so that $r$ is
	a timelike coordinate. Since this coordinate can only decrease when
	one crosses the outer horizon, $r$ is past-oriented. Region~$2$ is
	adjacent to region~$1$ through its lower right edge. Its lower left
	edge is adjacent to another asymptotic region which has the same
	structure as region~$1$, except that $t$ is past-oriented. We shall
	label it region~$3$. This second asymptotic region already exists in
	both the Schwarzschild and Reissner-Nordström metric, and so does
	region~$2$ except that it is bounded from above by the singularity in
	the Schwarzschild case\footnote{There does not seem to be any consensus
		on how to label these patches. For example,
		Ref.~\cite{hawking_ellis73} use Roman I for both our regions~$1$ and
		$3$, and II for our $2$ and $4$~(see below), whereas as
		Ref.~\cite{chandrasekhar83} use I, II, III, I', II', III' for our
		$1$, $2$, $5$, $3$, $4$, $6$ $\mod 8$, respectively.}.
	
	Region~2 upper edges correspond to $r = r_-$ and $t \to \pm \infty$
	and connect to two new patches. These both possess an inner ergoregion
	and both share similar properties, except that in one patch, $t$ is
	future-oriented outside the ergoregion, whereas it is past-oriented in
	the other. Since radial null geodesics from region~1 travel from right
	to left in the Carter-Penrose diagram, it is the left $r < r_-$ patch
	whose $t$ coordinate is future oriented out of its ergoregion. We
	shall label it region~$5$, its right counterpart being region~$6$.
	
	Each regions~$5$ and $6$, which are diamond-shaped, can further be
	split into two parts, one with $r > 0$, and one with $r < 0$, which is
	another asymptotic region, which we shall call $7$ and $8$,
	respectively. Since $t$ is future oriented in region~$7$, geodesics
	traveling there have positive $E$, whereas all geodesics of region~$8$
	coming from past null infinity have a negative $E$. Apart from the $r
	< 0$ part, regions~$5$ and $6$ also exist~(without ergoregion) in the
	Reissner-Nordstr\"om metric, where they are bounded from one side by
	the then uncrossable pointlike singularity at $r = 0$.
	
	Regions~$5$ and $6$ are bounded from above by a second inter-horizon
	region which has the same properties as region~$3$, except that $r$ is
	now future oriented. We shall label this region~$12$. The two upper
	edges of region~$12$ correspond to $r = r_+$, $t \to \pm \infty$ then
	connect to two new asymptotic region which we shall label $9$~(to the
	right) and $11$~(to the left). By a similar reasoning as above, we
	know that outside their respective ergoregions, $t$ is future-oriented
	in region~9 and past-oriented in region~11.
	
\begin{figure}[h]
		\centerline{\includegraphics*[clip,trim=275 50 150
			50,width=3.2in]{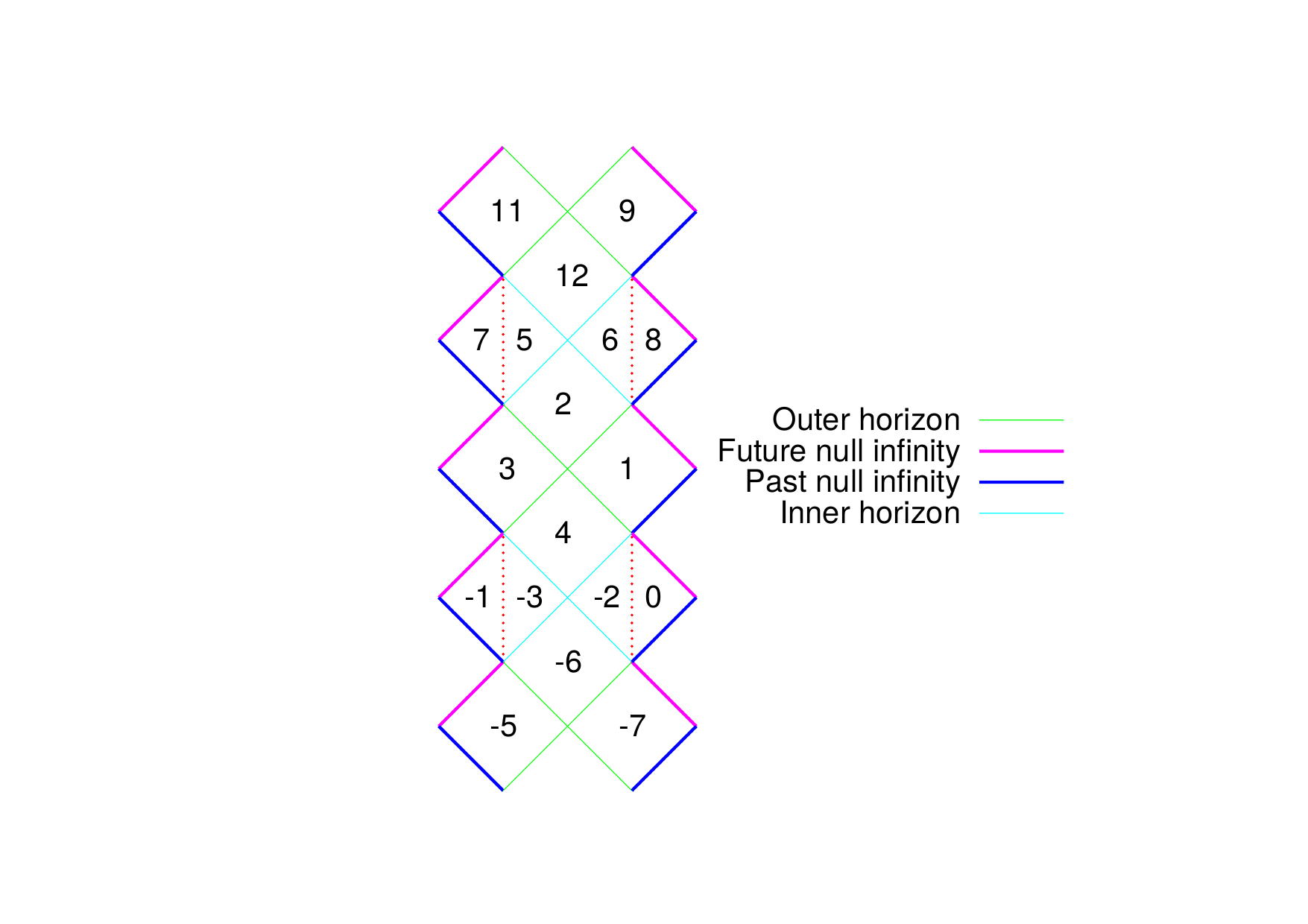}}
		\caption{Carter-Penrose diagram showing the causal structure of the
			full analytical extension of the Kerr metric. A bit more than two
			blocs of the diagram is shown here, although the full diagram is
			made of an infinite series of such blocks. Normal asymptotic
			regions~(i.e., $1$, $3$, and $1 + 8 k$, $3 + 8 k$ for any integer
			$k$) are bounded on one side by the outer horizon and from the other
			side by past and future null infinity. Other asymptotic region with
			negative $r$ exist as well, where wormhole gravity is
			negative. Those are regions~$7 + 8 k$, $8 k$. Normal asymptotic
			regions are partially filled with the outer ergoregion, whereas
			inner regions~(labeled $5 + 8 k$, $6 + 8 k$), which are bounded by
			$r = r_-$ and $r = 0$ also contain an inner ergoregion.}
		
		\label{fig_carter}
	\end{figure}

	The full analytic extension of the Kerr metric is then an infinite
	tower of six diamond-shaped patches whose labels will be deduced from
	those of the neighboring block by adding or subtracting $8$. The
	labels we choose are summarized in Figure~\ref{fig_carter}.
	It is also to be noted that dubbing this diagram as a causal one is
	slightly misleading as a small patch of the $r < 0$ region allows for
	causality violations~\cite{carter68} all the way to the neighboring
	$0 < r < r_-$ region by using some specific displacement along the
	$\varphi$ coordinate. This does not affect the discussion that follows
	nor the appearance of asymptotic regions seen from there, at least as
	long as we assume that the past null infinities of those regions can
	be considered as static, which is what we shall do there.
	
	\subsection{Dealing with horizon crossings by using Kerr-Schild coordinates}
	
	When leaving region~$1$ to enter region~$2$, $t$ goes to infinity and
	is therefore not a well-behaved coordinate. It has to be replaced by
	another coordinate, $T$~(together with $\varphi$ being replaced by
	$\tilde \varphi$), which are the so-called Kerr-Schild coordinates.
	Starting from the Boyer-Lindquist coordinates, they are defined as
	\begin{eqnarray}
	\label{def_T}
	\dot T & \equiv & \dot t + \epsilon \frac{2 M r}{\Delta} \dot r , \\
	\label{def_tphi}
	\dot {\tilde \varphi} & \equiv & \dot \varphi + \epsilon
	\frac{a}{\Delta} \dot r ,
	\end{eqnarray}
	where $\epsilon$ is defined up to its sign, i.e., $\epsilon = \pm 1$.
	The appropriate choice of sign depends on the horizon crossing one is
	interested in, which we shall study in a few paragraphs. The important point
	is that {\em both} values of $\epsilon$ are mandatory, depending on
	the geodesics and the horizon crossings we have to deal with.
	Therefore, there are, in practice, two distinct Kerr-Schild coordinate
	systems, $(T_-, \tilde \varphi_-)$ and $(T_+, \tilde \varphi_+)$
	depending on the value of $\epsilon$. We shall however drop tho $+,-$
	subscript as long as it does not induce any confusion on which
	coordinate system is being used.  Before dealing with the choice of
	$\epsilon$, we shall give the new version of the equation of motion in
	the Kerr-Schild coordinates.
	
	Firstly, regarding the variable themselves, one can express $T$ and
	$\tilde \varphi$ as a function of $t$, $r$, and $\varphi$, $r$,
	respectively. The transformation depends on the value of $M^2 - a^2$.
	When this quantity is positive~(i.e., the metric describes a black
	hole), one has
	\begin{eqnarray}
	T & = & t + \frac{\epsilon M^2}{\sqrt{M^2 - a^2}}
	\ln \left| \frac{r - r_+}{r - r_-}\right|
	+ \epsilon M \ln \left|\frac{\Delta}{M^2} \right| , \\
	\tilde \varphi & = & \varphi + \frac{\epsilon a}{2 \sqrt{M^2 - a^2}}
	\ln \left| \frac{r - r_+}{r - r_-}\right| .
	\end{eqnarray}
	In the extremal, $a = M$ case, one has
	\begin{eqnarray}
	T & = & t + 2 \epsilon M \ln \left| \frac{r - M}{M}\right|
	- 2 \epsilon \frac{M^2}{r - M} , \\
	\tilde \varphi & = & \varphi - \frac{\epsilon M}{r - M} .
	\end{eqnarray}
	
	Secondly, in both cases, the metric in term of the Kerr-Schild
	coordinates can be rewritten in the coordinate singularity free form,
	\begin{eqnarray}
	\ddd s^2
	& = &   \ddd T^2 - \ddd r^2 - \Sigma \ddd \theta^2
	+ 2 \epsilon a \SN^2 \ddd r\; \ddd \tilde \varphi
	- (r^2 + a^2) \SN^2 \ddd \tilde \varphi^2 \nonumber \\ & &
	- \frac{2 M r}{\Sigma} 
	\left(\ddd T + \epsilon \ddd r - a \SN^2 \ddd \tilde \varphi 
	\right)^2 .
	\end{eqnarray}
	Although it is not obvious, the first part of this expression~(the one
	without $M$) corresponds to a Minkowski metric expressed in spheroidal
	coordinates, so that the whole metric is in fact of the form $g_{ab} =
	\eta_{ab} - A l_a l_b$, where $\eta_{ab}$ is a Minkowski metric and
	$l_a = (1, \epsilon, 0, - a \SN^2)$ is a null vector\footnote{This null vector
		is the contravariant form of either $E^a$ or $S^a$ of Eq.~(\ref{def_E_a_S_a}).},
	either with
	respect to $g_{ab}$ or $\eta_{ab}$, and $A$ is a function given by the
	simple form $A = 2 M r / \Sigma$. A very useful quantity that we shall
	use afterward is $\Ldot$, defined as
	\begin{equation}
	\label{def_Ldot}
	\Ldot \equiv l_a u^a
	= \dot T + \epsilon \dot r - a \SN^2 \dot {\tilde \varphi}  ,
	\end{equation}
	where $u^a$ is the four-velocity/four momentum of the geodesic we are
	considering.  Using this $\Ldot$, a large number of expressions can be
	rewritten by getting rid of either $\dot T$ or $\dot{\tilde
		\varphi}$. In particular, the constants of motions $E$ and $L_z$
	defined in Eqns.~(\ref{def_E},\ref{def_Lz}) can be rewritten
	\begin{eqnarray}
	\label{Tdot_Ldot}
	\Sigma E & = &   (\Delta - a^2 \SN^2) \dot T
	+ 2 M r a \SN^2 \dot {\tilde \varphi}
	- \varepsilon 2 M r \dot r 
	\nonumber \\ & = &
	\Sigma \dot T - 2 M r \Ldot
	\nonumber \\ & = &
	(\Sigma - 2 M r) \Ldot
	- \Sigma \epsilon \dot r
	+ \Sigma a \SN^2 \dot{\tilde \varphi} , \\
	\Sigma \frac{L_z}{\SN^2}
	& = &   \left(\Sigma (r^2 + a^2) + 2 M r a^2 \SN^2 
	\right) \dot {\tilde \varphi}
	- 2 M r a \dot T - \epsilon (2 M r + \Sigma) a \dot r 
	\nonumber \\ & = &  
	\left((r^2 + a^2)^2 - \Delta a^2 \SN^2 \right) \dot {\tilde \varphi}
	- 2 M r a \dot T - \epsilon (2 M r + \Sigma) a \dot r 
	\nonumber \\ & = & 
	\Sigma (r^2 + a^2) \dot {\tilde \varphi}
	- 2 M r a \Ldot - \epsilon a \Sigma \dot r .
	\end{eqnarray}
	Several combinations of those expressions are useful. In particular,
	\begin{eqnarray}
	\label{tphidot_Ldot}
	a E - \frac{L_z}{\SN^2}
	& = &   a \dot T - (r^2 + a^2) \dot {\tilde \varphi} + \epsilon a \dot r
	\nonumber \\ & = &
	a \Ldot - \Sigma \dot {\tilde \varphi} , \\
	\label{r2a2E_aLz}
	(r^2 + a^2) E - a L_z & = & \Delta \Ldot - \epsilon \Sigma \dot r , \\
	\Sigma E - \Delta \Ldot & = &
	- a^2 \SN^2 \Ldot
	- \Sigma \epsilon \dot r
	+ \Sigma a \SN^2 \dot{\tilde \varphi} .
	\end{eqnarray}
	Thirdly, Eqns.~(\ref{Delta_dot_t},\ref{Delta_dot_varphi}) are
	rewritten as:
	\begin{eqnarray}
	\label{Delta_dot_T}
	\Delta \dot T & = &   \left( r^2 + a^2 + \frac{2 M r a^2 \SN^2}{\Sigma} \right)
	E
	- \frac{2 M r a}{\Sigma} L_z + 2 M r \epsilon \dot r , \\
	\label{Delta_dot_tvarphi}
	\Delta \dot {\tilde \varphi} & = &   \left(1 - \frac{2 M r}{\Sigma} \right)
	\frac{L_z}{\SN^2}
	+ \frac{2 M r a}{\Sigma} E
	+ \epsilon a \dot r .
	\end{eqnarray}
	Because these equations have a $\Delta$ factor in their left-hand side,
	there is nothing that guarantees that the Kerr-Schild coordinates are
	indeed regular at horizon crossing, but this can actually be the case.
	
	\subsection{Shortcut equations of motion in Kerr-Schild coordinates}
	
	Indeed this issue can be overcome by noting that Eq.~(\ref{def_R})
	can, rather obviously, be rewritten as
	\begin{equation}
	\label{Delta_C_kappar2}
	\left[(r^2 + a^2) E - a L_z - \epsilon \Sigma \dot r \right]
	\left[(r^2 + a^2) E - a L_z + \epsilon \Sigma \dot r \right]
	= \Delta (C + \kappa r^2) .
	\end{equation}
	This ensures that one of the terms in the left-hand side of this
	equation is zero at horizon crossing. Eq.~(\ref{r2a2E_aLz}) shows that it is
	the second one that
	cancels. Indeed, this equation can be rewritten
	\begin{equation}
	(r^2 + a^2) E - a L_z + \epsilon \Sigma \dot r = \Delta \Ldot .
	\end{equation}
	Combining the last two equations, the $\Delta$'s cancel out and 
	one can obtain an interesting closed form for $\Ldot$:
	\begin{equation}
	\label{def_L_shortcut}
	\Ldot = \frac{C + \kappa r^2}
	{(r^2 + a^2) E - a L_z - \epsilon \Sigma \dot r} .
	\end{equation}
	At horizon crossing, neither $\Sigma$ nor $\dot r$ can be $0$,
	therefore, there exists one choice of $\epsilon$ for which the
	denominator of the above equation is not $0$ and hence this equation
	is regular at horizon crossing. Then, using the second version of
	Eqns.~(\ref{Tdot_Ldot},\ref{tphidot_Ldot}), one obtains a regular,
	``shortcut'' version of the equations of motion for $T$ and $\tilde
	\varphi$:
	\begin{eqnarray}
	\label{def_T_shortcut}
	\dot T & = & E + \frac{2 M r}{\Sigma}
	\frac{C + \kappa r^2}
	{(r^2 + a^2) E - a L_z - \epsilon \Sigma \dot r}, \\
	\label{def_pht_shortcut}
	\dot{\tilde \varphi} & = &   \frac{1}{\Sigma} 
	\left(\frac{L_z}{\SN^2} - a E \right)
	+ \frac{a}{\Sigma}
	\frac{C + \kappa r^2}
	{  (r^2 + a^2) E - a L_z
		- \epsilon \Sigma \dot r} .
	\end{eqnarray}
	This extends a similar version found for the Schwarzschild for the $T$
	coordinate only.  metric~\cite{marck96}.
	
	Equation~(\ref{def_L_shortcut}) immediately allows to understand which
	is the proper choice of $\epsilon$ at horizon crossing. Firstly, from
	the value of $r$ and $\dot r$ at some given time, it is possible to
	know which is the next horizon crossing, if any: (i) if $r < r_-$,
	then the next horizon crossing~(if any) occurs at $r_-$, such as $\dot
	r_{\rm hor} > 0$, (ii) if $r > r_+$, the next horizon crossing~(in
	case it occurs) is at $r = r_+$ and will occur as $\dot r_{\rm hor} <
	0$, and (iii) if $r_- < r < r_+$, there is certainly a horizon
	crossing, which occurs at $r_+$ if $\dot r > 0$ and at $r_-$
	otherwise, and the sign of $\dot r$ will then be the same as it is at
	the current time. Secondly, knowing the value $r_{\rm hor}$ of the
	next horizon to be crossed, we can compute $(r_{\rm hor}^2 + a^2) E -
	a L_z$ and, most importantly its sign. Thirdly, we choose $\epsilon$
	so as to ensure that both terms of the denominator of
	Eq.~(\ref{def_L_shortcut}) are of same sign:
	\begin{equation}
	\label{choice_eps_gen}
	\epsilon = - \sgn \left(\left. \dot r \right|_{r_{\rm hor}} \right) 
	\sgn\left[(r^2_{\rm hor} + a^2) E - a L_z \right] .
	\end{equation}
	The same reasoning if of course valid if we integrate the geodesic
	backward in time. The only difference is then that the sign of
	$ \dot r_{\rm hor}$ has to be flipped in cases (i) and (iii).
	
	\subsection{Second order equation of motion for $T$ and
		$\tilde \varphi$}
	
	It is of course not mandatory to resort to using a first order
	equations such as the ``shortcut''
	Eqns.~(\ref{def_T_shortcut},\ref{def_pht_shortcut}) for the evolution
	of $T, \tilde \varphi$. It is also possible to find a second order
	differential equation for these variables by computing the proper
	time/affine parameter derivative of
	Eqns.~(\ref{Delta_dot_T},\ref{Delta_dot_tvarphi}) and by grouping terms
	so that everything is proportional to $\Delta$, which can be be
	canceled out.  This is actually more conveniently done by considering
	the derivative of $\Delta \Ldot$ starting from
	Eq.~(\ref{r2a2E_aLz}). This derivative possesses a term proportional
	to $\ddot r$ in the right-hand side, which, starting from
	Eq.~(\ref{def_Rp_2_v0}), must be recast into the largest possible
	amount of terms that are proportional to $\Delta$. After a few
	manipulations, we obtain
	\begin{equation}
	\label{def_Rp_2}
	\frac{R'}{2}
	=   \Delta X
	+ \epsilon \Sigma \dot r \left( - 2 r E + 2 \Ldot (r - M) \right) ,
	\end{equation}
	where we have defined $X$ as
	\begin{equation}
	\label{def_X}
	X = 2 r E \Ldot + (M - r) \Ldot^2 - \kappa r .
	\end{equation}
	In order to obtain this rather compact form in Eq.~(\ref{def_Rp_2}),
	we made use of the following equality:
	\begin{equation}
	C + \kappa r^2 =   \Ldot \left(\Delta \Ldot 
	- 2 \epsilon \Sigma \dot r \right) ,
	\end{equation}
	which is easily deduced from Eqns.~(\ref{r2a2E_aLz}) and~(\ref{def_R}).
	
	Once all this is set, proper time/affine parameter derivative of
	Eq.~(\ref{r2a2E_aLz}) takes the very simple form
	\begin{equation}
	\label{def_Lddot}
	\Sigma \Lddot = \epsilon X ,
	\end{equation}
	from which we obtain
	\begin{eqnarray}
	\Sigma^2 \ddot{\tilde \varphi} 
	& = & - 2 \dot \theta \Sigma \frac{\CS}{\SN} \frac{L_z}{\SN^2}
	+ \epsilon a X
	- \Sigma \dot{\tilde \varphi} \dot \Sigma , \\
	\Sigma^2 \ddot T
	& = &   2 M \dot r \Sigma \Ldot
	+ 2 M r \epsilon X
	- 2 M r \Ldot \dot \Sigma .
	\end{eqnarray}
	Although the three previous equations look regular regardless the
	value of $\epsilon$, this is not the case. The reason comes from the
	presence of the term proportional to $\Ldot^2$ in
	Eq.~(\ref{def_Lddot}) through variable $X$~(see Eq.~(\ref{def_X})). If
	we forget about all the other terms and assume that we are close to
	horizon crossing, so that $r$ and $\Sigma$ can be considered as
	constant, the~(very) simplified form of Eq.~(\ref{def_Lddot}) is
	\begin{equation} 
	\dot \Ldot \sim \epsilon \frac{M - r_{\rm hor}}{\Sigma_{\rm hor}} \Ldot^2 ,
	\end{equation}
	whose solution is of the form, after defining
	$\alpha \eqdef (M - r_{\rm hor}) / \Sigma_{\rm hor}$,
	\begin{equation}
	\frac{\Ldot_0}{\Ldot} = 1 - \epsilon \alpha \Ldot_0 (p - p_0) ,
	\end{equation}
	where the subscript $0$ denotes the value at the start of
	integration~(i.e., a short time before horizon crossing) and $p$ is
	the geodesic affine parameter. This equation has some chance to remain
	regular~(i.e.  $\Ldot$ will not blow up) only if $1 / \Ldot$ does not
	go to $0$, i.e. if $\epsilon$ has an opposite sign to that of $\alpha
	\Ldot_0$, which indeed explains why only one choice of $\epsilon$ can
	be valid at horizon crossing.
	
	\subsection{Values of $\epsilon$ in the causal diagram}
	
	We now want to address which values of $\epsilon$ are necessary for
	the Kerr-Schild coordinates to be suitable for each horizon crossing
	of the causal diagram. For this purpose, it suffices to know which
	value of $\epsilon$ has to be chosen for one geodesic.
	
	Starting from the definition of the radial null vectors $E^a$ and
	$S^a$~(Eq.~(\ref{def_E_a_S_a}), it is clear that the components of $E^a$ are
	regular when we choose $\epsilon = 1$. Since this vector corresponds to
	trajectories that travel at $45^\circ$ from right to left, all the horizon lines
	that are perpendicular to it necessitate to be dealt with the Kerr-Schild
	coordinates with the choice $\epsilon = 1$. A similar reasoning with vector
	$S^a$ shows that all the other horizon crossings have to be dealt by using
	$\epsilon = - 1$. We therefore have the following choices of $\epsilon$ for all
	the possible horizon crossings: 
	\begin{eqnarray}
	\epsilon_{1 \to 2} = 	\epsilon_{2 \to 5} = 	\epsilon_{6 \to 12} = 	\epsilon_{12
		\to 11} & = & 1 ,\\
	\epsilon_{3 \to 2} = 	\epsilon_{2 \to 6} = 	\epsilon_{5 \to 12} = 	\epsilon_{12
		\to 9} & = & - 1 .
	\end{eqnarray}
	And of course, the same applies for any transform whose both starting
	and ending regions are shifted by $8 k$, where $k$ is an integer,
	i.e. the crossing $4 \to 1$ is made with $\epsilon = - 1$.
	
	\subsection{The different types of geodesics}
	\label{diff_type_geod}
	
	If we want to visualize the Kerr metric, we need to solve the geodesic
	equation. Given the complexity of the full analytic extension of the
	metric, and given the fact that different coordinate systems may~(and,
	actually, have to) be used when crossing several regions, it is hardly
	possible to propagate a geodesic by solving the geodesic equation by
	brute force if we do not know in advance which regions will be crossed
	by a geodesic whose position $x_0^\mu$ and four-velocity/wavevector
	$u_0^\mu, k_0^\mu$ are known at some event of the metric. Since there
	exists a closed form for $\dot r$~(see Eq.~(\ref{def_R})), a geodesic
	can have at most two turning points, depending on the roots of $R(r)$
	and on where $r_0$ is situated with respect to them. There are
	essentially five possible configurations, some of them possessing
	several sub-cases.
	
	\begin{enumerate}
		
		\item \pmb{Geodesics starting from $r = + \infty$ and with one turning
			point}.  Those are the only geodesics that exist when there is not
		black hole~(or wormhole, in this context). In the presence of a
		wormhole, such geodesics can be divided into several sub-types,
		depending on where their turning point $r_{\rm t}$ lies.
		
		\begin{enumerate}
			
			\item $\pmb{r_{\rm t} > r_+}$. The turning point is above the~(outer)
			horizon, so that the geodesic never leaves its region of
			origin~(region~$1$, say). These geodesics exist in the black hole
			case~(and even in the absence of black hole). They may be called
			flyby geodesics\footnote{It seems that there is no
				consensus about this naming. For example, Ref.~\cite{hackmann10}
				use the term of ``flyby'' for any geodesic with a turning point
				situated of some positive $r$ even though it occurs within the
				horizon. In our opinion, such convention is ambiguous since,
				although it is a flyby of the singularity itself, it is not
				necessarily a flyby of the wormhole.}.
			
			\item $\pmb{0 < r_{\rm t} < r_-}$. These geodesics cross the outer
			then inner horizon, bounce at some positive $r$ and then return back
			to another asymptotic region. We shall call them crossing
			geodesics. The geodesic trajectory within the outer horizon is not
			uniquely defined by the above constraint. Indeed,
			Eqns.~(\ref{def_T_shortcut},\ref{def_pht_shortcut}) show that the
			coordinate change from Boyer-Lindquist $(t, \varphi)$ to Kerr-Schild
			$(T, \tilde \varphi)$ can be made in order to deal with horizon
			crossing only by choosing the suitable value of $\epsilon = \pm 1$.
			These equations immediately show that when leaving their asymptotic
			region of origin~(region~$1$, say) toward the inter-horizon
			region~(here, region~$2$), the fact that $\dot r$ is then negative
			implies that the only acceptable choice of $\epsilon$ is $\epsilon =
			\sgn((r^2 + a^2) E - a L_z)$, which, at horizon is rewritten
			\begin{equation}
			\label{def_eps_oi}
			\epsilon_{\rm outer, ingoing} = \sgn(2 M r_+ E - a L_z) .
			\end{equation}
			The very same reasoning says that when leaving region~$2$ in order to
			cross the inner horizon, the correct choice of $\epsilon$ is then
			\begin{equation}
			\epsilon_{\rm inner, ingoing} = \sgn(2 M r_- E - a L_z) .
			\end{equation}
			The same reasoning shows that when crossing out the two horizon the
			choice of $\epsilon$ is the opposite:
			\begin{eqnarray}
			\epsilon_{\rm inner, outgoing}
			& = & - \sgn(2 M r_- E - a L_z) = - \epsilon_{\rm inner, ingoing} , \\
			\epsilon_{\rm outer, outgoing} 
			& = & - \sgn(2 M r_+ E - a L_z) = - \epsilon_{\rm outer, ingoing}.
			\end{eqnarray}
			However, there is no reason that $\epsilon_{\rm outer, ingoing} =
			\epsilon_{\rm inner, ingoing}$ for geodesics originating from region
			$1$ past null infinity, $E$ is positive, so that this equality is
			satisfied when $a L_z / E < 2 M r_-$ or $a L_z > 2 M r_+$ and is not
			satisfied when $2 M r_- < L_z / E < 2 M r_+$. These two cases are met
			for geodesics crossing the horizon as can be seen from the reasonably
			well-known case of null equatorial geodesics which, in the extremal $a
			= M$, do cross the horizon when $-7 M < L_z / E < 2 M$~(see
			Ref.~\cite{chandrasekhar83} or next Section), a situation that
			encompass both cases mentioned above. Moreover, a geodesic leaving
			region~$1$ must end into another asymptotic region where $t$ is a
			future-oriented coordinate without experiencing more than one turning
			point, which leaves region~$9$ as its only possible
			destination. Consequently, these geodesics either travel through the
			$1, 2, 5, 12, 9$ or the $1, 2, 6, 12, 9$ sequences. In the first case,
			the same value of $\epsilon$ must be used for the first two~(ingoing)
			horizon crossings and must be changed after these, whereas in the
			second case, the sign of $\epsilon$ must be changed at each horizon
			crossing.
			
			\item $\pmb{r_{\rm t} < 0}$. Those geodesics that we may dub as
			adventurous have their turning point within a region of negative
			$r$. Because there is no ergoregion in any of the $r < 0$ region,
			the region into which the geodesic can enter must have its
			$t$-coordinate that has the same orientation as its asymptotic
			region of origin, which, in the case the starting point is region
			$1$ makes region~$7$ as its only negative $r$ region it can enter
			into. Therefore, the geodesic follows the sequence $1, 2, 5, 7, 5,
			12, 9$. According to the previous discussion, there is no change of
			$\epsilon$ when traveling region~$1$ to region~$2$ and from region
			$2$ form region~$5$, however, as we shall explain later it is
			necessary to switch to Cartesian Kerr-Schild coordinate during the
			two crossings $5 \to 7$ and $7 \to 5$, the choice of $\epsilon$
			being arbitrary at this stage~(but a change of sign of $\epsilon$
			will be necessary for the $5 \to 12$ crossing as compared to the $2
			\to 5$ one, see previous sub-case above). Whether it is performed
			before, after or in between the two $5 \leftrightarrow 7$ crossings
			does not matter, however.
			
		\end{enumerate}
		
		\item \pmb{Geodesics starting from $r = - \infty$ and with one turning
			point}. Those are the analog of the above, except that they start
		and end on the negative $r$ region~(region~$-1$, say). Although
		there may be several sub-cases as in the above case depending on
		where the turning point lies, there is actually only one possible
		configuration, where the turning point is situated at some negative
		$r$. The reason for this is that should the geodesic be allowed to
		enter into a positive $r$ region and leave it afterward, this would
		mean that is lowest real root of $R(r)$ is positive. From
		Eq.~(\ref{def_R}), it is clear that the sum of the roots is zero,
		therefore there cannot be four positive roots, and in the case we
		are considering, there must be two positive roots and two complex
		conjugate roots with a negative real part. Let us call the positive
		roots $X - \delta$, $X + \delta$ and the two others $-X \pm i Y$,
		with $0 \leq \delta < X$. The coefficient of $R(r)$ that is
		proportional to $r$ is, in this case, $- 2 X (Y^2 +
		\delta^2)$. However, this coefficient is also equal to $2 M C$,
		where the Carter constant must be positive for $\dot \theta$ to be
		defined in Eq.~(\ref{def_Th}), which implies that $X$ must be
		negative, which contradict the initial statement. Therefore, null
		geodesics starting and ending at $r = - \infty$ are of flyby type
		with a turning point at some negative $r$. They do not cross the ring
		singularity and remain in the same region they originate from.
		
		\item \pmb{Ingoing transit geodesics}. These geodesics start from
		$r = +\infty$ and end to $r = - \infty$. In order to do so, their
		initial and final region must at the same time orientation for the
		$t$-coordinate and cannot have any turning point. Therefore, if they
		start from region~$1$, they will experience the sequence
		$1, 2, 5, 7$. Their mirror analogue stating from region~$3$ will go 
		through the sequence $3, 2, 6, 8$.
		
		\item \pmb{Outgoing transit geodesics}. Those are the affine
		parameter-reversed of the above. From the previous discussion, those
		that are seen by an observer in region~$1$ have gone through the
		sequence of regions~$-1, -3, 4, 1$. For an observer in region~$3$,
		they must originate from region~$0$.
		
		\item \pmb{Bounded geodesics}. These geodesics exist when the
		polynomial $R (r)$ admits four real roots, the geodesics being
		bounded between the two intermediate roots. We shall devote a
		thorough analysis of these geodesics in the next Section, but we may
		already summarize the results: those geodesics can be seen in a
		rather short interval of $r$, that cannot exceed $[0, 4 M]$. For the
		same reason that is explained in the case of crossing geodesics, a
		bounded geodesic that travels at some point within region~$1$ can
		further go, after outer then inner horizon crossing, either in
		region~$5$ or $6$. Therefore, such geodesics can cross four types of
		infinite sequences: $1, 2, 5, 12, 9, ... [\mod 8]$,
		$1, 2, 6, 12, 9, ... [\mod 8]$, $3, 2, 5, 12, 11, ... [\mod 8]$ and
		$3, 2, 6, 12, 11, ... [\mod 8]$.
		
	\end{enumerate}
	
	The above discussion allows to determine which regions are seen by an
	observer as a function of its position. This is summarized in
	Table~\ref{table_reg}. One~(we think unexpected) consequence of the
	global structure of geodesics is that the richest patch of the Kerr
	wormhole is the outgoing inter-horizon region~(i.e., region~$12$,
	($\mod 8$, of course) which allows to see both two negative $r$
	regions as well as two positive $r$ asymptotic regions, a situation
	which is not symmetric with respect to ingoing inter-horizon region,
	where no geodesic starting from any past null infinity of negative $r$
	region penetrate into.
	
	\begin{table}[pht]
		\tbl{Summary of the asymptotic regions that can be seen as a
			function of the region of the Carter-Penrose diagram where the
			observe lies. For each of the region that lie one one side of the
			diagram, we have also included the view from its mirror
			counterpart of the diagram. All the label of given line can be
			simultaneously shifted by $8 k$. BG means that some bounded
			geodesics can be seen by the observer~(see next Section). \newline
			${}^1$: Bounded geodesics are seen for sufficiently small
			$r$; \newline
			${}^2$: Bounded geodesics are always seen; \newline
			${}^3$: Up to two bounded geodesic patches can be seen, depending
			on the observer's coordinates $r$ and $\theta$. }
		{\begin{tabular}{|c|c|c|} \hline Region where the observer lies
				& Type & Region that are seen \\ \hline $1$ (resp. $3$) &
				Asymptotic, positive $r$ & $1$, $-7$, $-1$ (resp. $3$, $-5$,
				$0$) + BG${}^1$ \\ $2$ & Ingoing inter-horizon & $1$, $3$ +
				BG${}^2$ \\ $5$ (resp. $6$) & Within inner horizon, positive
				$r$ & $1$, $3$, $7$ (resp. $3$, $1$, $8$) + BG ($2
				\times$)${}^3$ \\ $7$ (resp. $8$) & Negative $r$ & $7$, $1$
				(resp. $8$, $3$) \\ $12$ & Outgoing inter-horizon & $1$,
				$3$, $7$, $8$ + BG${}^2$ \\ \hline
			\end{tabular} \label{table_reg}  }
	\end{table}
	
	\section{Bounded geodesics}
	\label{sec_bound}
	
	In our opinion, the most overlooked aspect of geodesics in the Kerr
	metric deals with bounded null geodesics. Those are essential to be
	taken into account when visualizing the metric in region where they
	exist. The problem they pose becomes necessary to address for any
	observer sufficiently close to the black hole outer horizon as we
	shall see.
	
	The function $R (r)$ is a fourth degree polynomial and therefore can
	admit up to four roots. Consequently, it is possible that a geodesic,
	whether timelike or null, is bounded. For visualization purpose, we
	shall only consider null geodesics which share this property. For null
	geodesics, $\kappa = 0$, and the three other constants of motions are
	defined up to some arbitrary overall constant since the geodesic
	affine parameter is as well. Unless we have to deal with a very
	peculiar geodesic with $E = 0$, it is convenient to rescale the
	constants of motion so as to reduce them to two: $L_z / E$ and $C /
	E^2$. Moreover, the latter is more conveniently replaced by $(C - (a E
	- L_z)^2) / E^2$. Keeping the notations of Ref.~\cite{chandrasekhar83},
	we therefore define
	\begin{eqnarray}
	\xi & \eqdef & \frac{L_z}{E}, \\
	\eta & \eqdef & \frac{C}{E^2} - (a - \xi)^2 .
	\end{eqnarray}
	With these notations, $R (r)$ can be rewritten in the case of null
	geodesics as
	\begin{equation}
	\label{def_R_null}
	\frac{R (r)}{E^2}
	= r^4 - \left[\eta + \xi^2 - a^2 \right] r^2
	+ 2 M \left[\eta + (a - \xi)^2 ) \right] r - a^2 \eta .
	\end{equation}
	In the $(\xi, \eta)$ parameter space, the edge of the bounded geodesic
	region arises when $R(r)$ admits a double root that we shall label
	$e$, i.e., when one has simultaneously $R (e) = 0$ and $R' (e) = 0$.
	Writing these two equations and using one of them to express $\eta$ as
	a function of $\xi$ and $e$ and further solving the second order
	equation for $\xi$ yields a parametric equation for both $\xi$ and
	$\eta$ as a function of the double root $e$
	\begin{eqnarray}
	\label{def_xi_par}
	\xi
	& = & \frac{1}{a(e - M)} \left[M (e^2 - a^2) - e \Delta (e) \right] , \\
	\label{def_eta_par}
	\eta
	& = & \frac{e^3}{a^2 (e - M)^2} \left[ 4 M a^2 - e (e - 3 M)^2 \right] .
	\end{eqnarray}
	By definition, these two equations separate the loci of some pairs of
	geodesic types among the five of \S\ref{diff_type_geod} of the
	previous Section.  This is of course the case for any type of metric
	where a closed form for $\dot r$ exists. For example, in the case
	where $a$ tends to $0$, $\eta$ and $\xi$ remain defined only when
	their numerator tends to $0$ as well, which occurs only for $e = 3
	M$. We recover the critical case of null geodesics in the
	Schwarzschild metric which separates between the three types of
	geodesics that exist in this metric~(flyby geodesics, which have a
	turning point at some $r > 3 M$, transit geodesics which go from
	infinity to $r = 0$ or vice-versa and geodesics starting and ending at
	the singularity). What we need to do now is to narrow this constraint
	to consider when they refer to bounded geodesics only.
	
	Although not very illuminating, Eq.~(\ref{def_xi_par}) for $\xi (e)$
	can be rewritten in several ways which can occasionally shed more light on these
	expressions. For example, we have
	\begin{eqnarray}
	\label{def_xi1}
	\xi
	& = & a - \frac{e}{a (e - M)} \left[e^2 - 3 M e + 2 a^2 \right] , \\
	\xi - a 
	& = & - \frac{e}{a (e - M)}\left[e (e - 3 M) + 2 a^2 \right] , \\
	\xi + a 
	& = & - \frac{1}{a (e - M)}\left[e^2 (e - 3 M) + 2 M a^2 \right] .
	\end{eqnarray}
	The last two equations can be combined with Eq.~(\ref{def_eta_par}) to obtain
	\begin{eqnarray}
	\eta + \xi^2 - a^2
	& = & 2 e^2 + \frac{4 M e \Delta (e)}{(e - M)^2} , \\
	& = & \frac{2 e}{(e - M)^2} \left[e^3 - 3 M^2 e + 2 M a^2 \right] ,
	\end{eqnarray}
	together with
	\begin{eqnarray}
	\eta + (\xi - a)^2 & = & \frac{4 e^2 \Delta (e)}{(e - M)^2} , \\
	\label{eta_xi_p_a}
	\eta + (\xi + a)^2 & = & \frac{4 M}{(e - M)^2} \left[2 e^3 - 3 M e^2 + M a^2
	\right] .
	\end{eqnarray}
	Finally, Eq.~(\ref{def_eta_par}) alone yields
	\begin{equation}
	\label{eta_par_2}
	a^2 \eta + e^4 = \frac{4 M e^3}{(e - M)^2} \Delta (e) .
	\end{equation}
	The sum of the roots of $R$ is $0$ and their product is $- a^2 \eta$,
	therefore, since when $R$ admits $e$ as a double root, it can
	necessarily be rewritten according to
	\begin{equation}
	\label{R_22}
	R (r) = (r - e)^2 (r^2 + 2 r e - a^2 \eta / e^2) .
	\end{equation}
	
	In order to have bounded geodesics, there must be two other roots in
	$R$, otherwise the double root only separates between geodesics that
	start from $\pm \infty$ and have a finite turning point before going
	back to $\pm \infty$, and geodesics that cross the whole interval of
	$r$, i.e. from $- \infty$ to $+ \infty$ or vice versa. The
	discriminant of the second order polynomial in the right-hand side of
	Eq.~(\ref{R_22}) is $4 (e^4 + a^2 \eta) / e^2$.  Using
	Eq.~(\ref{eta_par_2}), it is therefore clear that the two mandatory
	other roots are defined only when $e \Delta(e) > 0$, something that
	happens either when $0 \leq e \leq r_-$ or $r_+ \leq e$. When one of
	these to conditions is satisfied, we shall label $s$ the largest of
	the two roots, the other being $- s - 2 e$.  These roots are given by
	\begin{equation}
	\label{def_s_m2ems}
	s, - 2 e - s = - e \pm \frac{2}{|e - M|} \sqrt{M e \Delta (e)} ,
	\end{equation} 
	where the plus sign of the right-hand side corresponds to $s$. For $e >
	M$, one immediately sees that $s < e$, which comes from the evident
	inequality $(e - M)^3 + M (M^2 - a^2) > 0$. When $e < M$, which in
	this context happens when $0 \leq e \leq r_-$, $s$ is smaller than $e$
	in a fairly limited interval whose lower bound is given by the
	equality $s = e$, which occurs when $(M - e)^3 = M (M^2 - a^2)$ whose
	unique real solution is
	\begin{equation}
	\label{def_emin}
	e_{\rm min} = M - [M (M^2 - a^2)]^{\frac{1}{3}} .
	\end{equation}
	Equivalently, this $e_{\rm min}$ can be found by noting that the
	double root remains a local maximum as long as $R''(e) < 0$. When
	$e < s$, $e$ becomes a local minimum, the transition between the two
	occurring when $R''(e_{\rm min}) = 0$, whose solution, according to the
	definition of Eq.~(\ref{def_R_null}) occurs when
	$12 e_{\rm min}^2 = 2 (\eta(e_{\rm min}) + \xi^2(e_{\rm min}) - a^2)$;
	whose solution is given by the same equation as that of
	Eq.~(\ref{def_emin}).
	
	What we know need is to determine whether 
	Eqns.~(\ref{def_xi_par},\ref{def_eta_par}) can actually correspond to 
	geodesics, that is to check which values of $\theta$~(if any) are
	compatible with these requirements. By virtue of
	Eq.~(\ref{def_Th}), we have, for null geodesics,
	\begin{equation}
	\label{def_pol_mu2}
	- \sin^2 \theta \frac{\Sigma^2 \dot \theta^2}{E^2} 
	= P(\mu^2)
	= a^2 \mu^4 + (\eta + \xi^2 - a^2) \mu^2 - \eta < 0 ,
	\end{equation}
	where we have set, following the usual convention,
	$\mu = \cos \theta$.  This inequality can be satisfied only when the
	corresponding second order polynomial in $\mu^2$ admits real roots,
	i.e., when the corresponding discriminant is positive.
	
	This discriminant, $\Delta '$, can be written
	\begin{equation}
	\Delta ' = (\eta + \xi^2 - a^2)^2 + 4 a^2 \eta
	= (\eta + (\xi - a)^2) (\eta  + (\xi + a)^2 ) .
	\end{equation}
	Let us consider the function $g_3 (e) = 2 e^3 - 3 M e^2 + M a^2$,
	whose sign is the same as $\eta + (\xi + a)^2$ for each value of $e$, see
	Eq.~(\ref{eta_xi_p_a}).
	Whichever value of $a$, the function $g_3$ possesses two extrema at
	$e = 0, M$ and is therefore increasing for $e < 0$ and $e > M$, and
	decreasing for $0 < e < M$. Moreover, for any $a$ such that
	$0 < a^2 < M^2$, we have $g_3 (0) = M a^2 > 0$ and
	$g_3 (M) = - M (M^2 - a^2) < 0$. Consequently, $g_3$ admits three roots,
	$r_1$, $r_2$ and $r_3$, such that $r_1 < 0 < r_2 < M < r_3$. In
	the limit case $a = 0$, we have $r_1 = r_2 = 0$ and when $a^2 = M^2$, we
	have $r_2 = r_3 = M$.  Moreover, we also have
	$g_3 (r_\pm) = 2 r_\pm (M^2 - a^2) > 0$, so that we have in fact
	\begin{equation}
	\label{class_r_spec}
	r_1 < 0 < r_- < r_2 < M < r_3 < r_+ .
	\end{equation}
	Consequently, the discriminant $\Delta'$ is positive within the three
	intervals $]r_1, r_-[$, $]r_2, r_3[$ and $]r_+, +\infty[$, and it is
	zero at these five $r_i$'s.
	
	The fact that the discriminant is positive is not sufficient. In
	addition, the interval between the two roots of the polynomial
	$P(\mu^2)$ of Eq.~(\ref{def_pol_mu2}) must have some intersection with
	interval $[0, 1]$ where $\mu^2$ is defined.
	
	From Eq.~(\ref{def_pol_mu2}), it is clear that there are viable
	solutions as soon as the product of the roots is negative, that is,
	when $\eta$ is positive. Conversely, if the product of the roots is
	positive~(i.e., $\eta < 0$) whereas their sum is negative, then there
	are no solutions to Eq.~(\ref{def_pol_mu2}).
	
	For positive $e$'s, the sign of $\eta$ is given by that of the
	function $f_3 (e) = 4 M a^2 - e (e - 3 M)^2$ (see Eq.~(\ref{def_eta_par})).
	This function appears in
	a well-known context when one studies the Kerr metric: its roots
	correspond to the radii of equatorial, null geodesics. This
	functions admits two extrema at $e = M, 3 M$. Moreover, since $f_3(0)
	= 4 M a^2 > 0$, $f_3(M) = 4 M (a^2 - M^2) < 0$ and $f_3 (3 M) = 4 M
	a^2 < 0$, $f_3$ admits three roots, we shall label $r_{\rm i}$,
	$r_{\rm p}$ and $r_{\rm r}$ which lies within the intervals $]0, M[$,
	$]M, 3 M[$ and $]3 M, \infty[$, respectively. Using the fact that
	$r_\pm^2 = 2 M r_\pm - a^2$, it is easy to show that $f_3(r_\pm) =
	- r_\pm (M^2 - a^2) < 0$, which means that we have in fact $0 <
	r_{\rm i} < r_-$ and $r_+ < r_{\rm p} < 3 M$. Furthermore, if we
	use the definition of $e_{\rm min}$, we can easily show~(by
	canceling terms proportional to $a^2$) that $f_3(e_{\rm min}) = 3
	e_{\rm min} (e_{\rm min} - M)^2 > 0$, which ensures that
	$e_{\rm min} < e_{\rm i}$.
	
	All this enables to extend Eq.~(\ref{class_r_spec}) into
	\begin{equation}
	\label{class_r_spec_2}
	r_1 < 0 < r_{\rm i} < r_- < r_2 < M < r_3 < r_+ < r_{\rm p} < 3 M < r_{\rm r} .
	\end{equation}
	For the sake of completeness, we recall that the equatorial null
	geodesics occur  at
	\begin{equation}
	\label{def_r_eq}
	r_{\rm eq, null}
	= 2 M [1 + \cos(2 \arccos(a / M) / 3 + 2 k \pi / 3)] , 
	\end{equation}
	with $k = 0, 1, 2$ for $r_{\rm r}$, $r_{\rm i}$ and $r_{\rm p}$,
	respectively. The corresponding value of $\xi$ is given by
	\begin{equation}
	\label{def_xi_eq}
	\xi_{\rm eq, null}
	= - a - 6 M \cos\left[  \frac{1}{3} \arccos\left(\frac{a}{M}\right)
	- \frac{2 k \pi}{3} \right] ,
	\end{equation}
	where we have defined $k$ in a consistent way between
	Eqns.~(\ref{def_r_eq}) and~(\ref{def_xi_eq}).
	
	Let us now consider the quantity $\eta + \xi^2 - a^2$, which is the
	opposite of the sum of the roots of Eq.~(\ref{def_pol_mu2}). It is of
	same sign as $e (e^3 - 3 M e + 2 M a^2) \eqdef e h_3 (e)$. The third order
	polynomial $h_3 (e)$ admits two extrema at $e= \pm M$, and $h_3(-M) =
	2 M(M^2 + a^2) > 0$, $h_3(M) = -2 M (M^2 - a^2) < 0$. Also, $h_3 (0) =
	2 M a^2 > 0$.  Consequently, $h_3$ possesses three roots, one lower
	than $- M$ and two positive ones. Moreover, $h_3 (r_\pm) = r_\pm (M^2
	- a^2) > 0$, so that the two positive roots of $h_3$ are between $r_-$
	and $r_+$. If we consider $h_3 (r_1)$, where $r_1$ is the negative
	root of $g_3$, we have immediately $h_3 (r_1) = - r_1^3 + M a^2 > 0$,
	so that $r_1$ is larger that the negative root of $h_3$, and $h_3 (r)$
	is positive everywhere in the interval $[r_1, 0]$. Conversely, $\eta +
	\xi^2 - a^2$ is negative. We can now summarize the domain of existence
	of a double root $e$ of $R$ which delineates the edge of a bounded
	geodesic. It must fill the following requirements:
	\begin{enumerate}
		\item Some roots must exist to polynomial $P(\mu^2)$ in
		Eq.~(\ref{def_pol_mu2}), i.e. discriminant $\Delta'$ must be
		positive;
		\item The roots of $P(\mu^2)$ must be such that they span an interval
		that has a non zero intersection with physically allowed values for
		$\mu^2$, i.e., $[0, 1]$;
		\item The turning point $e$ must be outside the outer horizon or
		inside the inner horizon;
		\item There must be two other real roots, i.e., according to
		Eq.~(\ref{def_s_m2ems}), one must have $e \Delta(e) > 0$;
		\item The double root must be a local maximum.
	\end{enumerate}
	All these requirements are summarized into Table~\ref{table_allr},
	from which we see that acceptable values for $e$ are those which lies
	within $[0, r_{\rm i}]$ and $[r_{\rm p}, r_{\rm r}]$. Within these two
	intervals, both $\eta$ and $\eta + \xi^2 - a^2$ are positive, which
	means that Eq.~(\ref{def_pol_mu2}) admits one positive and one
	negative solution. Consequently, bounded null geodesics for which $e$
	is defined all oscillate around the equatorial plane and reach a
	maximum value of $\mu^2$ defined by
	\begin{equation}
	\label{eq_mu2_max}
	\mu_{\rm max}^2 = \frac{e}{a^2 (e - M)^2}
	\left[ - (e^3 - 3 M^2 e + 2 M a^2 )
	+ \sqrt{4 M \Delta(e) (2 e^3 - 3 M e^2 + M a^2)}
	\right] .
	\end{equation}
	One can check that this expression reaches $1$ only when $e$ is a root
	of Eq.~(\ref{def_xi1}), i.e. when $\xi = 0$ and for $e = r_{\rm pol}$. Another useful
	value is when $e = e_{\rm min}$, for which Eq.~(\ref{eq_mu2_max})
	simplifies considerably into
	\begin{equation}
	\label{mu2max_emin}
	\mu_{\rm max}^2 (e_{\rm min}) = 
	(2 \sqrt{3} - 3) \frac{e_{\rm min}^2}{a^2} .
	\end{equation}
	Although we already know that it is the case, we can check that this
	quantity is smaller than $1$ since the ratio $e_{\rm min} / a$ is
	already.
	\begin{table}[pht]
		\tbl{Intervals where double roots to $R$ exist and delineate edge of bounded
			geodesics, according to  Eq.~(\ref{def_pol_mu2}). Intervals where it is not
			necessary to define $\eta + \xi^2 - a^2$ are labeled with a question mark.
			\CHL
			1: No root since $\Delta' < 0$ \CHL
			2: No acceptable root as their sum ($- \eta - \xi^2 + a^2$) is negative and
			their product ($- \eta$) is positive \CHL
			3: No root allowed since no geodesic turning point can exist between $r_-$
			and
			$r_+$ and, equivalently,
			$C = \Sigma^2 \dot \theta^2 + (a E \sin \theta - L_z / \sin \theta)^2 < 0$
			\CHL	
			4: No other roots apart from the double root \CHL
			5: Double root is not a local maximum }
		{\begin{tabular}{|c|p{0.2cm}p{0.2cm}p{0.2cm}p{0.2cm}p{0.2cm}p{0.2cm}p{0.2cm}p{0.2cm}p{0.2cm}p{0.2cm}p{0.2cm}p{0.2cm}p{0.2cm}p{0.2cm}p{0.2cm}p{0.2cm}p{0.2cm}p{0.2cm}p{0.2cm}p{0.2cm}p{0.2cm}p{0.2cm}p{0.2cm}p{0.2cm}p{0.2cm}p{0.2cm}|}
				\hline
				& & $r_1$ & & $0$ & & $e_{\rm min}$ & & $r_{\rm i}$ & & $r_-$ & & $r_2$ & &
				$M$
				& & $r_3$ & & $r_+$ & & $r_{\rm p}$ & & $3 M$ & & $r_{\rm r}$ & & \\ \hline
				$\Delta (r)$ & $+$ & & $+$ & & $+$ & & $+$ & & $+$ & $0$ & $-$ & & $-$ & &
				$-$ &
				& $-$ & $0$ & $+$ & & $+$ & & $+$ & & $+$ & \\ \hline
				$\eta + (\xi - a)^2$ & $+$ & & $+$ & $0$ & $+$ & & $+$ & & $+$ & $0$ & $-$ &
				&
				$-$ & $||$ & $-$ & & $-$ & $0$ & $+$ & & $+$ & & $+$ & & $+$ & \\ \hline
				$\eta + (\xi + a)^2$ & $-$ & $0$ & $+$ & & $+$ & & $+$ & & $+$ & & $+$ & $0$
				&
				$-$ & $||$ & $-$ & $0$ & $+$ & & $+$ & & $+$ & & $+$ & & $+$ & \\ \hline
				$\Delta'$ & $-$ & $0$ & $+$ & $0$ & $+$ & & $+$ & & $+$ & $0$ & $-$ & $0$ &
				$+$
				& $||$ & $+$ & $0$ & $-$ & $0$ & $+$ & & $+$ & & $+$ & & $+$ & \\ \hline
				$\eta$ & $-$ & & $-$ & $0$ & $+$ & & $+$ & $0$ & $-$ & & $-$ & & $-$ & $||$
				&
				$-$ & & $-$ & & $-$ & $0$ & $+$ & & $+$ & $0$ & $-$ & \\ \hline
				$\eta + \xi^2 - a^2$ & $?$ & & $-$ & $0$ & $+$ & & $+$ & & $+$ & & $?$ & &
				$?$ &
				$||$ & $?$ & & $?$ & & $+$ & & $+$ & & $+$ & & $+$ & \\ \hline
				Some $\mu^2$ defined & No & & No & & No & & {\bf Yes} &  & No & & No & & No
				& &
				No & & No & & No & & {\bf Yes} & & {\bf Yes} & & No & \\ \hline
				Notes & ${}^{1,4,5}$ & & ${}^{4,5}$ & & ${}^5$ & & &  & ${}^2$ & &
				${}^{1,3,4}$
				& & ${}^{3,4}$ & & ${}^{3,4}$ & & ${}^{1,3,4}$ & & ${}^2$ & & & & & & ${}^2$
				&
				\\ \hline
			\end{tabular}
			\label{table_allr}}
	\end{table}
	
\begin{figure}[h]
		\centerline{\includegraphics*[width=4.8in]{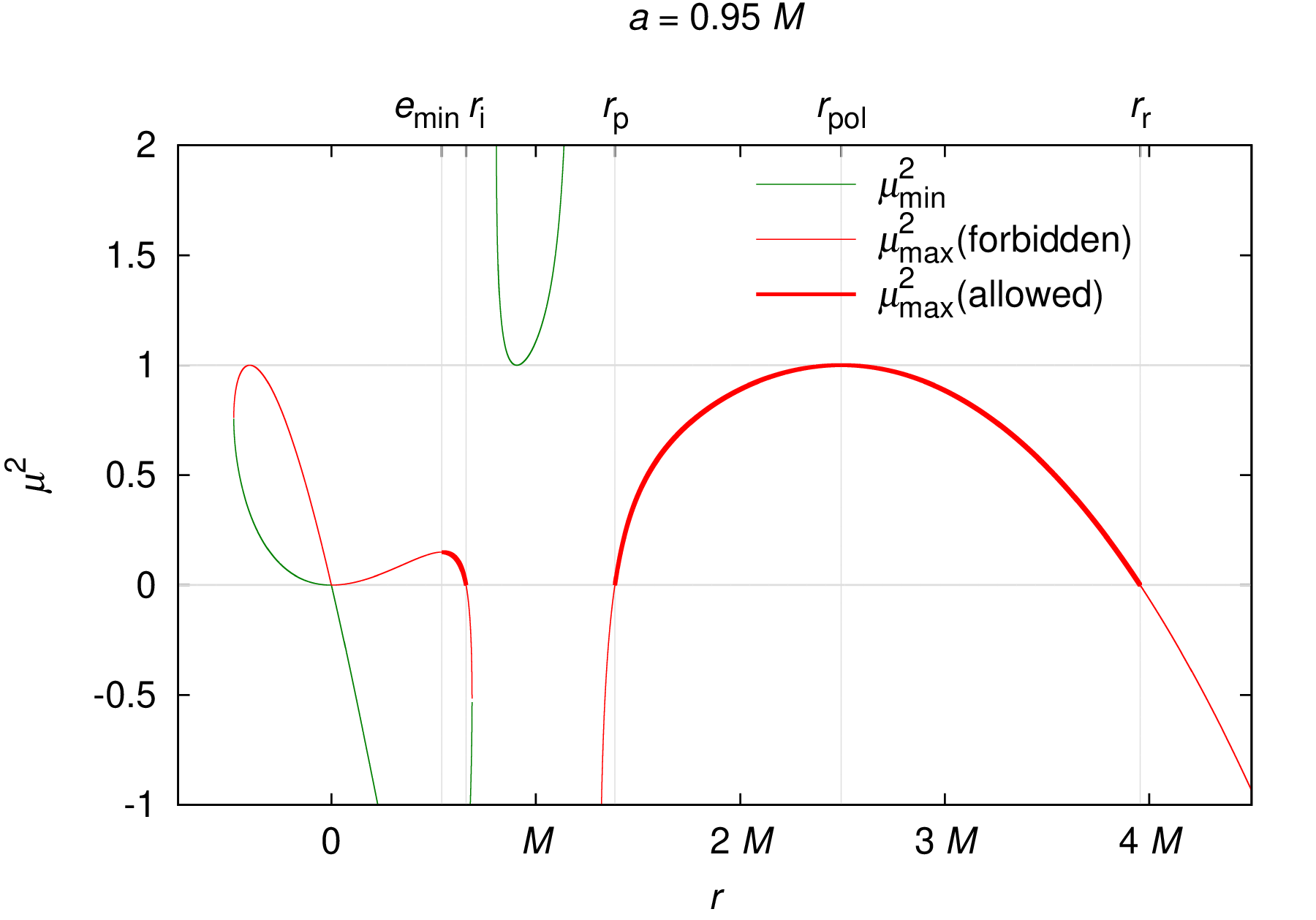}}
		\caption{Roots of polynomial $P(\mu^2)$ defined in
			Eq.~(\ref{def_pol_mu2}) as a function of the double root of $R (r)$.
			See text for the physical meaning of this function. The overall
			shape of the function does not depend on $a$~(as long as $a^2 <
			M^2$, of course), however, the small part that lies between $0$ and
			$r_{\rm i}$ is extremely small and has a low vertical extension and,
			hence, difficult to see unless one considers a value of $|a|$ close
			to $M$, which is the reason why we chose $a = 0.95 M$ here.}
		
		\label{fig_mu2}
	\end{figure}

	The value of the solutions to Eq.~(\ref{def_pol_mu2}) are shown in
	Figure~\ref{fig_mu2}.
	
	An equatorial observer will intersect bounded null geodesics at $r =
	r_{\rm r}$ and below, but this will be not the case for a non
	equatorial observer. Such observer lying somewhere in the interval $r
	\in [r_{\rm p}, r_{\rm r}]$ will actually intersect bounded null
	geodesics only if such geodesics can reach the observer given its
	colatitude $\theta$, that is if the observer's colatitude lies with
	the range allowed by Eq.~(\ref{def_pol_mu2}). For example, an observer
	which lies along the rotation axis of the black hole will intersect
	bounded null geodesics only if those are polar, that is if $\xi(e) =
	0$. Given that, according to Eq.~(\ref{def_xi_par}), $\xi (e) \propto
	j_3(e) = M (e^2 - a^2) - e \Delta (e) = - e^3 + 3 M e^2 - a^2 e - M
	a^2$, we need to find where the roots of $j_3$ lie.  From the first
	expression of $j_3$, one obtains that $j_3 (r_\pm) = M (r_\pm^2 - a^2)
	= 2 M (M r_\pm - a^2) = 2 M r_\pm (r_\pm - M)$, so that $j_3 (r_-) <
	0$ and $j_3 (r_+) > 0$. Since, in addition, $j_3 (e)$ is decreasing
	for sufficiently large $|e|$ and $j_3 (0) < 0$, $j_3$ evidently admits
	three roots, a negative one, one between $r_-$ and $r_+$ and one, we
	shall label $r_{\rm pol}$, above $r_+$. Direct investigation of
	Eq.~(\ref{def_xi_eq}) show that $\xi(r_{\rm p}) + a \in [3 M, 3
	\sqrt{3} M]$, whereas $\xi(r_{\rm r}) + a \in [- 6 M, - 3 \sqrt{3}
	M]$, which ensures that $r_{\rm pol}$ lies between $r_{\rm p}$ and
	$r_{\rm r}$, an unsurprising result since the latter corresponds to an
	equatorial, retrograde orbit~(hence with a negative $\xi$) and the
	former to an equatorial, prograde orbit~(hence with a positive $\xi$).
	
	Consequently, an observer falling toward the black hole will begin to
	intersect bounded null geodesics at some $r$ between $r_{\rm r}$ and
	$r_{\rm pol}$ depending on how its colatitude evolves with $r$. If the
	observer always lies along the equatorial plane, bounded null
	geodesics will be intercepted as early as $r = r_{\rm r}$, and as late
	as $r = r_{\rm pol}$ for a polar observer~(i.e., always along the
	black hole rotation axis). The value of $r_{\rm pol}$ is given by
	\begin{equation}
	\label{def_r_pol}
	r_{\rm pol} = M + 2 \sqrt{M^2 - a^2/3} \cos \left(\frac{1}{3} \arccos \left[M
	(M^2 - a^2) / (M^2 - a^2 / 3)^{\frac{3}{2}}\right]  \right) ,
	\end{equation}
	a quantity which is $3 M$ for $a = 0$~(as expected  since it
	corresponds to the unstable circular photon orbits of the
	Schwarzschild case) and which is otherwise always smaller than $3 M$.
	
	\section{Dark crescent, dark outgrowth, dark shell, dark bubble}
	\label{sec_dark}
	
	We now come to the visual translation of the above discussion
	regarding the aspect of bounded geodesics.
	
	When considering the black hole case, these bounded null geodesics
	are, from a visual point of view, indistinguishable from the black
	hole silhouette itself. Actually, they do delineate the black hole
	silhouette because all these geodesics have crossed out the horizon
	before reaching the observer, so that they can be considered as
	originating from a perfectly black~(actually infinitely redshifted)
	surface. However, if we consider the full analytical extension of the
	metric, there is a fundamental distinction between bounded and
	unbounded geodesics.
	
	Let us first consider an observer situated on the equator of the
	coordinate system at coordinate distance $r$. The observer's motion
	does not matter here, since switching from a given observer to another
	observer endowed with another four-velocity will translate into an
	aberration transformation due to the corresponding Lorentz boost, and
	because such distortion of the celestial sphere does not affect the
	topology of the patches that are on that celestial sphere. Let us suppose that
	the observer
	starts from a large value of $r$ of its region of origin,
	region~$1$~(say). As long as $r > r_{\rm r}$, there are two type of
	null geodesics the observer can cross. Either the null geodesic
	originates from past null infinity of the observer's region, in which
	case they will show some part of region~1, or the geodesic originates
	from some region that lies in the past lightcone of the observer,
	which, according to the discussion of \S\ref{sec_carter} is either
	$-7$ or $-1$. Forgetting about the latter, the edge between direction
	where regions~1 and $-7$ are given by unstable null geodesics, each of
	which lie at some fixed $e$ whose value is given by the double roots
	of $R (r)$ which are a local maximum.
	
	Let us consider geodesics showing region~1 close to this
	boundary. These geodesics have constants of motion close to $\xi(e)$,
	$\eta(e)$, the departure from these values being such that the local
	extremum of $R$ is no longer a double root but a~(very) slightly
	positive extremum. Those geodesics are seen after they have reached
	their lowest approach $r$ coordinate~(that is, $e$) and they are now
	receding away from the wormhole. The same applies for geodesics
	originating from region~$-7$ except that now the corresponding
	constants of motion, although also close to $\xi(e)$, $\eta(e)$ are
	this time such that the local extremum of $R$ is no longer a double
	root but a~(very) slightly negative extremum, which allowed these
	geodesics, when ingoing from region~$-7$ to spend a long time close to
	$r = e$~(in region~$-7$), then crossed the two outer, then inner
	horizons, bounced at some $r$ close to $s(e)$ either in region~$-3$ or
	$-2$, exited the inner then outer horizon through region~$4$ and spent
	a large amount of time close to $r = e$ in region~$1$ when they were
	outgoing. Those are therefore seen as well when they are outgoing.
	
	When the observer reaches some $r$ only a bit smaller than $r_{\rm
		r}$, the situation changes. Focusing on equatorial geodesics, the
	equatorial null geodesic from region~1 past null infinity that has $e
	= r_{\rm m}$ as a double root is no longer seen by the observer,
	however the one with slightly different value of $\xi$ $\eta$, for
	which $r = e$ is a negative local maximum of $R$, is seen whereas it
	is ingoing. Conversely, its analog from region~$-7$ is still seen
	whereas it is, of course, outgoing. This means that these two
	geodesics are no longer seen along the same direction, because
	although they have identical~(or almost identical) constants of
	motion, the former is seen as it is still ingoing, whereas the latter
	is seen when it is outgoing~(i.e., if we consider Boyer-Lindquist
	coordinates, each of their $k^\mu$ components are equal to that of the
	other geodesic up to an overall normalization factor, except for the
	$r$ component, which changes sign). In the segment~(along sky
	directions) that joins these two geodesics, we now see bounded
	geodesics. Consequently, the interior of the wormhole silhouette is
	going to be split into two parts: one which, as before, shows
	region~$-7$, and the other that corresponds to bounded geodesics which
	we assume to the devoid of any photon, and hence perfectly black\footnote{One may dispute this choice. Photons travelling along bounded null geodesics may originate from any asymptotic region of the observer's past lightcone. One may therefore consider that looking towards these directions should instead show unpredictable patterns instead of a black patch.}. The
	angular size of this patch is exactly $0$ as long at the observer lies
	at $r \geq r_{\rm r}$ and starts growing from this point. In the usual
	representation where the black hole is spinning counterclockwise if
	seen from above the equator, the dark patch appears on the right
	within the wormhole silhouette since the rightmost point of this
	silhouette is delineated by clockwise equatorial geodesics. Again when
	looking at the black hole or wormhole silhouette at some distance, the
	furthest to the left to point of this silhouette we are looking at,
	the lowest its closest approach point $e$ is in term of the
	$r$-coordinate. The upper and lower edge of the silhouette correspond
	to the polar null unstable geodesics which lies at $r = r_{\rm
		pol}$~(see Eq.~(\ref{def_r_pol})), the leftmost part to
	equatorial prograde geodesics with turning point at $r_{\rm p}$.
	
	As the observer decreases its $r$ coordinates, it is going to
	intercept bounded null geodesics which have excursions outside the
	equatorial plane according to the relation $\mu_{\rm max} (e)$, see
	Figure~\ref{fig_mu2}. Visually this will translate into the fact that
	the patch of bounded null geodesics will increase in size upward and
	downward along the inner part of the wormhole silhouette, while
	increasing in horizontal thickness. Indeed, if we consider for example
	the geodesics coming from regions~$1$ and $-7$ both endowed with
	constant of motion $\xi(r_{\rm r}), \eta(r_{\rm r})$ they will be seen
	further and further from their $r$-coordinate ``loitering point'', so
	that that their opposite $k^r$ component will be increasingly
	different~(opposite to another but with a larger absolute
	value). Conversely, the interior of the wormhole silhouette occupied
	by geodesics originating from region~$-7$ will decrease in size
	accordingly. Numerical investigation show that the wormhole silhouette
	remains reasonably close to circular even when one reaches $r_{\rm
		r}$, and so does the now different patch that shows region
	$-7$. Consequently, the patch of the celestial sphere spanned by
	bounded geodesics takes the shape of some crescent~(of first crescent
	type if we refer to Moon phases seen from the northern hemisphere).
	
	The topology of the scenery will change when the observer reaches $r =
	r_{\rm p}$. There and from now on, all the geodesics starting from
	region~$1$ past null infinity that delineate the wormhole silhouette
	are seen as they are ingoing, whereas all the geodesics that come from
	region~$-7$ are seen outgoing. Therefore, none of the geodesics of the
	first set are seen, in term of angular distance, close to geodesics of
	the second set. In between those two sets, null geodesics that
	intersect observer's trajectory are bounded, therefore, they form a
	thick shell inside which region~$-7$ is seen, a situation which is
	qualitatively similar of the Reissner-Nordstr\"om
	metric~\cite{riazuelo19b}, except that here the patch showing region
	$-7$ is off-center\footnote{Although it can be put in-center by
		performing a Lorentz transformation.}.
	
	The above situation will be also qualitatively the same for a non
	equatorial observer, the only difference being the value of the
	$r$-coordinate at which the different steps will begin. Assuming for
	simplicity that this new observer has a trajectory of decreasing $r$
	but constant $\theta$, then the dark crescent will start of appear at
	the $r_{\rm cresc}$ such that $\mu_{\rm max} (r_{\rm cresc}) = \cos^2
	\theta$, where we choose the largest value of $r$ satisfying this
	constraint~(see Fig.~\ref{fig_mu2}). The dark crescent will transform
	into a dark shell at the second largest root, $r_{\rm shell}$ of the
	equation $\mu_{\rm max} (r_{\rm shell}) = \cos^2 \theta$. One special
	case arises in this context: a polar observer will see an immediate
	transition between region~$-7$ and region~$1$ patches stuck together
	to them being separated by a dark shell at $r = r_{\rm pol}$, just as
	this was the case for the Reissner-Nordstr\"om metric.
	
	When going toward lower values of $r$, three other events plus two
	optional ones are worth mentioning. Among the first three, one occurs
	at outer horizon crossing, the two others after inner horizon
	crossing. The optional events occur also after inner horizon
	crossing~(in case they actually do occur), prior to the last two
	certain ones, which we shall focus on first.
	
	Firstly, when approaching the outer horizon, the patch showing region
	$-7$ is going to shrink to $0$. This can be understood as an extreme
	example of aberration, and/or as a consequence of the fact that
	entering into the outer horizon toward region~$2$ is equivalent to
	leaving region~$1$ which geodesics from region~$-7$ could reach,
	whereas it is not the case for region~$2$. As soon as the observer
	enters region~$2$, region~$3$ enters into its past lightcone and
	hence become visible. Where region~$3$ appears is very easy to
	compute: if we consider a null geodesic coming from region~$-7$ with
	given constants of motion $E$, $L_z$, and $C$, the direction along
	which it is seen by an observer of region~$1$~(where the geodesic is
	outgoing) is uniquely determined by the reduced constants of motion
	$\xi$ and $\eta$. Let us now consider a null geodesic coming from
	region~$3$. Its constant of motion $E'$ is now negative, but
	obviously, the null geodesic with constants of motion $E' = -E$, $L_z'
	= - L_z$ and $C' = C$ possesses the {\it same} reduced constants of
	motion, $\xi$, $\eta$ as the above mentioned geodesic coming from
	region~$-7$. The former, seen just before outer horizon crossing is
	therefore seen along the same direction as the latter, seen
	immediately after horizon crossing. Region~$3$ is therefore going to
	occupy a patch whose size is initially $0$ and that shall grow
	afterward within the patch of bounded geodesics.
	
	Secondly, there is a moment where the observer will reach $r_{\rm
		bubble} \equiv s(r_{\rm shell})$. Since $r_{\rm shell}$ was the
	starting point of the shell-type configuration, with patch of region
	$-7$ and then region~$3$ fully surrounded by the shell of bounded null
	geodesics, $s(r_{\rm shell})$ corresponds to the disappearance of the
	configuration, that is the~(actually temporary) disappearance of
	region~$3$. Such a configuration is not unexpected: except when the
	observer remains within the equatorial plane, decreasing $r$ while
	keeping its $\theta$ fixed makes the observer exit the inner
	ergoregion, within which region~$3$ could be seen from region
	$5$. Within region~$5$ but outside the inner ergoregion, there is no
	possibly to see region~$3$. The patch of bounded null geodesics
	which was at this stage like a shell therefore transforms into a
	bubble, whose name is inspired by the fact that we noticed from the
	numerical simulations that this region was reasonably close to a
	disk~(although slightly oblate for the configurations we studied).
	
	Thirdly, we going even closer to $r = 0$, the observer will reach
	$r_{\rm dis} \equiv s(r_{\rm shell})$, which means that it will exit
	the region that can be reached by bounded null geodesics. In this case
	the patch on the celestial sphere which showed them will simply
	disappear.
	
	As an option, if the observer is sufficiently close to the equatorial
	plane, more precisely if its~(assumed constant) $\theta$ is such that
	$\cos^2 \theta < \mu^2_{\rm max} (e_{\rm min})$~(see
	Eq.~(\ref{mu2max_emin})), then the observer will intersect a second
	set of null bounded geodesics, as early as $r_{\rm i}$ if it lies on
	the equatorial plane, and as late as $r = e_{\rm min}$ if it has the
	highest allowed latitude in order to do so, given by
	Eq.~(\ref{mu2max_emin}).
	
\begin{figure}[h]
		\centerline{\includegraphics*[width=4.8in]{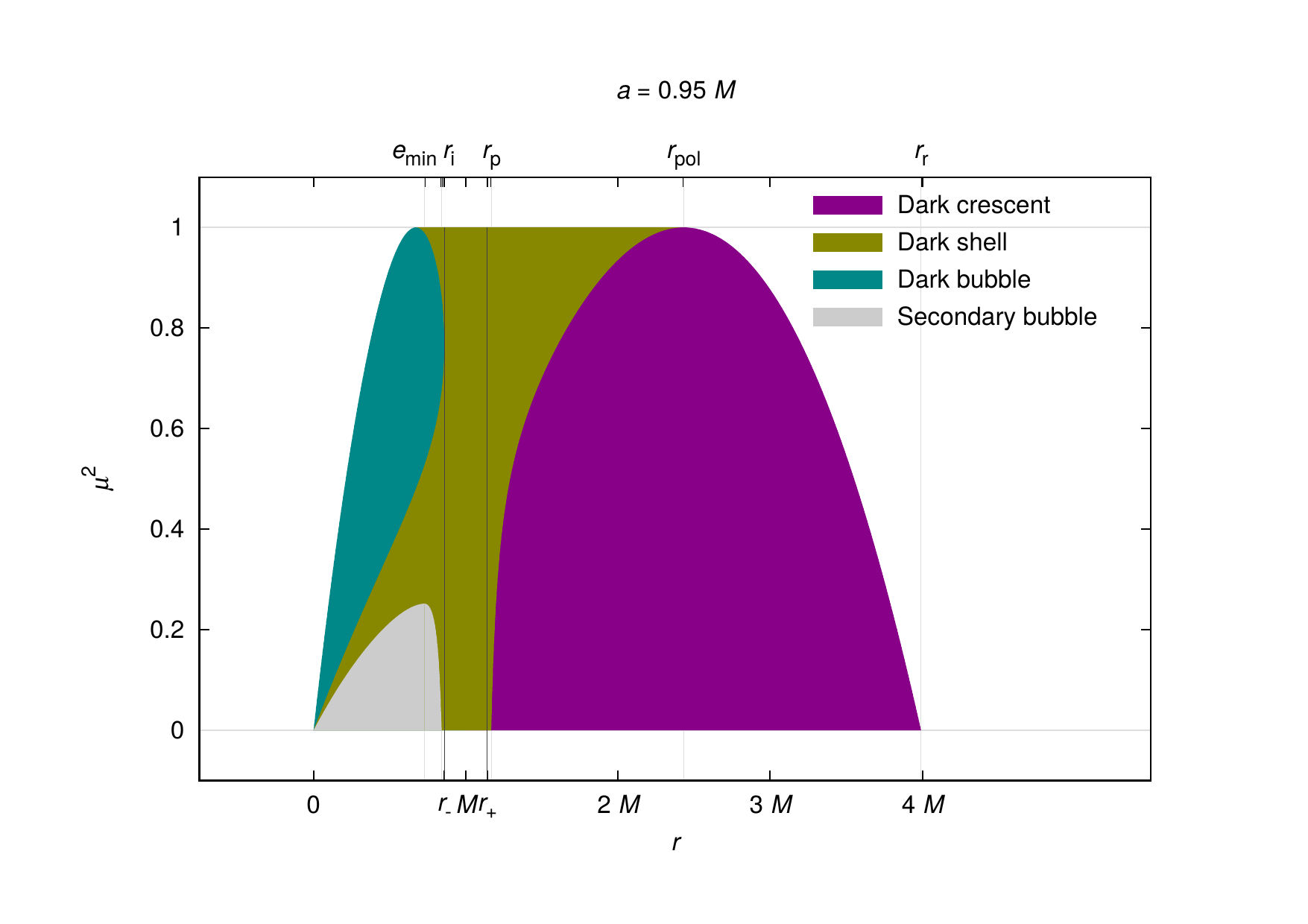}}
		\caption{Regions in the $(r, \theta)$ plane of the observer's position
			which show different types/topologies of the patch(es) occupied by
			bounded null geodesics. For any $r$ larger than $r_{\rm r}$, no such
			geodesics are seen. They then occupy a crescent-shaped zone in
			between patches showing regions~$1$ and $-7$~(assuming the observer
			lies in region~$1$) at least till $r = r_{\rm pol}$~(purple
			zone). Then the crescent becomes a thick shell separating
			regions~$1$ and $-7$ and, after outer horizon crossing, regions~$1$
			and $3$~(dark yellow zone). After inner horizon crossing, several
			possibilities arises, allowing the bounded region to take the shapes
			of a bubble, because region~$3$ is no longer visible~(dark blue
			zone), or, alternatively, a secondary bubble within region
			~$3$~(gray zone). In any case, the bounded geodesic cease to be
			visible for sufficiently small $r$, unless the observer lies on the
			equatorial plane. The figure is made for $a = 0.95 M$. Other values
			of $a$ give the same structure~(as long as $a^2 < M^2$, of course),
			however some zones quickly become hard to see even for moderately
			smaller values of $|a|$. The two horizons are indicated for
			convenience as dark Grey vertical lines.}
		\label{fig_mu22}
	\end{figure}

	The overall possibilities regarding the shape and topology of the
	bounded geodesic patches are summarized in Fig.~\ref{fig_mu22}.
	By looking at this Figure, it appears that for some $\mu^2 = \cos^2
	\theta$ of the observer, the topology of the bounded geodesics
	immediately changes from the dark shell to the dark bubble
	configuration at inner horizon crossing. Given the previous
	discussion, this amounts to say that among the bounded geodesics,
	i.e., null geodesics which admit a double real root at their largest
	root of $R (r) = 0$, there is one whose turning point lies exactly at
	$s (e) = r_-$. Although it might seem unexpected since one may think
	that a geodesic crossing the inner horizon takes some time to bounce
	back from a negative $\dot r$ to a positive $\dot r$, visualization of
	the Carter-Penrose diagram~(see Figure~\ref{fig_carter}) tells that is
	is likely to happen in case some geodesic never enters the inner
	horizon region, i.e. regions~$5$ and $6$ ($\mod 8$, of course), which
	amounts to say that the geodesic passes directly from the ingoing
	inter-horizon region~$2$ to the outdoing inter-horizon
	region~$12$. This happens if $R (r)$ admits $r = r_-$ as a turning
	point. Given the form of $R (r)$, which includes a term proportional
	to $\Delta (r)$~(see Eq.~(\ref{def_R})), this shall happen as soon as
	the remaining term, $(r^2 + a^2) E - a L_z$ cancels out at $r = r_-$,
	which happens for $\xi = L_z / E = (r_-^2 + a^2) / a = 2 M r_- / a$.
	There are many such null geodesics since fixing $\xi$ is not enough,
	as $\eta$ is kept arbitrary. Stating that there exists a bounded null
	trajectory whose apoastron is $e_-$ and
	``periastron''~(notwithstanding the fact that this periastron occurs
	within the object of study) is $s (e_-) = r_-$ is equivalent to
	solving Eq.~(\ref{def_s_m2ems}) for this value, which yields~(using
	the fact that we have of course $a^2 = 2 M r_- - r_-^2$),
	\begin{equation}
	e^4_- - 2 e^3_- (3 M - r_-) + e^2_- (r_-^2 - 4 M r_-+ 9 M^2)
	+ 2 e_- M (r_-^2 - 3 M r_-) + r_-^2 M^2 = 0 .
	\end{equation}
	Setting $f = e_- - (3 M - r_-) / 2$ then leads to a significant
	simplification of the odd terms in powers of $f$ leading to a
	biquadratic equation:
	\begin{equation}
	f^4 - \frac{1}{2} (r_-^2 - 10 M r_- + 9 M^2) f^2
	+ \frac{1}{16} (81 M^4 - 180 M^3 r_-+ 118 M^2 r_-^2 - 20 M r_-^3 + r_-^4) = 0 .
	\end{equation}
	The discriminant of this equation happens to be $0$, so that the
	solution is
	\begin{equation}
	f^2 = \frac{1}{4} (9 M^2 - 10 M r_- + r_-^2) .
	\end{equation}
	Among the two solutions in term of $e$, we must keep those such that
	$e_- > r_{\rm p}$ which necessitates to keep the positive value of $f$,
	so that in the end, we have
	\begin{equation}
	e_{s(e) = r_-} = \frac{3 M - r_- + \sqrt{(9 M - r_-) (M - r_-)}}{2} .
	\end{equation}
	(The same reasoning does not apply to $r_+$ because although the above
	derivation remains valid, the corresponding value of $f^2$ is now
	negative.)  Knowing this value of $e_-$ and already knowing the value
	of $\xi$, we can deduce the values of $\eta$ through
	Eq.~(\ref{def_eta_par}) or any other equation involving $e$ and $\eta$
	and $\xi$ and then deduce through Eq.(\ref{def_Th}) by imposing $\dot
	\theta = 0$ the corresponding value of $\mu$. We did not find any
	simple expression for this, however.
	
	\section{Cartesian Kerr-Schild coordinates}
	\label{sec_KScart}
	
	\subsection{Why these coordinates are needed}
	
	The last bit of implementation we need in order to simulate view deals
	with geodesics passing though the ring singularity, whether they are
	of transit or adventurous type. These geodesics are not mere
	academic features. There {\em always} within reach of an observer
	unless it is situated in the ingoing inter-horizon region~($2 \mod
	8$). Unfortunately, the Kerr-Schild coordinates are not only singular
	along the polar axis~(as any spherical -- or, here, spheroidal --
	coordinates), but also at $r = 0$. The reason is that it is now the
	$\theta$ coordinate that is not regular.
	
	This can be seen by switching from the $\theta$ coordinate to the
	pseudo Cartesian coordinate $z \eqdef r \cos \theta$. Passing through
	the ring singularity means that if we are in, says, region~$5$ with
	positive $z$, we shall end in region~$7$ with negative $z$, just as if
	we went though a $z = 0$ plane in Cartesian coordinates. But in region
	$7$, $r$ is negative, so that {\em both} $z$ and $r$ change sign in
	the process. Consequently, if $z$ varies linearly in the process,
	$\cos \theta$ does not since it does not change sign. In other words, $\theta$
	varies like
	\begin{equation}
	\theta (p) \sim \frac{\pi}{2} + Z \left|p - p_0\right| ,
	\end{equation}
	where $Z$ is some constant and $p_0$ is the value of affine parameter
	$p$ at ring singularity crossing. Consequently, although $\theta (p)$
	is continuous at this event, $\dot \theta$ is not, so that
	Boyer-Lindquist coordinates cannot be used there.
	
	This has to be overcome by switching to Cartesian-like coordinates
	$(r, \theta, \tilde \varphi) \to (x, y, z)$. The fact that when $M =
	0$, the $g_{\tilde \varphi \tilde \varphi}$ coefficient is of the form
	$(r^2 + a^2) \sin^2 \theta$ suggests that an good definition of $x$
	and $y$ is such that $x^2 + y^2 = (r^2 + a^2) \sin^2 \theta$. It then happens
	that a definition using complex numbers is more convenient, although
	is does depend on the choice of $\epsilon$. After some tries, one
	obtains that assuming
	\begin{eqnarray}
	\label{def_xiy}
	x + i y & = & (r + i \eta a) \SN \exp(i \tilde \eta \tilde
	\varphi) , \\
	z & = & r \CS ,
	\end{eqnarray}
	and assuming that the two quantities $\eta$, $\tilde \eta$ are both
	equal to $\pm 1$, the new coordinate system is simply a Cartesian one
	in the case $M = 0$ if the following constraint is satisfied:
	\begin{equation}
	\eta \tilde \eta = \epsilon .
	\end{equation}
	In what follows, we shall shall keep the usual orientation
	convention~(despite $r$ being negative) and therefore impose that
	$\tilde \eta = 1$, so that $\eta = \epsilon$.
	
	When performing this transform, four-velocity/wavevector components
	also have to be changed. This is done using the Jacobian whose
	components are very easy to compute. One has for example
	\begin{equation}
	V^x = \frac{\partial x}{\partial r^\mu} V^\mu ,
	\end{equation}
	where $V^\mu$ are the three spatial components expressed in
	Kerr-Schild spherical coordinates $r^\mu = (r, \theta, \varphi)$. The
	inverse transform is then obtained by inverting this matrix, which
	does not seem to possess a simple form\footnote{Consequently, when we
		need this inverse matrix, we compute the direct matrix and invert it
		numerically.}. One extra difficulty arises from the fact that even
	if we do not perform any coordinate change, we must keep track of the
	value of $r$. However, the relation between $r$ and the pseudo
	Cartesian coordinates $x$, $y$, $z$ involve only even powers of each
	coordinates:
	\begin{equation}
	\label{def_r2_cart}
	r^4 - r^2 (x^2 + y^2 + z^2 - a^2) - a^2 z^2 = 0 .
	\end{equation}
	This means that when we compute $r$ back from the $(x, y, z)$, we must
	keep track on which was the previous value of $r$ and whether on has
	passed though the ring singularity. The way this will be done in the
	case of null geodesics we are considering will be explained later,
	after we explain the properties of null geodesics that are of interest
	here.
	
	\subsection{New equations of motions}
	
	Having defined the Cartesian Kerr-Schild coordinates, we now need to
	find the equations of motions in such coordinate system. The
	coordinate that most easily allows to compute its equation of motion
	is $z$. After short manipulations, we find
	\begin{equation}
	\label{ddot_z_KS}
	\frac{\Sigma^2}{\CS} \ddot z
	=   \frac{R'}{2} - \frac{2 r}{\Sigma} R
	- r \frac{\Theta'}{2} \frac{\SN}{\CS}
	- r \Theta \left( 1 + \frac{2 a^2 \SN^2}{\Sigma} \right) .
	\end{equation}
	This equation of motion can further be expressed into a variety of
	ways, depending on whether one chooses to group first derivatives of
	the coordinates into constants of motion. Keeping in mind that in the
	case where $M = 0$, the metric reduces to Cartesian coordinates, we
	expect that the equation of motion can be cast such that all terms in
	the right-hand side are proportional to $M$, just as this would be the
	case in a Newtonian framework and as it was also shown to be the case
	for the Schwarzschild metric~\cite{marck96}. In this case, the next, most
	straightforward, step of the derivation is to expand the above
	expression by using
	Eqns.~(\ref{def_R},\ref{def_Th},\ref{def_Rp_2_v0},\ref{def_Thp_2}).
	
	All the term including powers of $E$ and $L_z$ cancel away, and so do
	terms proportional of $C$ and $\kappa$, except those among these that
	are proportional to $M$. We then obtain
	\begin{equation}
	\Sigma^2 \ddot z
	= - M z \left[  \kappa r \left(\frac{4 r^2}{\Sigma} - 3 \right) 
	+ \frac{C}{r} \left(\frac{4 r^2}{\Sigma} - 1 \right) \right] .
	\end{equation}
	
	The equation of motion for $x$ and $y$ is more difficult to
	obtain. Starting from the definition~(\ref{def_xiy}), and performing
	two derivative with respect to proper time/affine parameter, we obtain
	straightforwardly
	\begin{eqnarray}
	\label{ddotxiy}
	\Sigma^2 (\ddot x + i \ddot y)
	& = & 4 i M r a \Ldot (\dot x + i \dot y) \\
	\nonumber & &
	+ \frac{x + i y}{r^2 + a^2} 
	\left[  (r^2 + a^2) F + r A + a B
	+ i \epsilon \left( (r^2 + a^2) D + r B - a A\right) \right] ,
	\end{eqnarray}
	where the quantities $A$, $B$, $F$ and $D$ are defined by
	\begin{eqnarray}
	A & = & \frac{R'}{2} - \frac{2 r}{\Sigma} R , \\
	B & = &   2 \Sigma^2 \dot{\tilde \varphi} \epsilon \dot r 
	- 4 M r a \Ldot \epsilon \dot r , \\
	F & = &   \CS^2 \frac{\Theta'}{2 \CS \SN}
	+ \left(1 - \frac{2 r^2}{\Sigma} \right) \Theta
	- \Sigma^2 \dot{\tilde \varphi}^2
	+ 4 M r a \Ldot \dot{\tilde \varphi} , \\
	D & = & a X - 2 \Sigma \dot {\tilde \varphi} r \epsilon  \dot r .
	\end{eqnarray}
	The next step is to expand these four quantities, which transform into
	\begin{eqnarray}
	A
	& = &   \frac{2 r}{\Sigma}
	\left( (r^2 + a^2) E - a L_z \right)
	\left( a L_z - a^2 \SN^2 E \right) \\ \nonumber & & 
	+ r \left[\frac{2 (r^2 + a^2)}{\Sigma} - 1 \right] (C + \kappa r^2 )
	- r (r^2 + a^2) \kappa \\ \nonumber & & 
	- M C \left(\frac{4 r^2}{\Sigma} - 1 \right)
	- M \kappa r^2 \left(\frac{4 r^2}{\Sigma} - 3 \right) \\
	B & = &   2 \Sigma^2 \dot{\tilde \varphi} \epsilon \dot r 
	- 4 M r a \Ldot \epsilon \dot r , \\
	F & = &   \left(\frac{2 r^2 \SN^2}{\Sigma} - 1 \right) 
	\left(a E - \frac{L_z}{\SN^2} \right)^2
	- \CS^2 \left(a E - \frac{L_z}{\SN^2} \right) 2 \frac{L_z}{\SN^2}
	\\ \nonumber & &
	+ \left(1 - \frac{2 r^2}{\Sigma} \right) (C + \kappa r^2)
	+ r^2 \kappa , \\
	D & = &   a r \Ldot^2 - 2 a r \Ldot \epsilon \dot r
	+ 2 r a^2 \SN^2 \Ldot \dot{\tilde \varphi}
	- 2 \Sigma r \epsilon \dot r \dot{\tilde \varphi}
	- a r \kappa
	- a M \Ldot^2 \left( \frac{4 r^2}{\Sigma} - 1 \right) .
	\end{eqnarray}
	Grouping these terms as in Eq.~(\ref{ddotxiy}) then gives
	\begin{eqnarray}
	(r^2 + a^2) F + r A + a B
	& = & - \left(\frac{4 r^2}{\Sigma} - 1 \right) M r C 
	- \left(\frac{4 r^2}{\Sigma} - 3 \right) M r \kappa r^2        
	\\ \nonumber & &
	- 4 M r a^2 \Ldot (\epsilon \dot r - a \SN^2 \dot{\tilde \varphi}) 
	+ 2 M r a^2 \Ldot^2 , \\
	(r^2 + a^2) D + r B - a A
	& = &    \left(\frac{4 r^2}{\Sigma} - 1 \right) M a (C - a^2 \Ldot^2)
	\\ \nonumber & &
	+ \left(\frac{4 r^2}{\Sigma} - 3 \right) M a (\kappa r^2 - r^2 \Ldot^2)
	\\ \nonumber & &
	- 4 M r^2 a \Ldot (\epsilon \dot r - a \SN^2 \dot{\tilde \varphi}) .
	\end{eqnarray}
	A large number of terms appears in both expression, and it is easy to
	group them according to
	\begin{equation}
	\begin{aligned}
	(r^2 + & a^2) F + r A + a B
	+ i \epsilon \left( (r^2 + a^2) D + r B - a A \right) ={} \\ &
	- \left(\frac{4 r^2}{\Sigma} - 1 \right) M (C - a^2 \Ldot^2)
	(r - i \epsilon a) \\ &
	- \left(\frac{4 r^2}{\Sigma} - 3 \right) M r (\kappa r + i \epsilon a \Ldot^2)
	(r - i \epsilon a) \\ &
	- i 4 M r a \Ldot (r - i \epsilon a) 
	(\dot r - \epsilon a \SN^2 \dot{\tilde \varphi}) .
	\end{aligned}
	\end{equation}
	This being done, one obtain a rather compact equation of
	motion for $x$ and $y$:
	\begin{eqnarray}
	\Sigma^2 (\ddot x + i \ddot y)
	& = & 4 i M r a \Ldot \left( \dot x + i \dot y - \frac{x + i y}{r + i \epsilon
		a}
	\left(\dot r - \epsilon a \SN^2 \dot{\tilde \varphi}\right) \right)   \\
	\nonumber & &
	- M \frac{x + i y}{r + i \epsilon a} 
	\left[  \left(\frac{4 r^2}{\Sigma} - 1 \right) (C - a^2 \Ldot^2)
	+ \left(\frac{4 r^2}{\Sigma} - 3 \right) r 
	(\kappa r + i \epsilon a \Ldot^2) 
	\right] .
	\end{eqnarray}
	A last rearrangement further allows to find a more similar-looking form as
	compared to the $z$ equation by transforming the last $\Ldot^2$ into a $\kappa$
	and by putting the corresponding difference in the first line. The whole,
	final,
	set of equations then reads:
	\begin{eqnarray}
	\label{eq_xiy_final}
	\ddot x + i \ddot y
	& = & 4 i M a \frac{r}{\Sigma^2} \Ldot \left[ \dot x + i \dot y - \frac{x + i
		y}{r + i \epsilon a}
	\left\lbrace \dot r - \epsilon a \SN^2 \dot{\tilde \varphi} +  \left(\frac{4
		r^2}{\Sigma} - 3 \right) \epsilon  \frac{\Ldot^2 - \kappa}{4 \Ldot }   \right
	\rbrace \right]  \\
	\nonumber & &
	- M (x + i y) \frac{r}{\Sigma^2}
	\left[ \left(\frac{4 r^2}{\Sigma} - 3 \right) \kappa  
	+  \left(\frac{4 r^2}{\Sigma} - 1 \right) \frac{C - a^2 \Ldot^2}{r (r + i
		\epsilon a)} 
	\right] , \\
	\label{eq_z_final}
	\ddot z
	& = & - M z \frac{r}{\Sigma^2} \left[ \left(\frac{4 r^2}{\Sigma} - 3 \right) 
	\kappa 
	+ \left(\frac{4 r^2}{\Sigma} - 1 \right)  \frac{C}{r^2} \right] .
	\end{eqnarray}
	As a first crosscheck, we can
	notice that when setting $a = 0$, these equations reduce to those that
	are valid for the Schwarzschild metric~(see Ref.~\cite{marck96}). The term
	proportional to $\kappa$ corresponds to the Newtonian term and  after
	performing
	the substitution $C \to L^2$, where $L^2$ is the particle total angular
	momentum~(or angular momentum per unit of mass for timelike
	geodesics), this term corresponds to the Schwarzschild term. Moreover,
	expanding
	this expression in powers of $a$ allows to recover the gravitomagnetic term for
	non relativistic particles. We give some details about this
	in~\ref{app_gravitom}.
	
	We can now go back to the problem of determining how to handle ring
	singularity crossing. This is done as follows. Firstly, we care about
	this issue only for geodesics that have been identified as crossing
	the ring singularity, see \S\ref{diff_type_geod}. Secondly, it is
	known that null geodesics either oscillate around the equatorial plane
	or exist in a limited interval of $\theta$ which is comprised
	somewhere within $]0, \pi/2[$~(or within $]\pi/2, \pi[$). Thirdly, it
	has been shown~\cite{chandrasekhar83} that null geodesics crossing
	the ring singularity only belong to this latter category.
	Consequently, geodesics crossing the ring singularity will
	experience simultaneously a sign flip for both $z$ and $r$, and,
	according to the third point above, the flip in $z$ occurs only
	once, at ring singularity crossing. Therefore, the evolution of
	the $z$ quantity suffices to detect the ring singularity crossing
	at which the sign of $r$ must be flipped when solving
	Eq.~(\ref{def_r2_cart}).
	
	Incidentally, this means that there are not one set of Kerr-Schild
	coordinates, but {\em four}. First because we must choose among the
	two values of $\epsilon$~(although the choice of value does not matter
	at ring singularity crossing), and second we must consider the choice with the
	product $z r$ being either positive or negative. The choice $z r > 0$
	is suitable when we go from the $z , r > 0$ region to the $z, r < 0$ one,
	that is we enter into the negative $r$ region by diving into the ring
	singularity from above. This same coordinate system~(or these same
	coordinate systems, since there are two such systems depending on the
	value of $\epsilon$) can also be used when going back to the $r > 0$
	side by springing upward from the ring singularity. The two other sets
	of Kerr-Schild coordinates~(with $z r < 0$ whichever value of
	$\epsilon$) must be used for the two other crossings~(going into the
	negative $r$ region from below the equatorial plane, etc.).
	
	In the problem of horizon/ring singularity crossings, the last but
	crucial step is to find a procedure for the choice of the affine
	parameter so that the jumps in $r$ at each integration step are not
	too large so that we have time to switch from one coordinate system
	to another when necessary. For example, if we are in region~$1$ and integrate
	backward a geodesic coming from the outer horizon, then this geodesic
	originates further back from the inner horizon. If the value of
	$\epsilon$ must be changed when crossing these two horizons, then our
	step of integration must send $r$ first in the inter-horizon
	region~($4$ in this case) before switching to the other set of
	Kerr-Schild coordinate an continue the integration for the inner
	horizon crossing. This is done by finding upper bounds on the
	possible variations of $r$ as a function of the affine parameter by
	fiddling with Eq.~(\ref{def_R}). We then use standard numerical rounites~\cite{press92} to solve the geodesic equations.

	\section{Simulated views, 2nd part}
	\label{views_2}
	
	We now come to the graphical illustrations of the previous discussions
	of \S\ref{sec_class}--\ref{sec_KScart}. Details of our raytracing
	software has been given elsewhere~\cite{riazuelo19a}, so that we shall
	only very briefly summarize the features that are of interest here.
	
	Since in many situations we need to observe our scenery along very
	different directions~(in term of angular distance), we need to compute
	simulation on a very large field of view. The most convenient
	projection we found for this purpose is the azimuthal equidistant
	projection, which is now commonly used in any digital planetarium
	under the nickname of DomeMaster format\footnote{This is actually the
		reason why we implemented this projection in our software.}.
	
	Our simulations take into account aberration due to observer's motion
	and can also implement Doppler effect. They naturally take into
	account lensing of extended objects by showing their pixel by pixel
	distortion. We can also compute lensing of individual stars,
	i.e. magnification. However, in order to focus on the aspects of
	visualization that have been described above, we chose to either
	strongly attenuate or even remove the Doppler shift from our simulated
	views, as those induced far too contrasted images. We also removed
	stars from our simulations, the main reason being that computing the
	distortion of a celestial sphere amounts to solve backward the
	geodesic equations from the observer position to some past null
	infinity, which can be done in one shot. Adding star necessitates to
	further perform direct raytracing, which is much more
	time-consuming. Moreover, some extreme lensing phenomena prevented
	from obtaining a satisfactory rendering.
	
	The main issue we had to address was the distortion pattern of the
	different asymptotic regions our fiducial observer would see as a
	function of its position. Our first guess was to use simple coordinate
	grids for each asymptotic regions, but those happened to be too
	complicated to interpret in most contexts. Moreover, the different
	asymptotic regions were too difficult to distinguish from another in
	this case. Therefore, it was was mandatory to use several different
	celestial spheres:
	\begin{itemize}
		
		\item Observer's initial region, region~$1$ is covered by the
		starless Milky Way seen in the near infrared as observed by the
		Two-Micron All-Sky Survey~\cite{2mass06}.
		
		\item Region~$1$ twin, i.e., region~$3$, used the Milky Way seen by
		Planck satellite High Frequency instrument~(mostly dominated by the
		857~GHz map) after a one-year full-sky survey~\cite{planckhfi}.
		
		\item Negative $r$ regions~$7$ and~$8$ show CMB full-sky maps coming
		from Planck full mission CMB-only map~\cite{planck16} and WMAP
		9~year Internal Linear Combination Map~\cite{wmap}, respectively. We
		also use the same Planck map for region~$-1$ whom a tiny portion is
		seen from region~$1$. This cannot introduce any confusion since
		regions~$-1$ and $7$ cannot be seen simultaneously nor immediately
		one after the other.
		
		\item We made an exception regarding region~$-7$ whose celestial
		sphere is a coordinate grid with $5^\circ \times 5^\circ$ grayscale
		patches~(whose colors actually match some black body). Polar regions
		are orange, and we used different color for four meridians situated
		$90^\circ$ apart from each others~(whitish, light gray, dark and
		dark red). We also used a different colors for the northern and
		southern equatorial bands~(very light gray, and dark, respectively).
		
	\end{itemize}
	
\begin{figure}[ht]
		\centerline{
			\includegraphics*[width=3.2in]{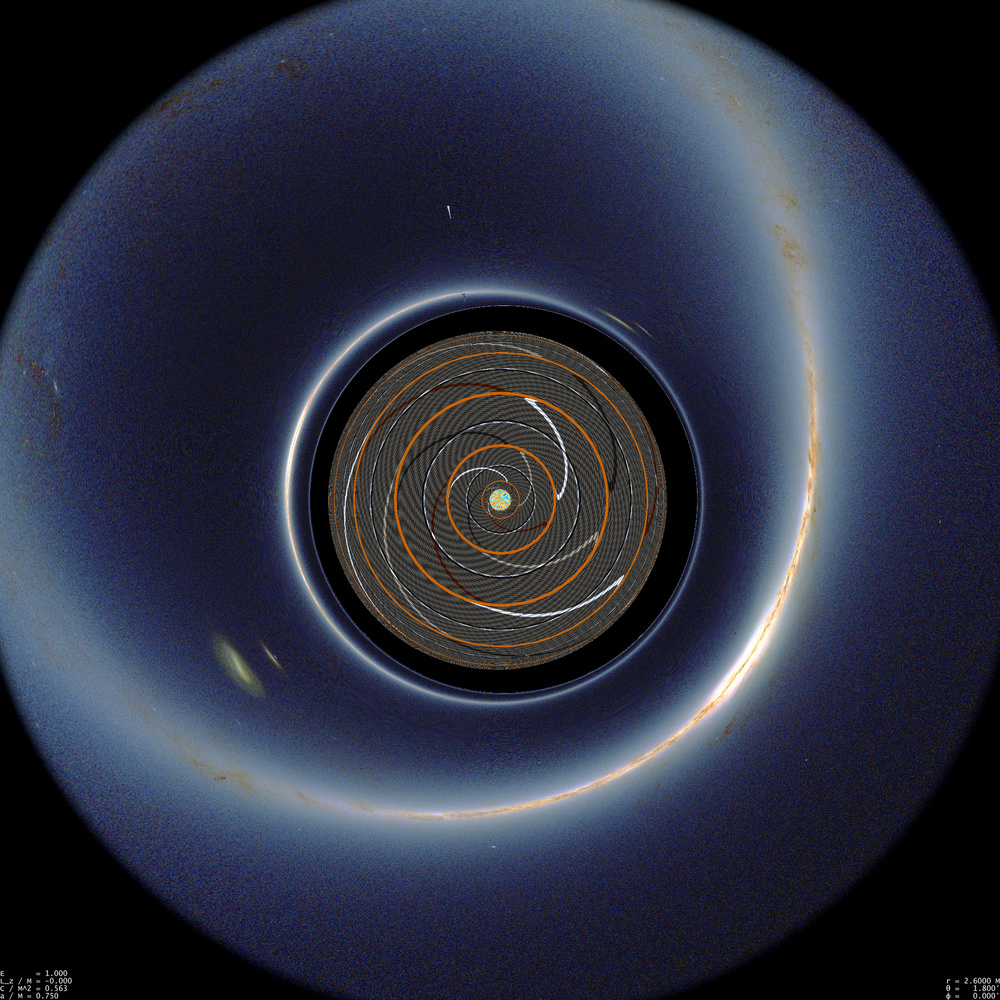}
			\includegraphics*[width=3.2in]{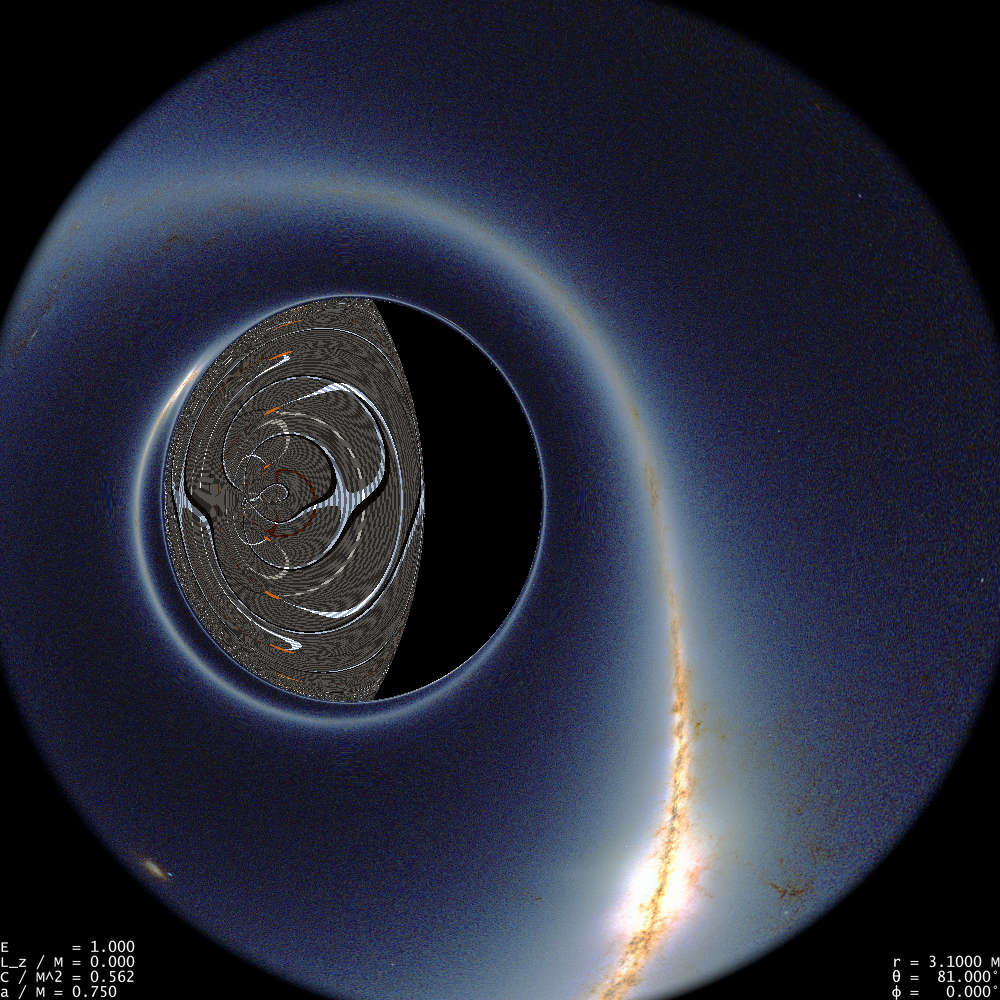}}
		\centerline{
			\includegraphics*[width=3.2in]{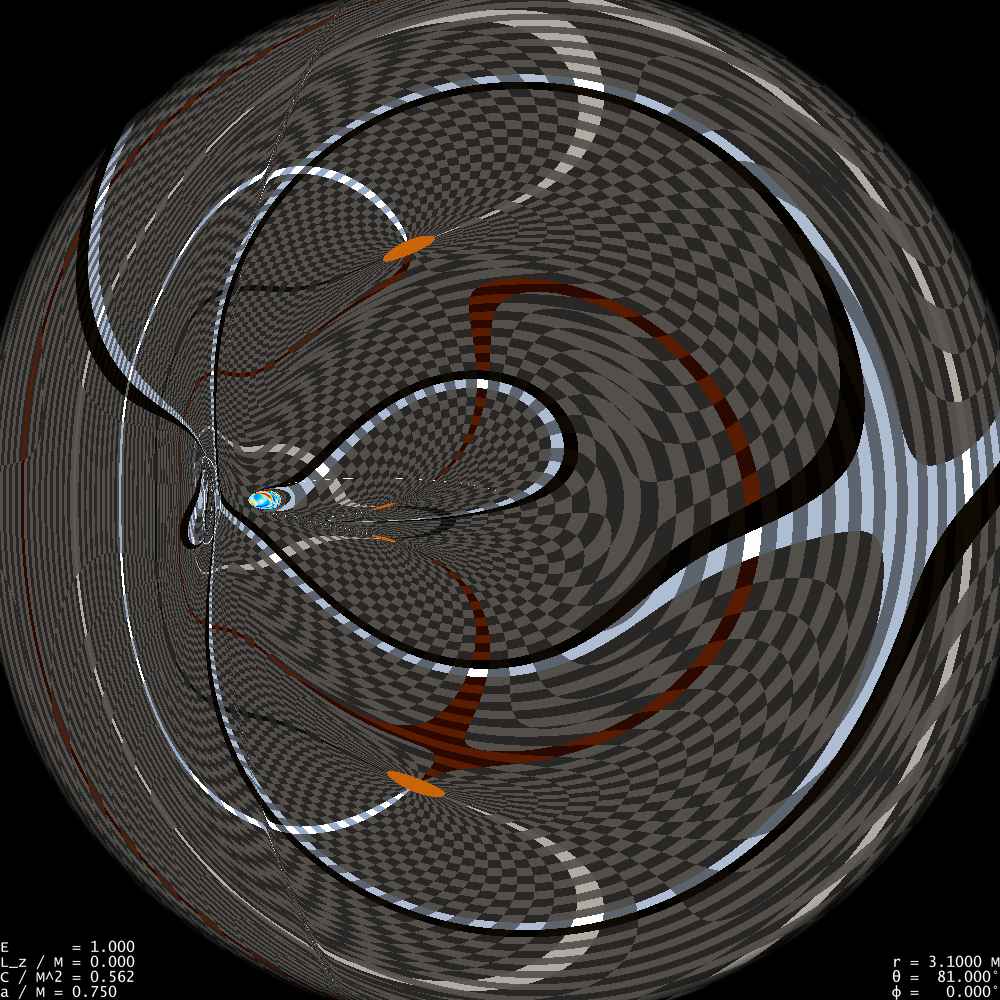}
			\includegraphics*[width=3.2in]{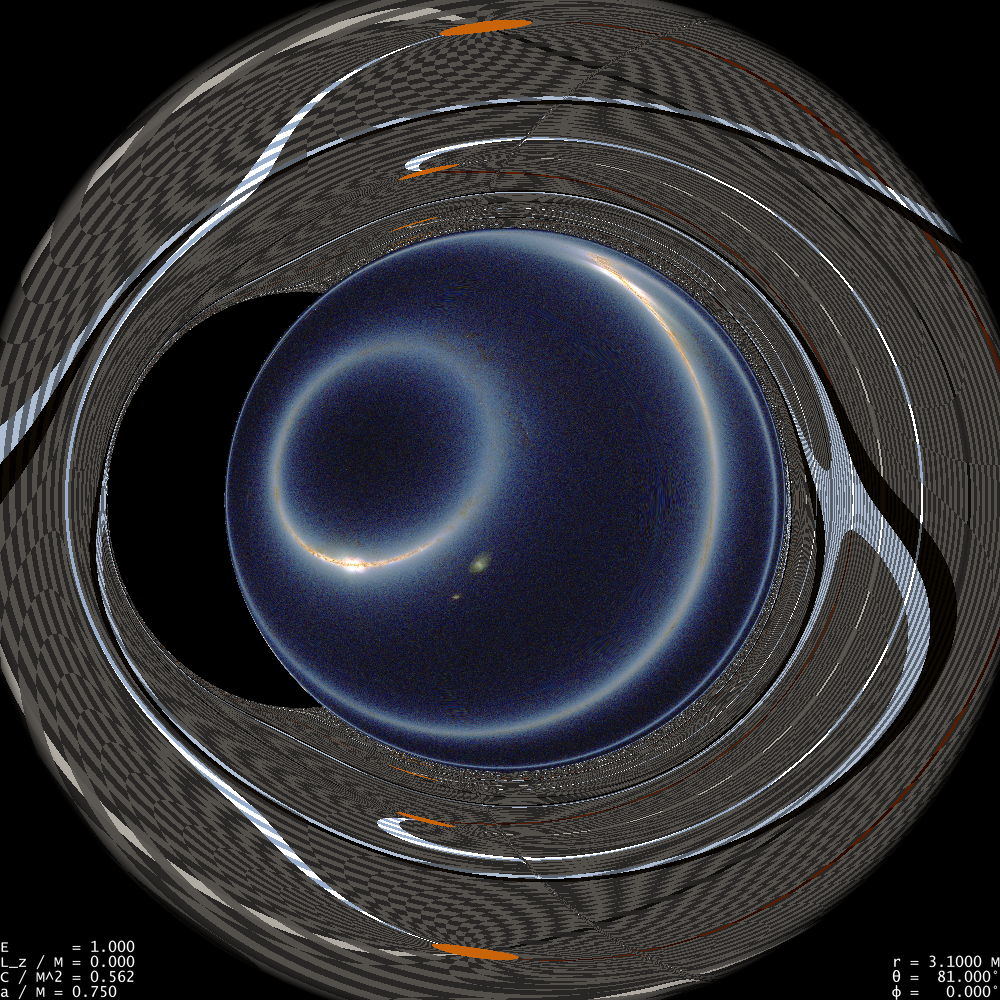}}
		\caption{Observer in three situations: (i) polar, infalling
			trajectory at $r = 2.6 M$~(top left), close-to-equatorial~($\theta
			= 81^\circ$) infalling at $r = 3.1 M$~(top right) and
			close-to-equatorial~($\theta = 81^\circ$) outgoing at $r = 3.1
			M$~(bottom left). This view correspond to the same position as in
			the infalling one, except that we have flipped the sign of the
			observer's $\dot r$. The bottom left view shows the opposite
			direction of that of bottom right view.}
		\label{fig_div}
	\end{figure}

	We first want to illustrate how different the wormhole silhouette can
	look as a function of the observer's latitude and
	velocity~(Fig.~\ref{fig_div}). Along a polar, infalling trajectory,
	the wormhole silhouette appears perfectly circular. Region~$-7$
	occupies almost all of the space within in but only if $r$ is
	sufficiently large. For smaller $r$, it is surrounded by the dark shell
	phenomenon described in the previous Section and which also occurs in
	the Reissner-Nordstr\"om case. However, contrarily to the latter, the
	Kerr case induces a rotational deformation of region~$-7$ as can be
	seen by following the swirling patterns of the meridians. An almost
	polar trajectory from region~$-7$ does not experience the
	Lense-Thirring effect, but so do trajectories with increasing impact
	parameter. Since we are here looking at a counterclockwise rotating
	wormhole from above, the swirls are also counterclockwise from center
	to edge~(Fig.~\ref{fig_div}, top left). Note that in the middle of
	region~$-7$, a small patch of region~$-1$ is also seen. When near the
	equator, the dark shell occurs, but before, the~(infalling) observer
	will see the dark crescent phenomenon~(Fig.~\ref{fig_div}, top right),
	which appear on the retrograde side of the wormhole. Region~$-1$ is
	also seen but this time along a flattened ellipse since the
	singularity is seen much more edge-on. If we now consider the same
	observer but who is now outgoing~(by flipping the $r$-component of
	its four-velocity), a huge aberration ``unfolds'' the whole
	wormhole silhouette which now occupies most of the celestial sphere,
	with an angular diameter significantly larger than
	$\pi$~(Fig.~\ref{fig_div}, bottom left). If one looks along the
	opposite direction of the wormhole~(Fig.~\ref{fig_div}, bottom
	right), the current region the observer is in appears within a small
	patch of the sky. The dark crescent is still there, but it now looks
	like a sort of ``outgrowth'' protruding over region~$1$'s patch.
	
	After this warm-up, we now show a more detailed sequence seen by an observer travelling from region~$1$ to region~$7$.
We shall consider an observer freely-falling on the
	Kerr wormhole starting from a zero velocity and angular momentum at
	infinity, i.e., $E = 1, L_z = 0$. The Carter constant is chosen as $C
	= a^2$ so as $\theta$ to remain constant. If we want to see through
	the ring singularity, we cannot consider the case of an observer being
	on the equatorial plane, but a too large latitude will prevent us from
	seeing all the features outlined in Fig.~\ref{fig_mu22}. We therefore
	choose $\theta = 81^\circ$~(i.e., northern latitude of
	$9^\circ$). With these constants of motion, the observer will
	penetrate into the wormhole, cross the two horizon and then bounce
	when reaching the ring singularity~($r = 0, \theta = 81^\circ$). We then
	decide to push the observer in the negative $r$ region. This part of the
	journey
	will not be done along a geodesic. In order to do so we would need to
	start with a fairly high value of $E$, which would significantly
	shrink several zones of interest because of aberration. Therefore, the
	journey into the $r < 0$ region will be shown as seen by a quasistatic
	observer. Even with all these requirements, it is not very easy to find
	a value of $a$ which allows to see well all the features related to
	the null geodesics. After several tries, we found that the value of $a
	= 0.75 M$ was a good compromise. Also, the values of $\theta$ and $a$
	allow to intersect all the regions of interest regarding the bounded
	geodesics of Fig.~\ref{fig_mu22}: the complete wormhole, then the dark
	crescent, the dark shell, the dark shell together with the secondary
	bubble, then the dark shell alone again, the dark bubble and then
	nothing, followed by the crossing of the ring singularity.
	
	During the overall trajectory there are huge Doppler effects that are
	involved in many directions, thus making the images impossible to
	interpret. We therefore removed on purpose such Doppler shift in order
	to focus on the shape of the different patches of the sky that we can
	see. Lastly, as the observer is moving, its $\varphi$
	coordinate~(in the Boyer-Lindquist frame) varies a lot. However, we
	found that this variation led to more confusion than enlightenment as
	this drift in $\varphi$ led to large changes in the aspect of the
	various celestial spheres. Therefore we have~(rather disputably, we
	admit) chosen not to follow the evolution of the $\varphi$ coordinate,
	despite computing the view with the correct value of $\dot \varphi$,
	i.e., the one imposed by the constants of motion we have chosen.
	
	This being said, the next sets of images show what we think are the
	most interesting features of this numerical exploration.
	
\begin{figure}[ht]
		\centerline{
			\includegraphics*[width=3.2in]{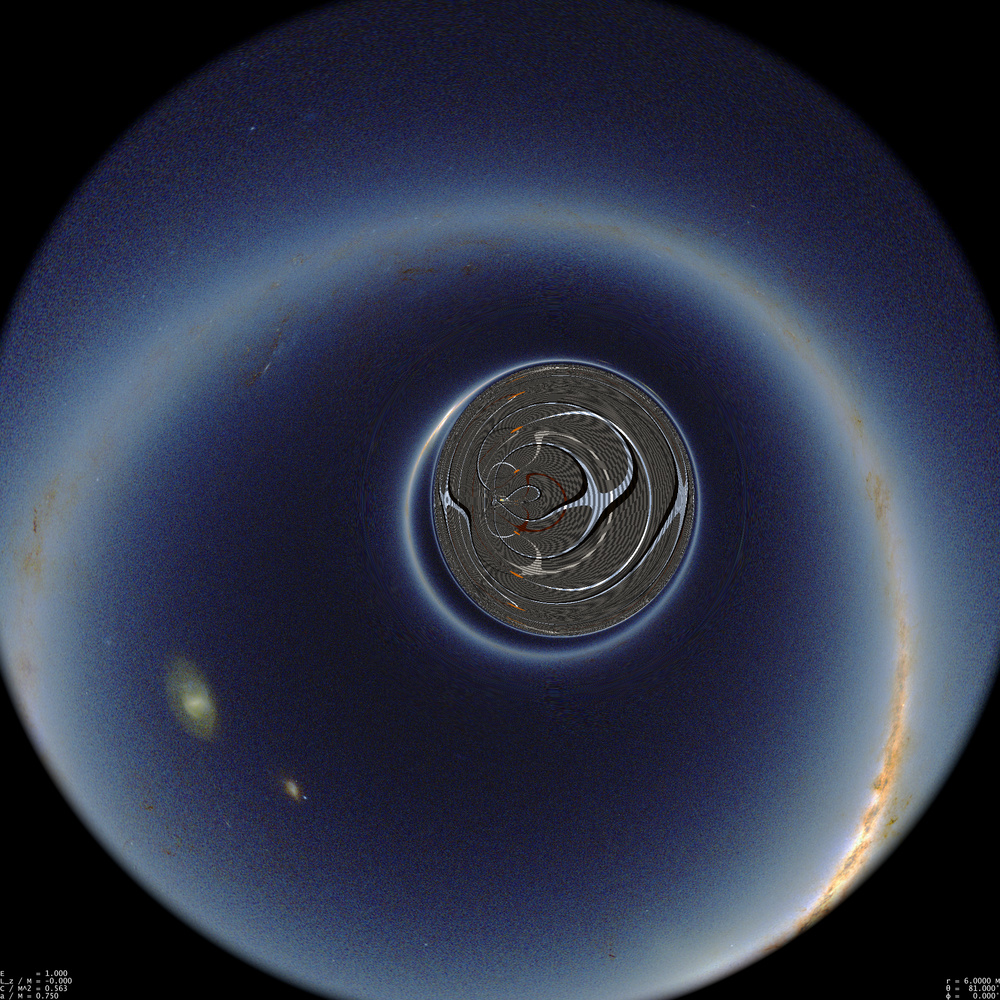}
			\includegraphics*[width=3.2in]{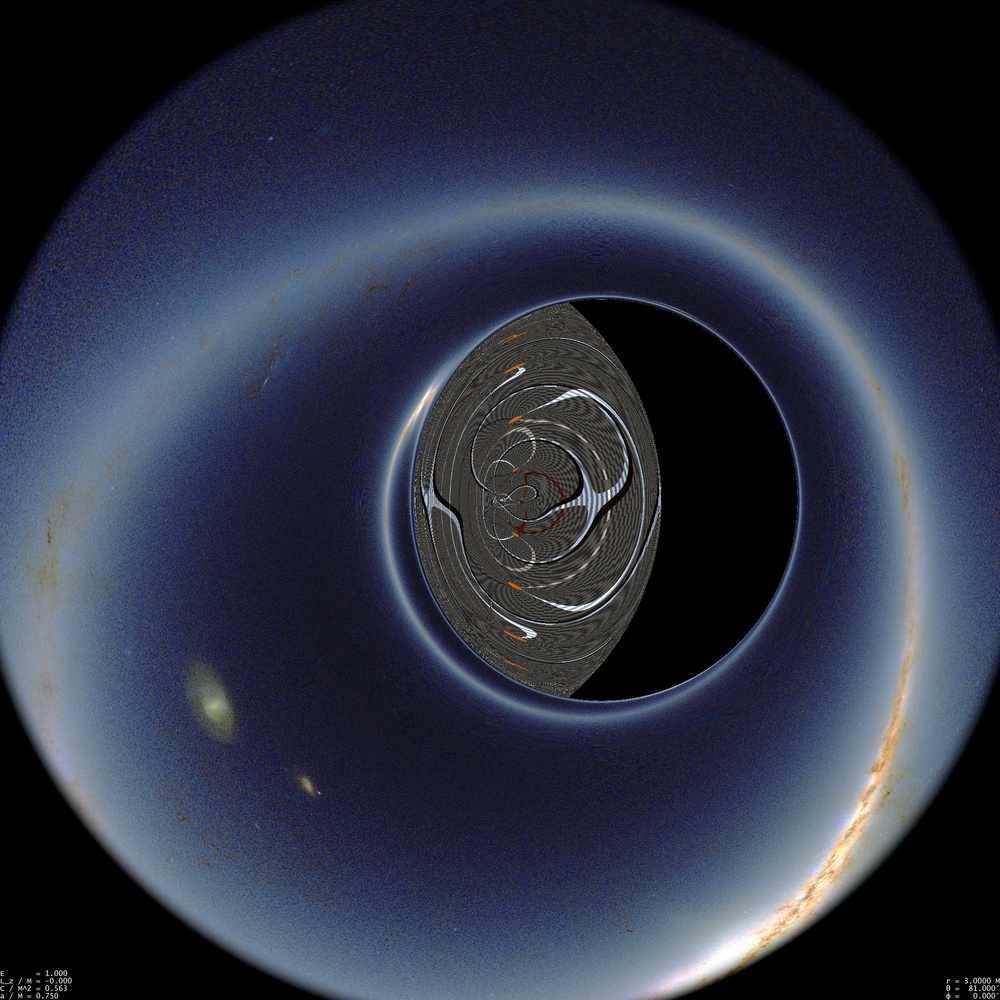}}
		\centerline{
			\includegraphics*[width=3.2in]{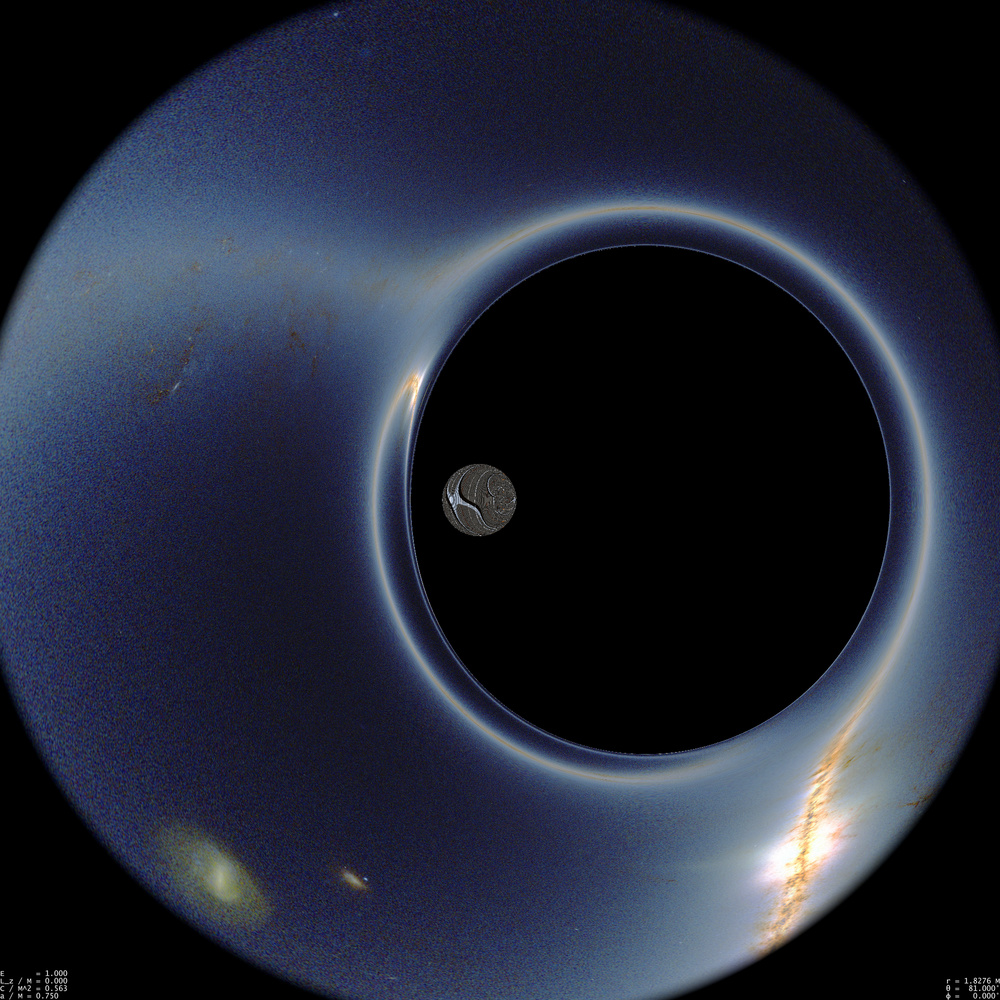}
			\includegraphics*[width=3.2in]{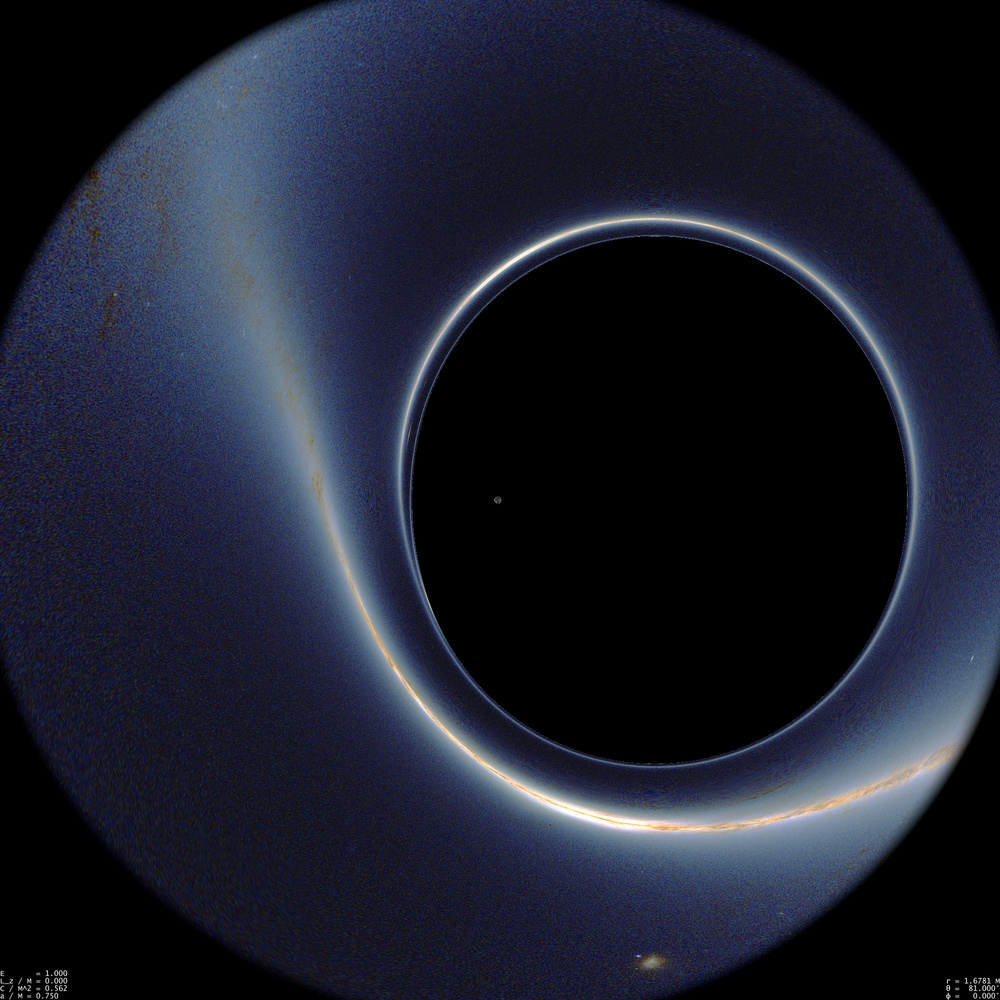}}
		\caption{Infalling observer at $r = 6 M$, $3 M$, $1.8276 M$ and
			$1.6781 M$.}
		\label{fig_seq1}
	\end{figure}

	The first set of images~(Fig.~\ref{fig_seq1}) shows the progress
	toward the wormhole outer horizon at coordinate distance at $r = 6 M$,
	$3 M$, $1.8276 M$ and $1.6781 M$. The first picture is well above the
	bounded geodesics regions. It therefore shown region~1 as well as,
	inside the wormhole silhouette, region~$-7$. Within this region, a
	very tiny bit~(almost impossible to see) of region~$-1$ where $r$ is
	negative. More will be seen later about negative $r$ region, so that
	it is not necessary to focus on it at this stage. The shape of the
	wormhole silhouette is slightly different than in Fig.~\ref{fig_r6m},
	first because the observer is slightly off the equatorial plane and
	mostly because the black hole is far from extremal for this
	purpose~(the black hole shape with a flattened prograde side occurs
	only when $a$ is very close to $M$. Inside the wormhole, we see a very
	distorted view of region~$-7$ The colored poles as well as the four
	main meridian and the equator form somehow regular but highly
	distorted patterns. At $r = 3 M$, the observer enters into the region
	where bounded geodesics are seen. The dark patch corresponding to
	those first appears on the retrograde side of the wormhole. This is
	what we had called the dark crescent. At $r = 1.8276 M$, the dark
	crescent has spread over the whole wormhole edge. However since we are
	still outside the outer horizon, region~$-7$ is still visible. At $r =
	1.6781 M$, the size of region~$-7$ has drastically diminished. This is
	because the observer is very close to the wormhole outer horizon~($r_+
	= 1.6614 M$ with our value of $a$).
	
\begin{figure}[ht]
		\centerline{
			\includegraphics*[width=3.2in]{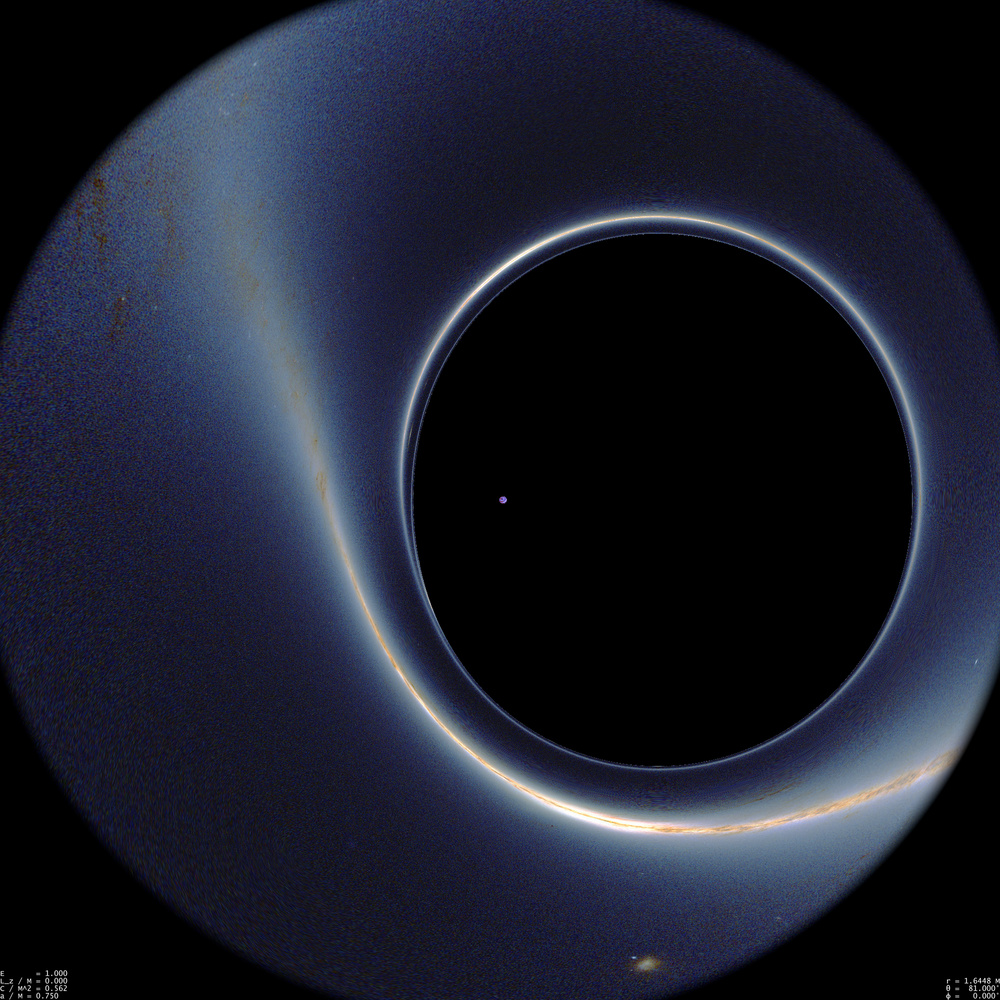}
			\includegraphics*[width=3.2in]{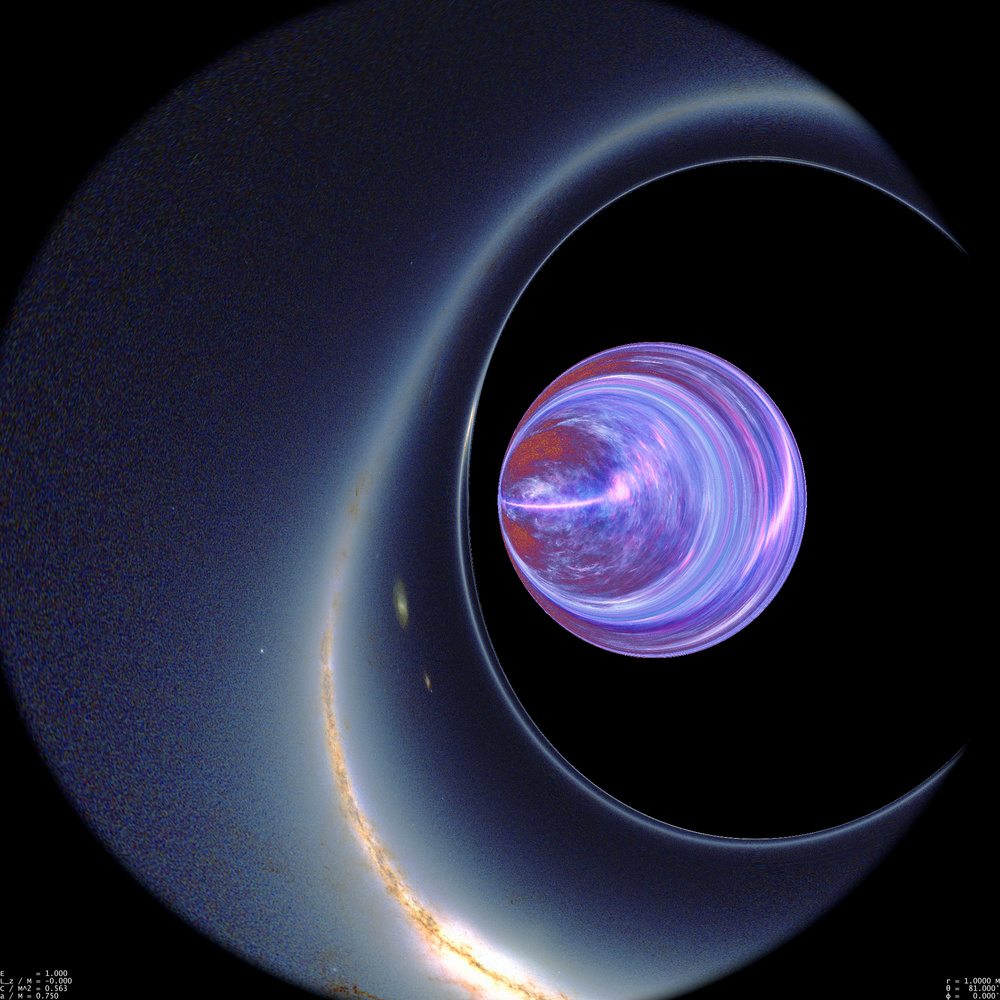}}
		\centerline{
			\includegraphics*[width=3.2in]{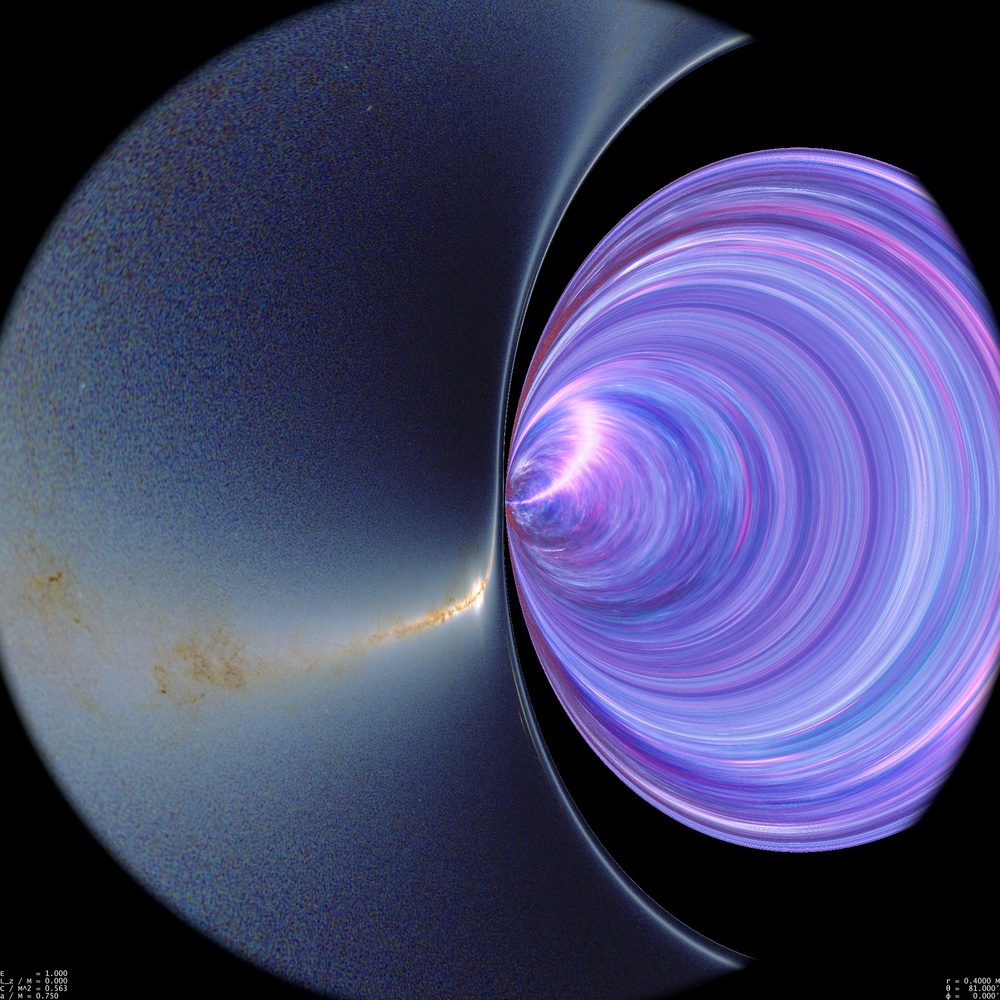}
			\includegraphics*[width=3.2in]{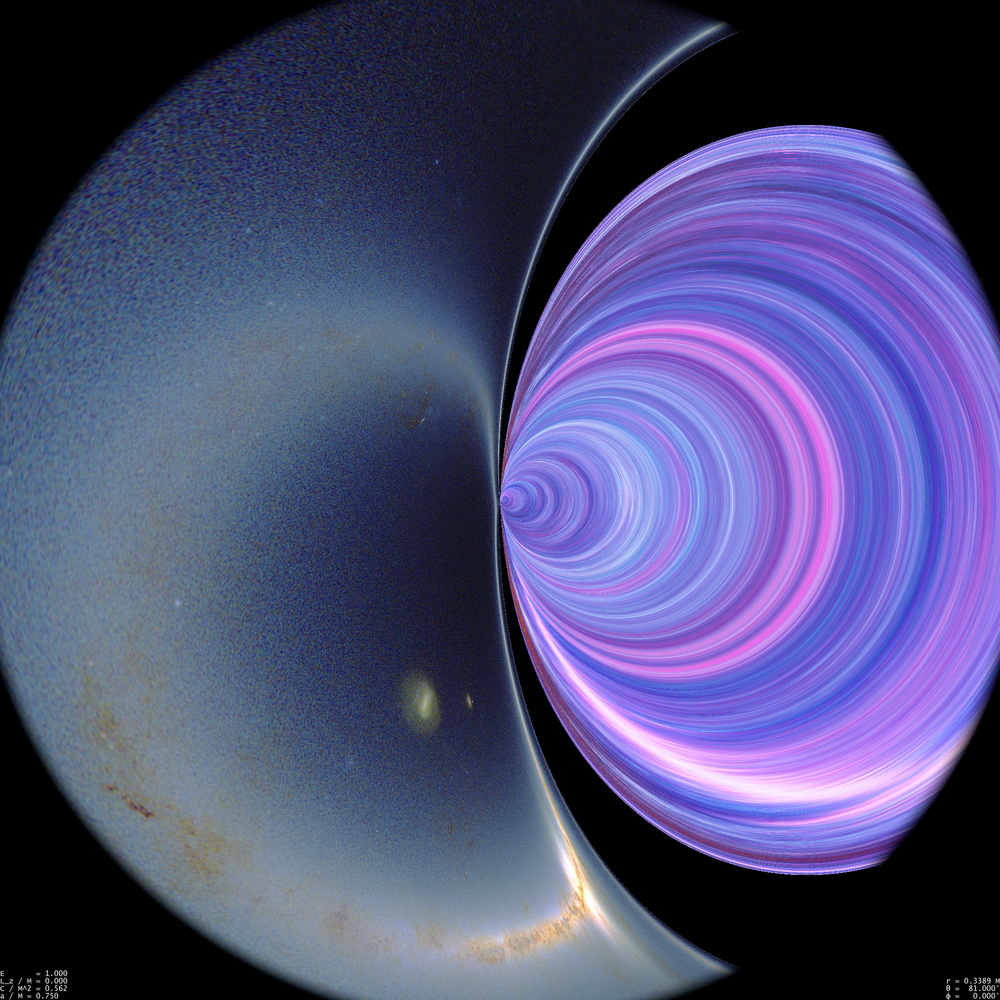}}
		\caption{Infalling observer at $r = 1.6448 M$, $M$, $0.4 M$ and
			$0.3389 M$.}
		\label{fig_seq2}
	\end{figure}

	The next set of images~(Fig.~\ref{fig_seq2}) shows the inter-horizon
	region. There, only region~$1$ and its mirror, region~$3$, are
	seen. The snapshots we show are situated at $r = 1.6448 M$, $M$, $0.4
	M$ and $0.3389 M$. Region~$3$ appears at the exact spot where region
	$-7$ had disappeared. As its angular size grows, we notice that more
	or less half of it appears reasonably undistorted~(one recognizes the
	Milky Way band), whereas of the right~(i.e., retrograde) part show a
	``whirlpool''-like pattern. The relative size of the highly
	distorted region as compared to the less distorted region increases at
	the observer gets closer and closer to the inner horizon, at $r_- =
	0.3386 M$.
	At first, it seemed to us that such whirlpool-like pattern was a
	little bit unexpected as region~$1$ appears fairly undistorted in
	comparison. There are two reasons for this. Firstly, the celestial
	sphere of region~$1$ does not possess many features as opposed to
	region~$3$, which makes the comparison difficult. The second~(and
	main) reason is more subtle. Once in region~$2$, regions~$1$ and $3$
	share very similar role~(as opposed to Fig.~\ref{fig_seq1} where
	region~$1$ and $-7$ when both were seen from region~$1$). The only
	difference between the two is that our observer has an $E$ equal to
	$1$ whereas a similar observer coming from region~$3$ would have an
	opposite $E$ and otherwise identical constants of motion. In other
	words, passing from our observer to its mirror analogue just
	amount to perform a Lorentz boost. But it this Lorentz boost is made
	in the direction opposite to that where there is a large distortion,
	then the angular size of the patch will enormously increase, giving
	the impression that the amount of distortion increases. But this is
	merely an artefact of the zooming power of aberration induced by large
	Lorentz boost. We give a few figures about this in~\ref{app_lazy}.
	Figure~\ref{fig_lazy} show the progressive
	distortion that occurs when performing a Lorentz boost in order to
	change the observer's $E$ from $1$ to $-1$. The symmetric case where
	$E = 0$~(sometimes dubbed as ``lazy geodesic'' clearly shows that both
	regions~$1$ and $3$ experience the same amount of distortion, although
	with our choice of celestial spheres the actual amount of distortion
	is more readily apparent in region~$3$, the detail of which will
	deserve further scrutiny.
	\begin{figure}[ht]
		\centerline{
			\includegraphics*[width=3.2in]{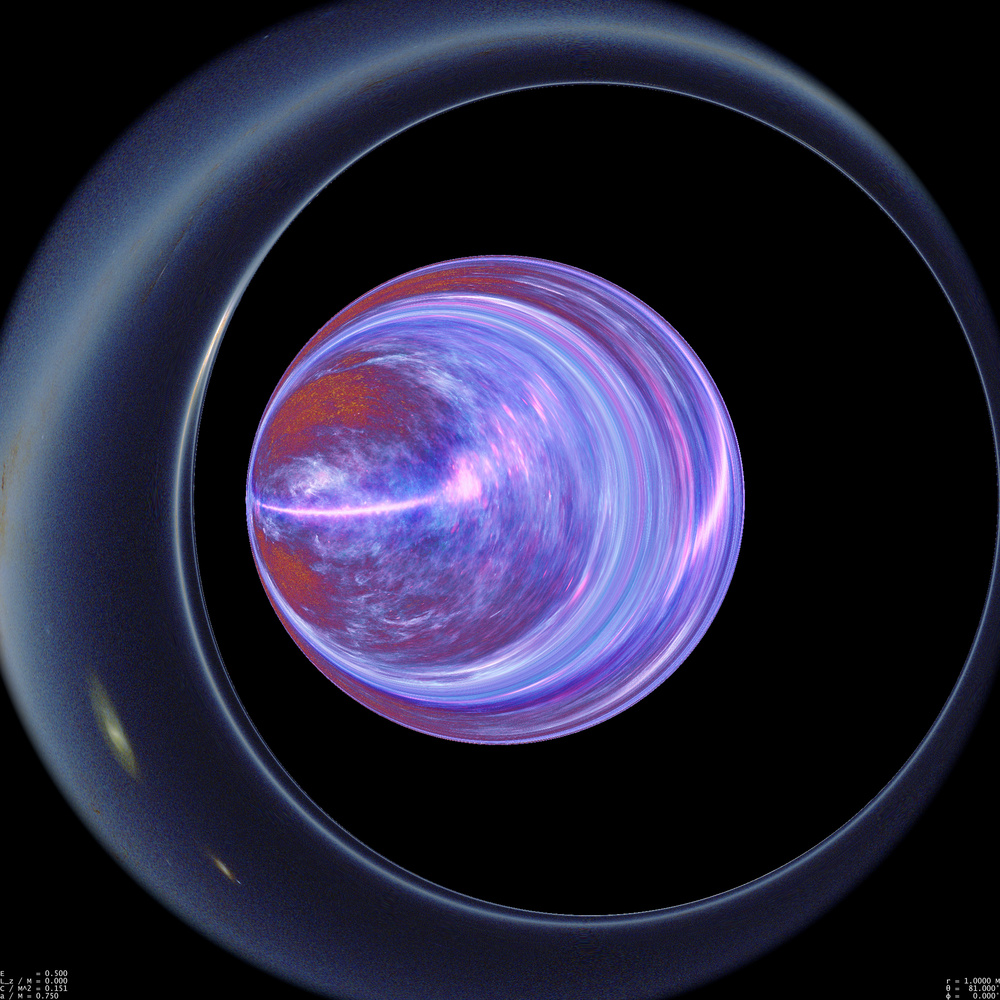}
			\includegraphics*[width=3.2in]{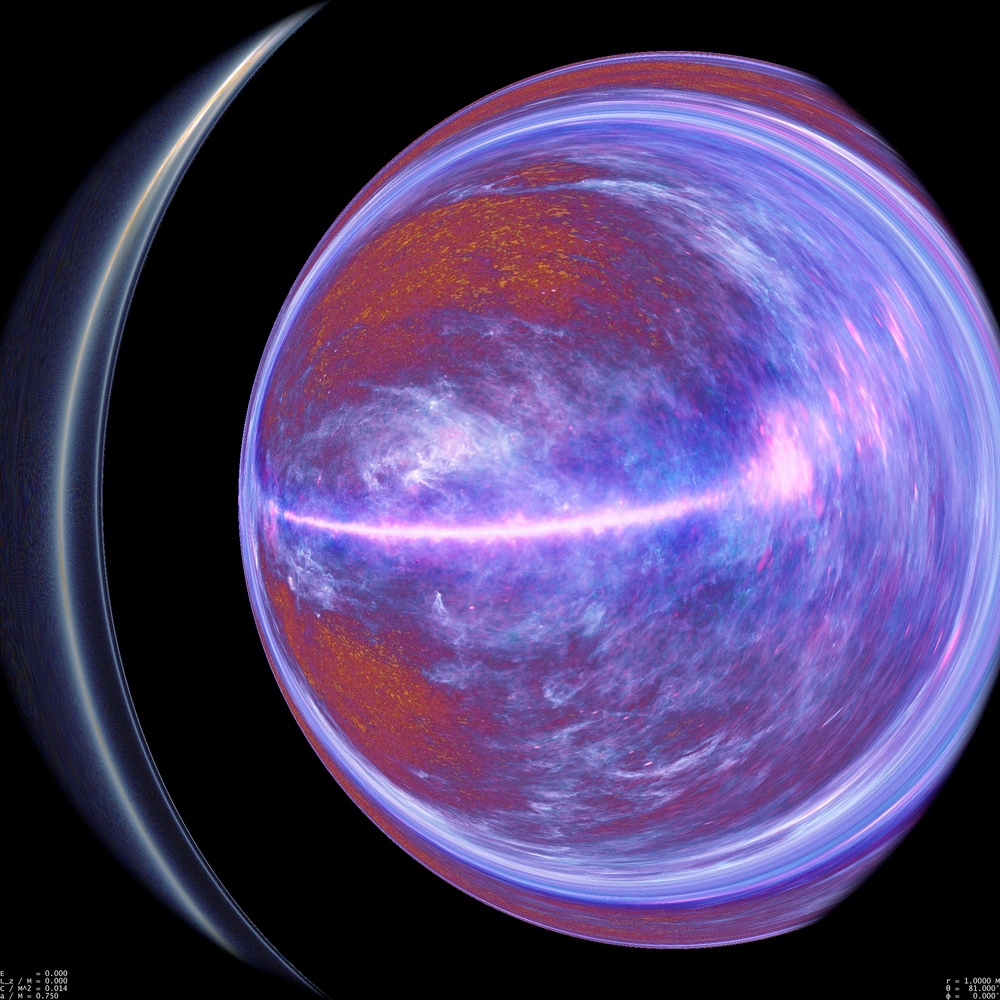}}
		\centerline{
			\includegraphics*[width=3.2in]{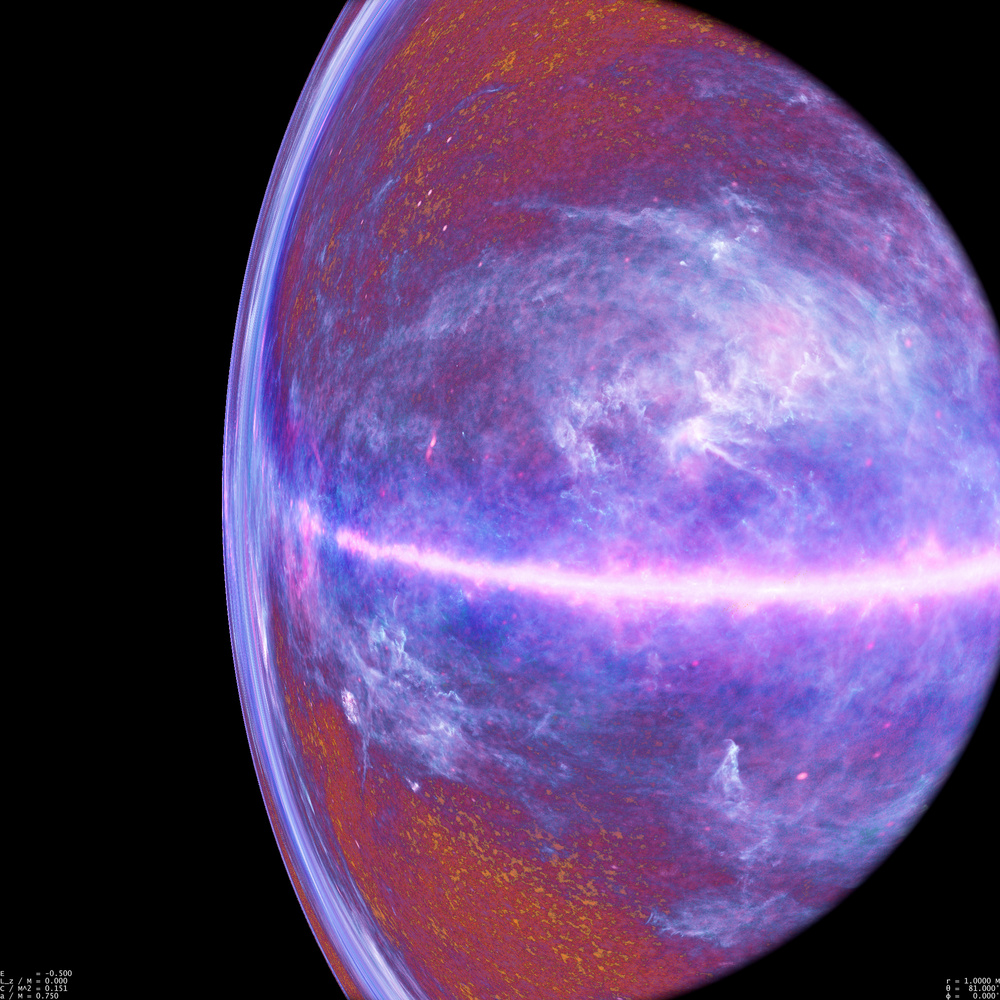}
			\includegraphics*[width=3.2in]{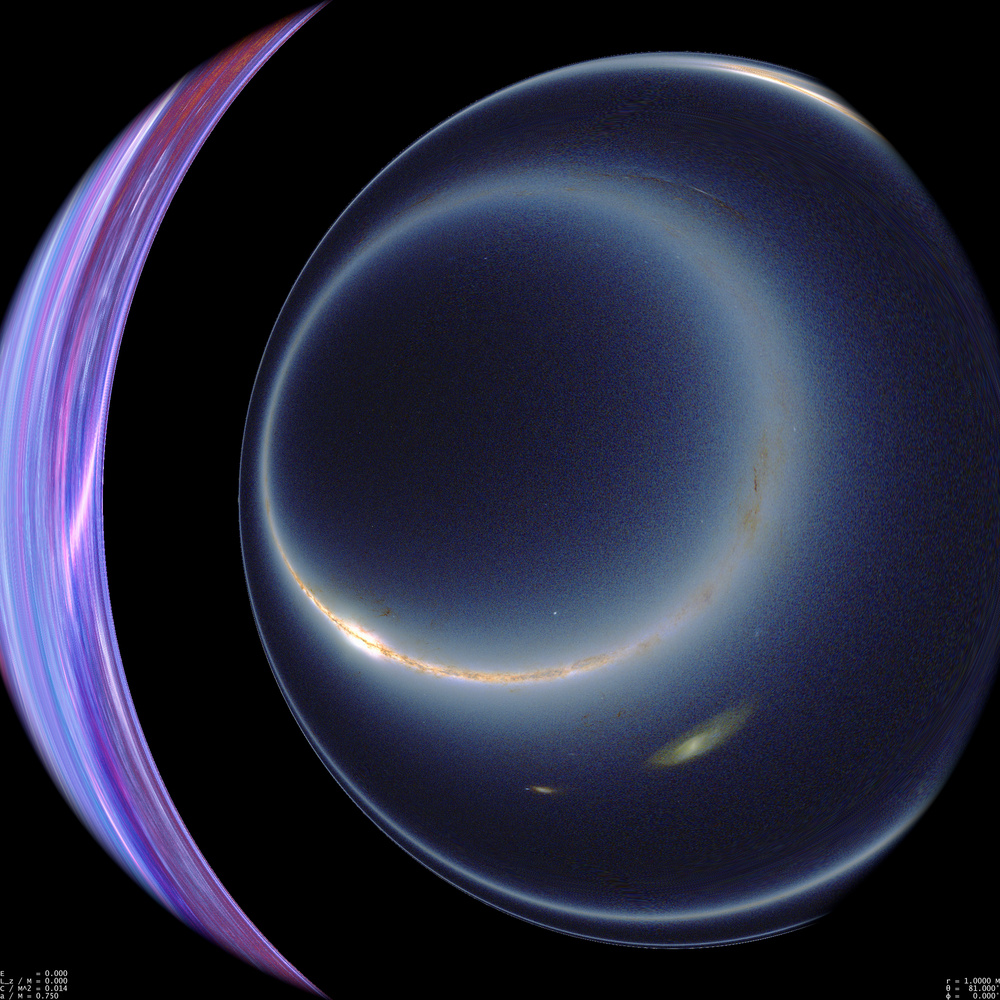}}
		\caption{Observer with varying $E$ at $r = M$. Starting from
			Fig.~\ref{fig_seq2} upper right image, we reduce $E$ from to
			$0.5$~(top left image), then $0$~(top right) and $-0.5$~(bottom
			left).  Bottom right image shows the opposite direction for $E =
			0$ and illustrates the perfectly symmetric role of the two
			regions.}
		\label{fig_lazy}
	\end{figure}
	
	Continuing the free-fall within the inner horizon~(Fig.~\ref{fig_seq3})
	allows to see several changes in the topology of the bounded geodesics
	regions. At $r = 0.3047$, i.e., very soon after inner horizon
	crossing, a tiny patch the the sky show the region~$7$, with negative
	$r$~(upper left image of Fig.~\ref{fig_seq3}). This region appears at
	the edge of the bounded geodesics shell, although at horizon
	crossing, the shell very temporarily gets pinched in one point by the
	region~$3$ patch. A new type of distortion within the region~$3$
	patch appears close to the pinch. Decreasing $r$ makes the patch of
	region~$7$ grow larger and also allows for the inner, secondary dark
	bubble of bounded geodesics to appear~(upper right panel, $r = 0.1
	M$). The secondary bubble does not last long and has already
	disappeared at $r = 0.02 M$. At this point, the patch of region~$7$
	keeps on increasing in size, but that of region~$3$~(within the dark shell)
	decreases In this image, we have outlined the ring singularity by
	white pixels. As already explained, within the ring singularity we
	see both region~$7$ and some bits of region~$1$ through the
	adventurous geodesics which are traveling back from their short trip
	in the negative $r$ region. At some point~($r = 0.120 M$, lower
	right image), the distortion of region~$3$ is completely dominated
	by the whirlpool-like pattern. Note at this point that since we are
	now within the inner horizon, regions~$1$ and $3$ are no longer
	equivalent. The dissymmetry in their distortion patterns is no
	longer an issue.
	\begin{figure}[ht]
		\centerline{
			\includegraphics*[width=3.2in]{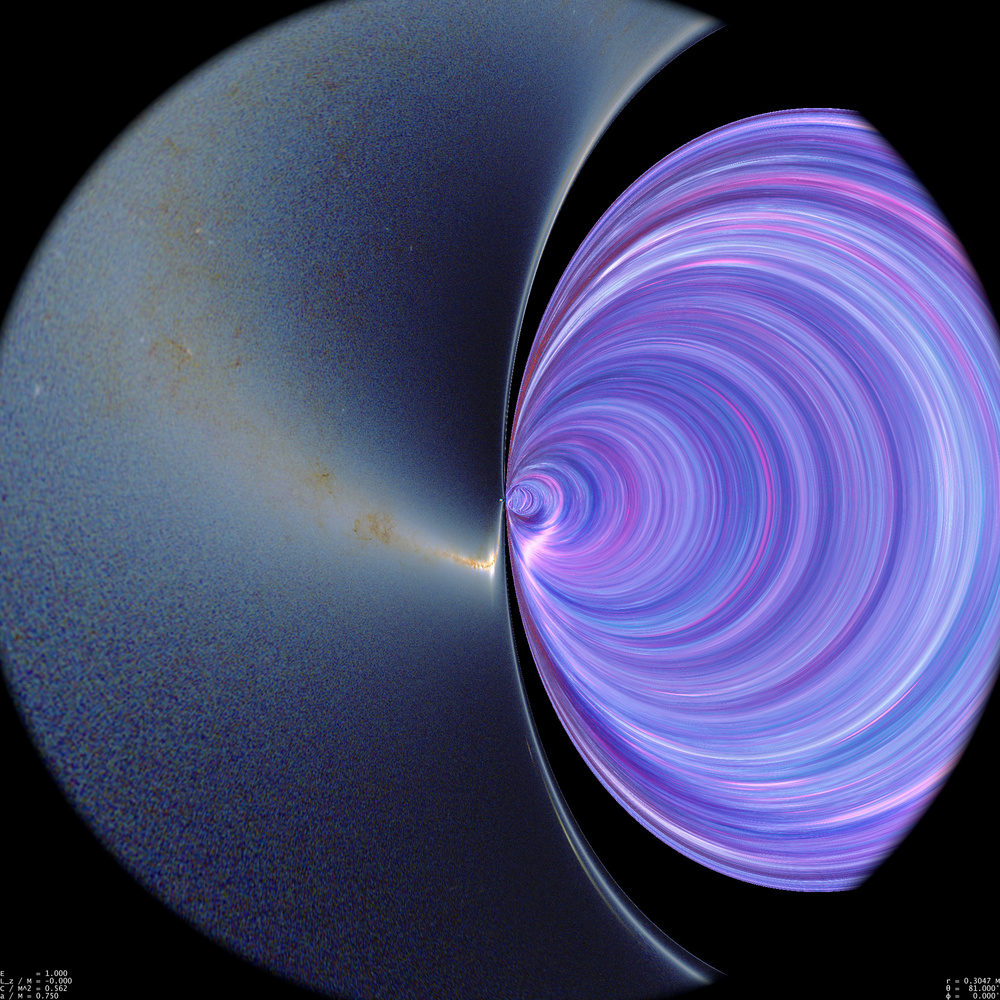}
			\includegraphics*[width=3.2in]{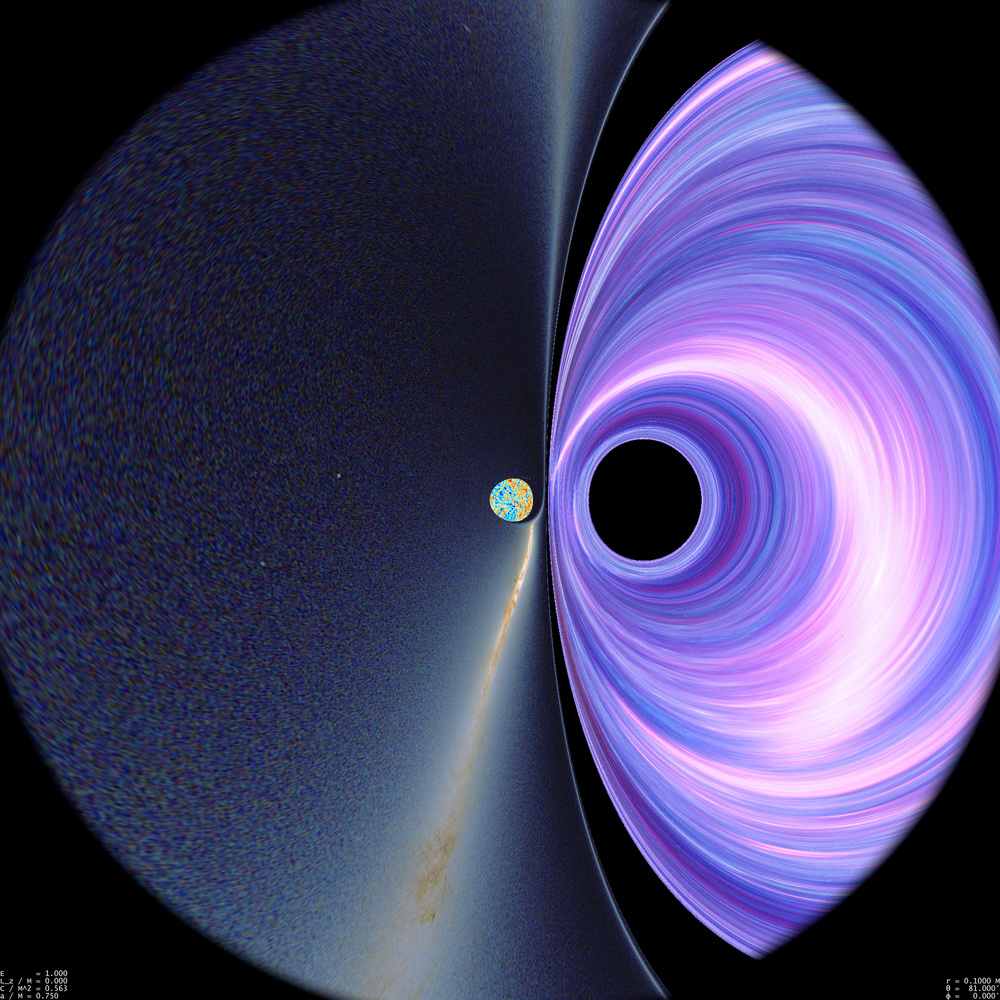}}
		\centerline{
			\includegraphics*[width=3.2in]{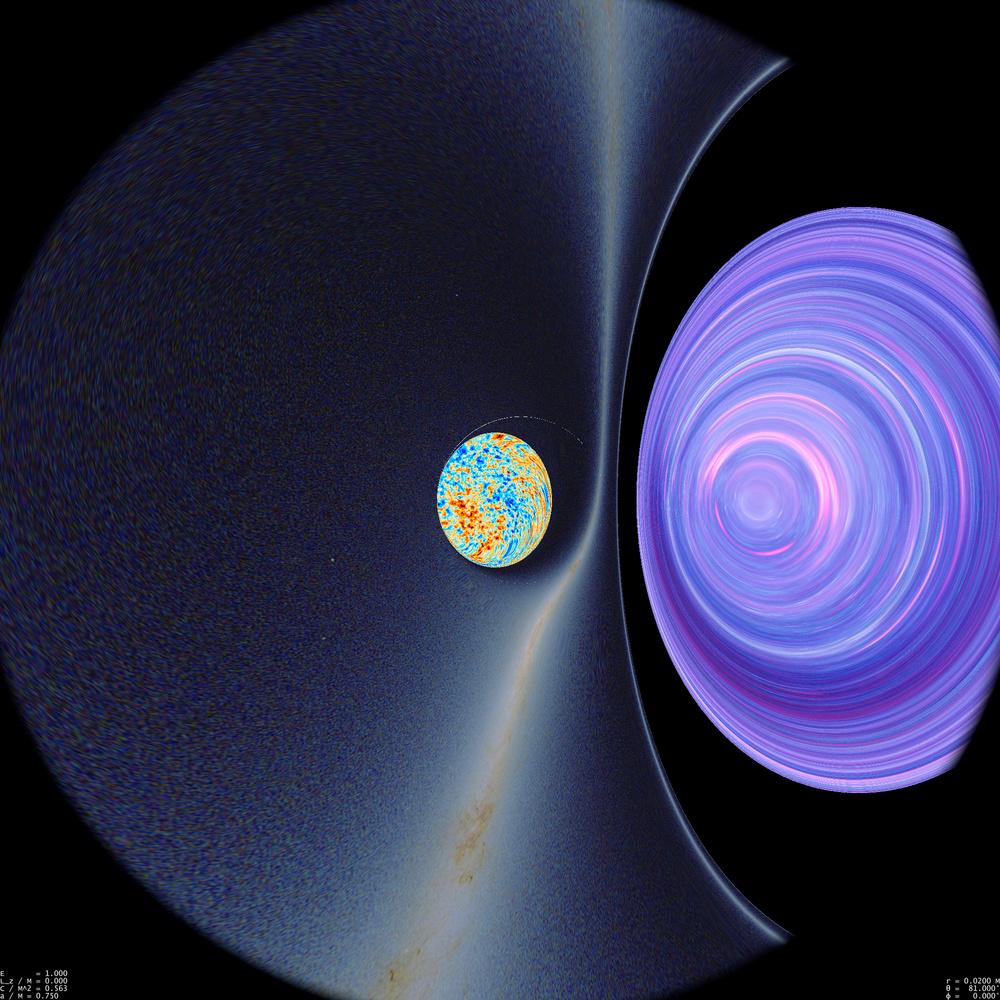}
			\includegraphics*[width=3.2in]{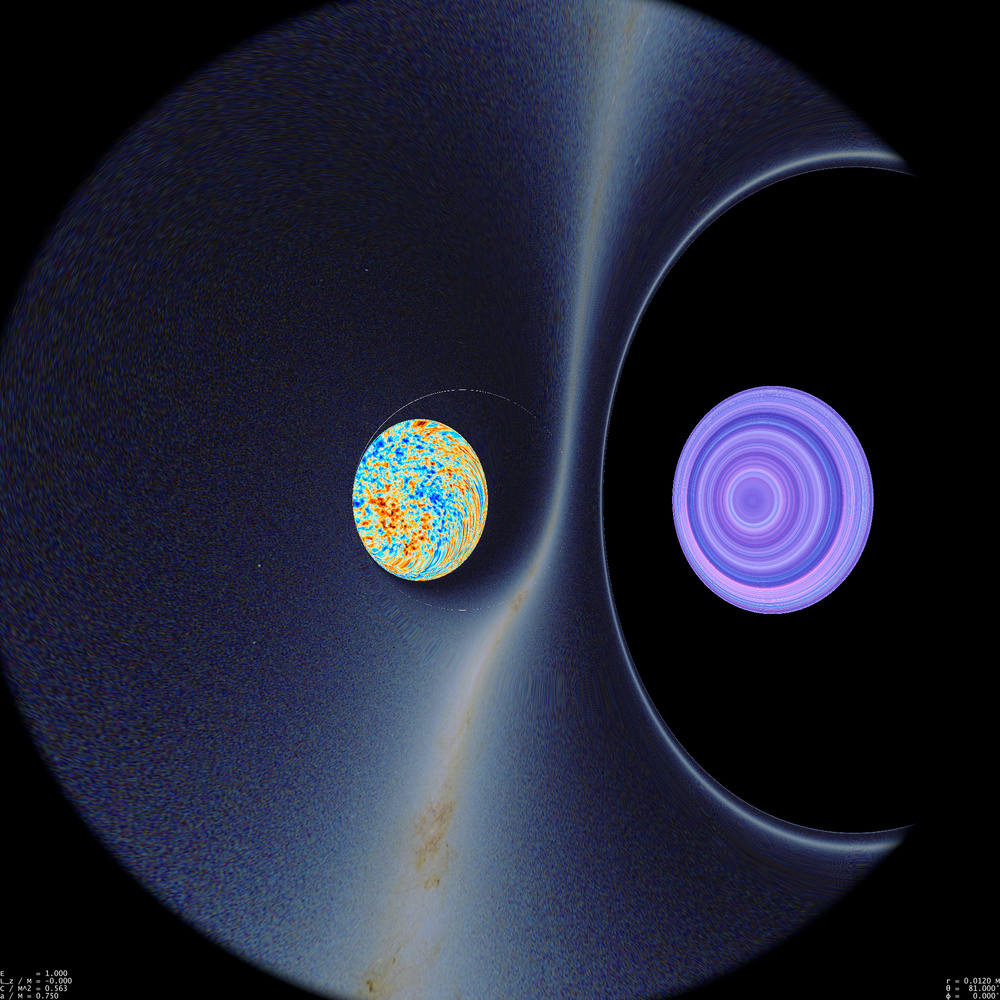}}
		\caption{Continuation of the free-fall, now within the inner
			horizon, at~(from left to right and top to bottom) $r = 0.3047 M$,
			$0.1000 M$, $0.0200 M$ and $0.0120 M$.}
		\label{fig_seq3}
	\end{figure}
	
	The next step of the journey consists in getting closer and closer to
	the ring singularity. According to Fig.~\ref{fig_mu22}, the various
	steps are first the disappearance of region~$3$ at latest when the
	observer leaves the inner ergoregion~(Fig.~\ref{fig_seq3}, top left
	image). At this point the ``dark shell'' transforms into a dark bubble
	which is also led to shrink~(upper right image) and then
	disappear~(bottom right image). We have found that a whirlpool-like
	pattern is briefly seen in the patch of region~$1$ in the direction
	where the dark bubble disappeared. From now on, only regions~$1$ and
	$7$ are seen, the latter having a growing angular size~(bottom right
	image). Two interesting features are worth mentioning at this
	stage. Firstly, we do not see the whole celestial sphere of
	region~$7$. This was already the case before but it now becomes more
	obvious. The reason is that transit geodesics experience a limited
	amount of variation of their $\theta$. Conversely, at a given
	observer's colatitude $\theta$, there is only a small interval around
	this $\theta$ from which geodesics can originate from the celestial
	sphere. Since transit geodesics are also limited in term of their $\xi$
	parameter~\cite{chandrasekhar83} the same remark is very likely to
	apply in term of longitude as well. The second obvious feature is that
	the ring singularity has a shape closer to that of a circle than an
	ellipse. This may look surprising since the observer's latitude is
	large, however, without even considering light deflection, it has to
	be noted that the ring singularity is visible only after inner horizon
	crossing, where the $r$ in Kerr-Schild coordinates is already
	small. Consequently, at least in term of coordinates, the observer is
	more ``above'' the singularity than close to its plane. This seems the
	most likely explanation of the visual roundness of the
	singularity. Let us add that this is not an artefact of aberration:
	aberration transforms circles into circles~\cite{penrose59} so that
	any observer situation at the same point but with different velocities
	would also see the singularity shaped close to a circle.
	\begin{figure}[ht]
		\centerline{
			\includegraphics*[width=3.2in]{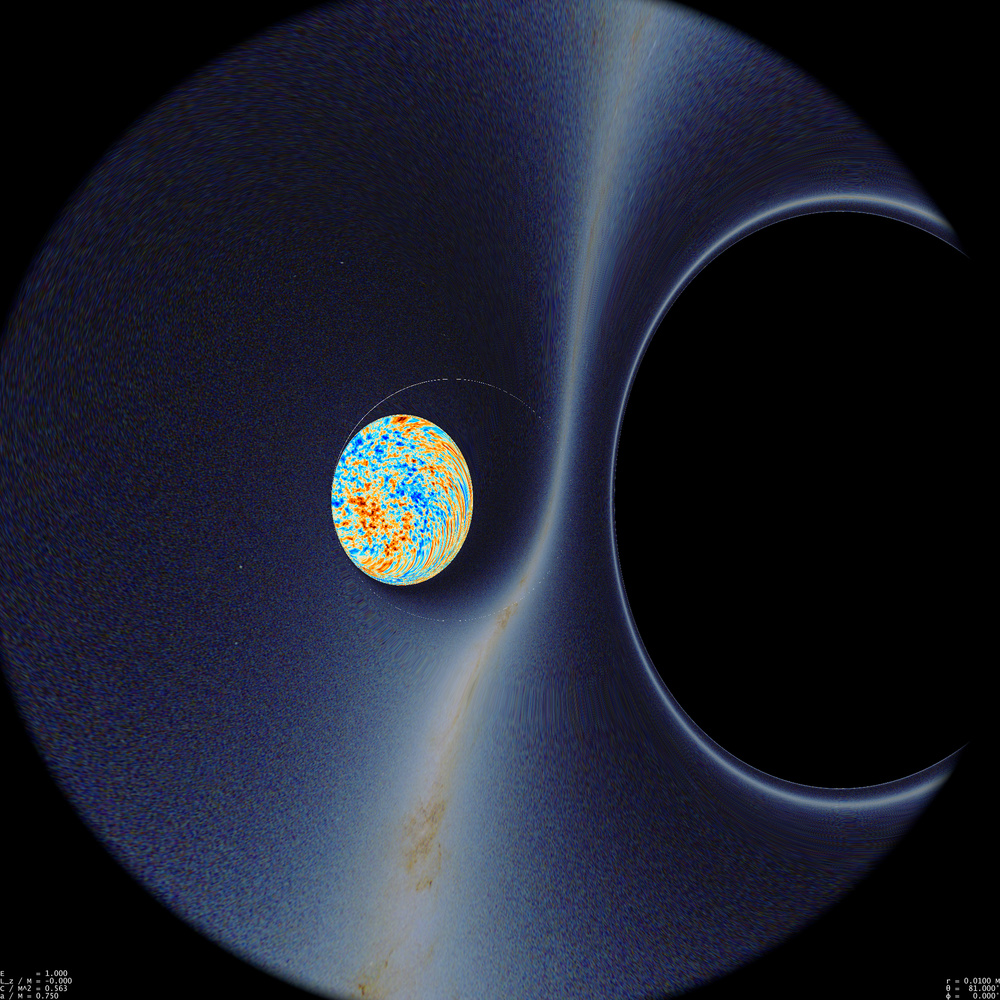}
			\includegraphics*[width=3.2in]{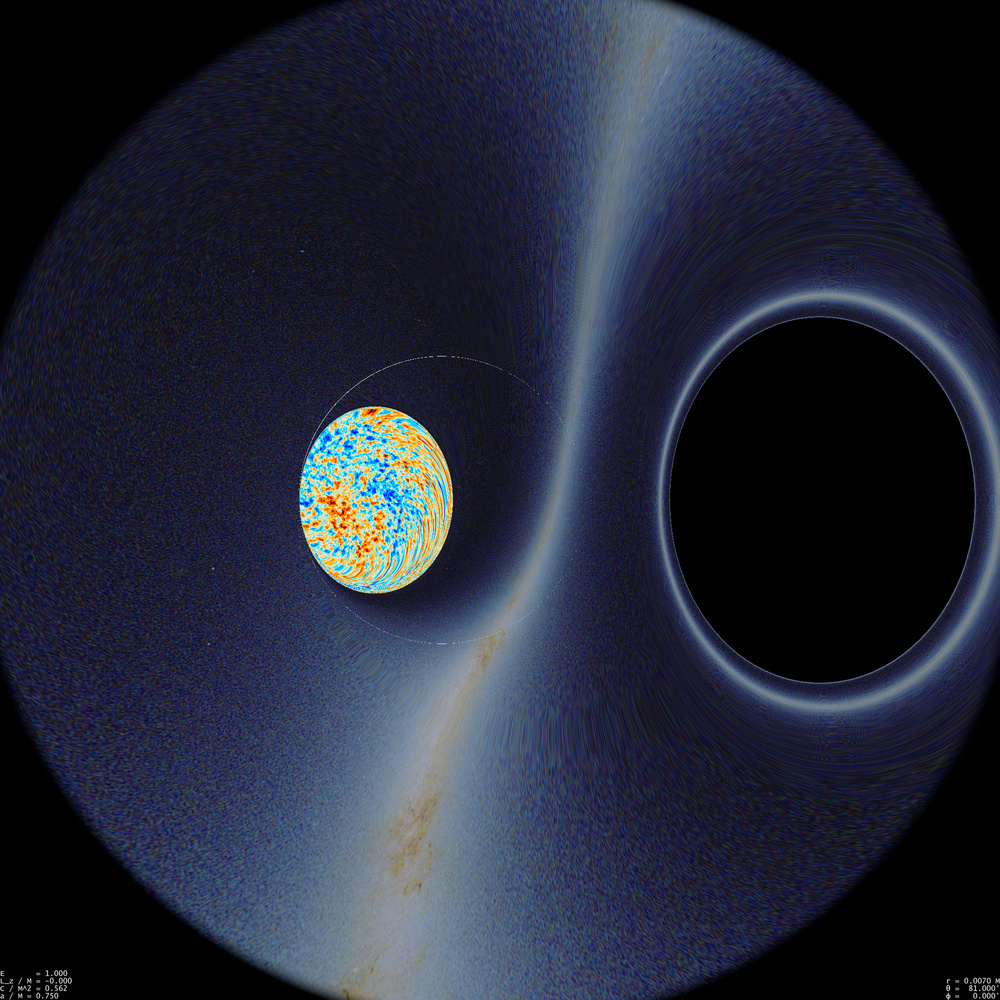}}
		\centerline{
			\includegraphics*[width=3.2in]{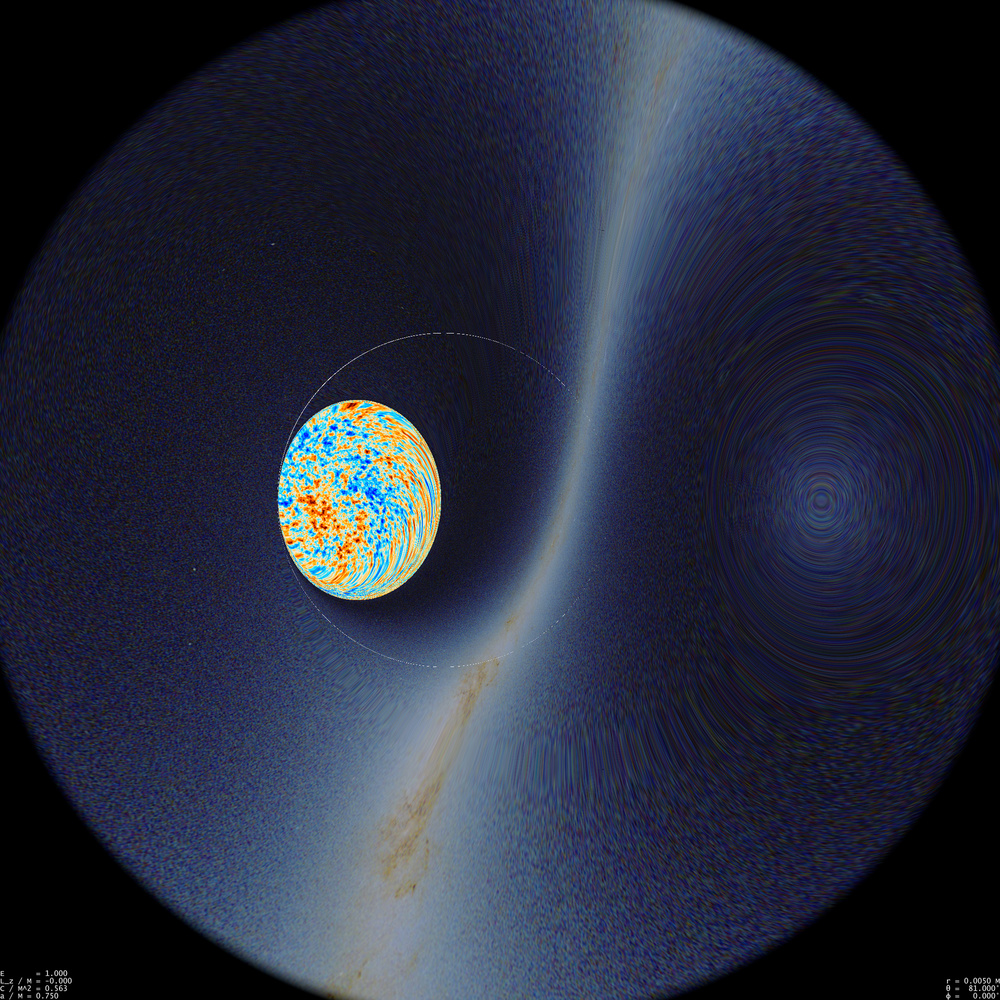}
			\includegraphics*[width=3.2in]{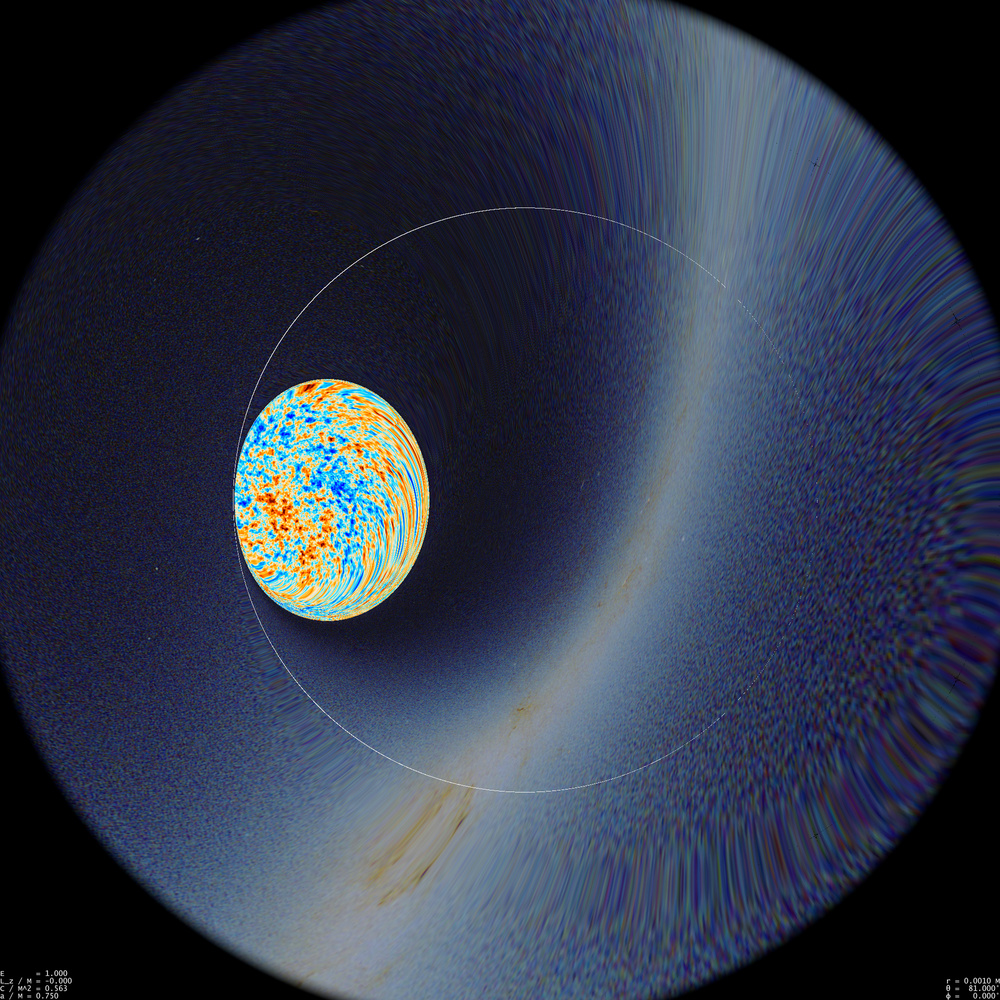}}
		\caption{Approaching the ring singularity at~(from left to right
			and top to bottom) $r = 0.0100 M$, $0.0070 M$, $0.0050 M$ and
			$0.0010 M$.}
		\label{fig_seq5}
	\end{figure}
	
\begin{figure}[ht]
		\centerline{
			\includegraphics*[width=3.2in]{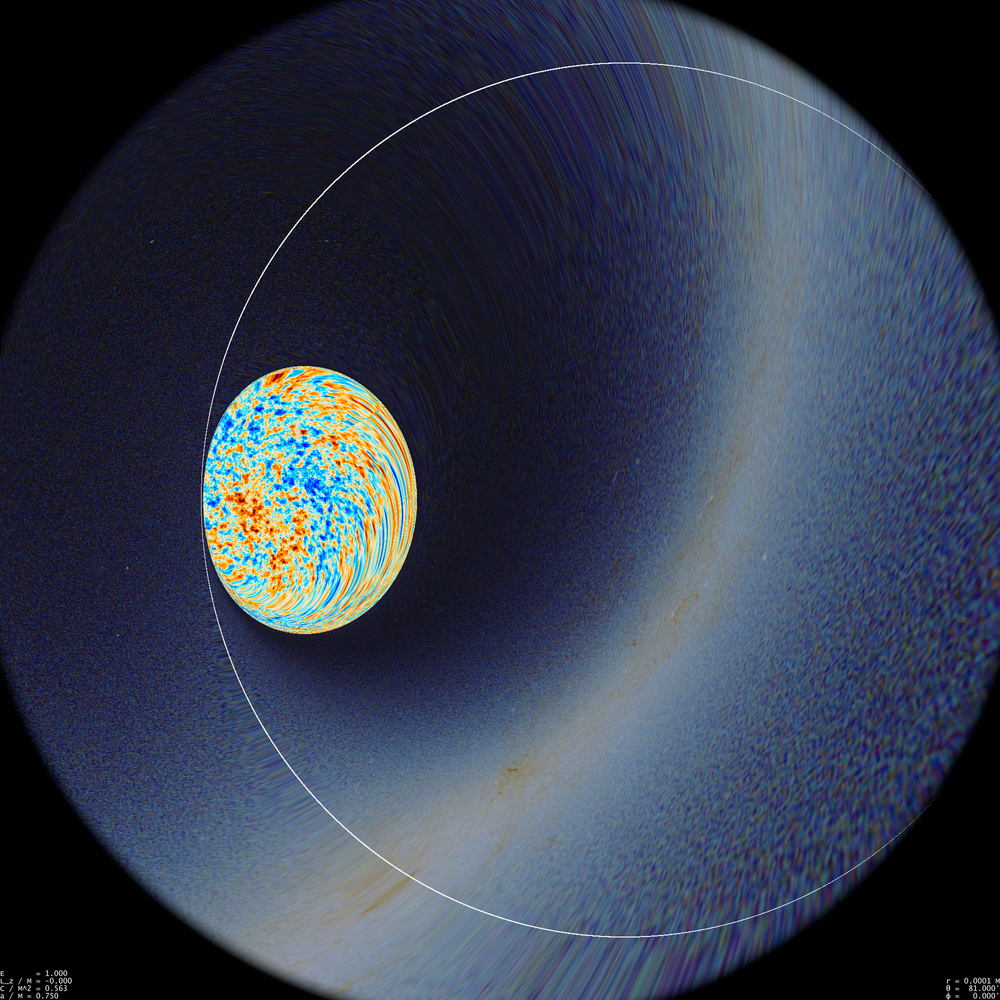}
			\includegraphics*[width=3.2in]{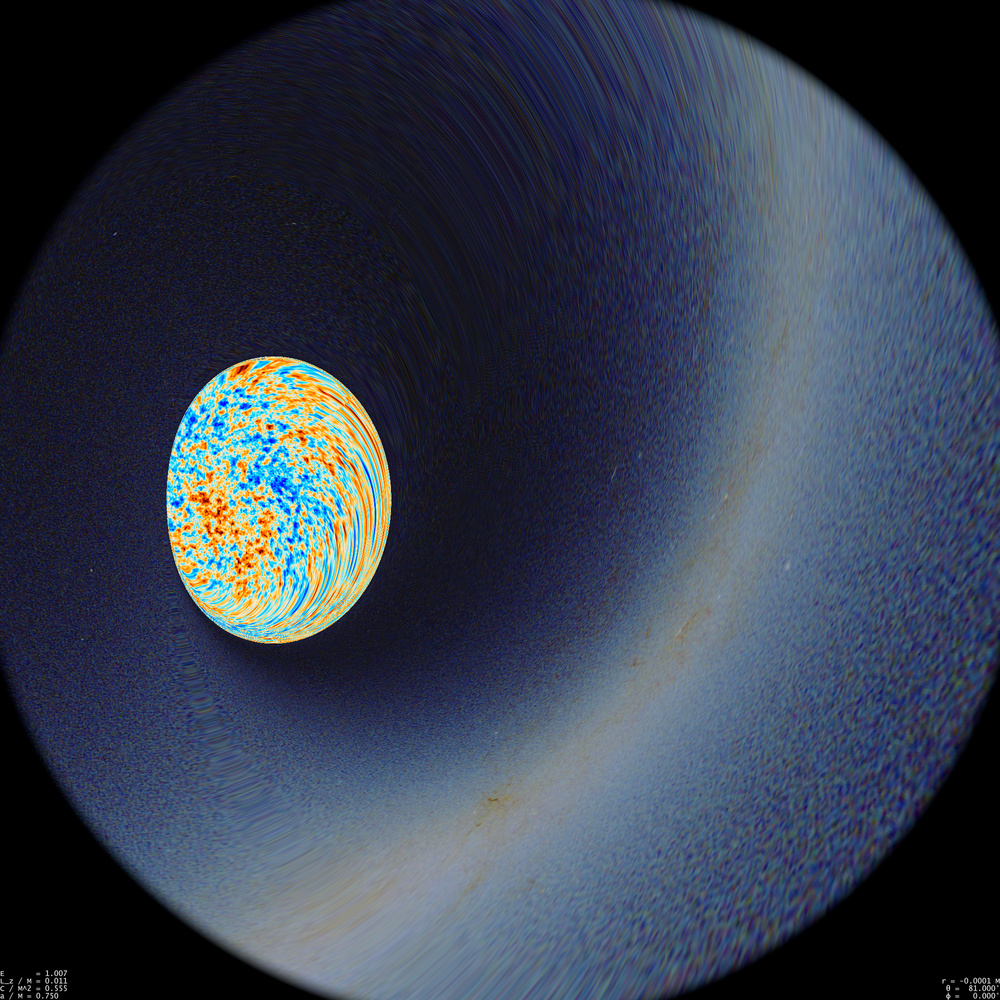}}
		\caption{Before and after crossing the ring singularity at $r =
			0.0001 M$ and $r = - 0.0001 M$.}
		\label{fig_seq6}
	\end{figure}

	Crossing the ring singularity does not go along with significant
	changes in what the observer sees~Fig.~\ref{fig_seq6}. This is not
	completely surprising since the actual ``crossing'' of the singularity
	cannot be associated with the crossing of some special hypersurface
	just as horizon crossing is. The fact that we have implicitly decided
	that singularity crossing occurs at $r = 0$ is a rather disputable
	choice motivated by the seemingly naturalness of the Kerr-Schild
	coordinate system. Another possibly more relevant criteria would be
	the envelope of the adventurous geodesics which would include part of
	the negative $r$ region. In any case, when $r = 0$, the ring
	singularity corresponds to a great circle from the point of view of our
	observer, which in this case is a static observer. Whether or not this
	result could be expected also deserves a further study. However, if we
	are interested in the angular size of the negative $r$ region, then it
	still occupies a limited patch in the sky: from the observer's point
	of view, most of the angular area what originates from the negative
	$r$ region corresponds to adventurous geodesics that are on their way
	back to the positive $r$ region.
	
\begin{figure}[ht]
		\centerline{
			\includegraphics*[width=3.2in]{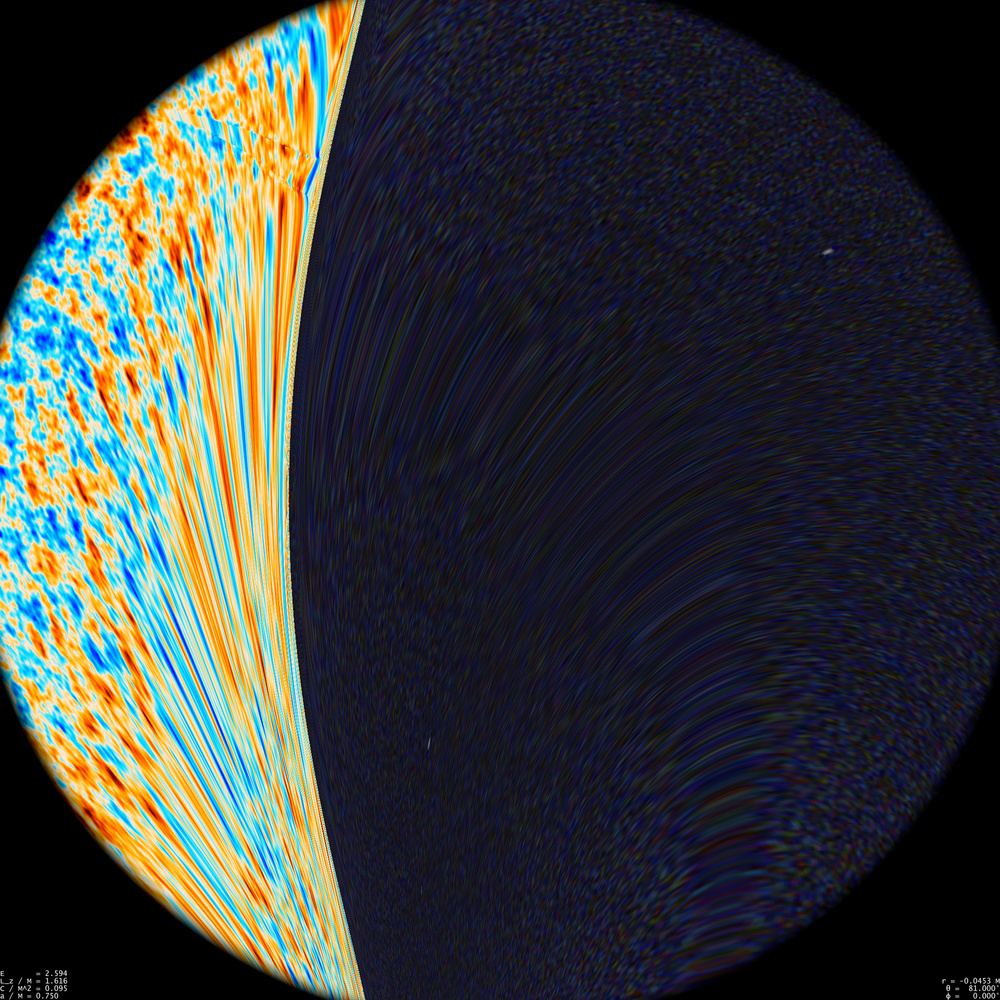}
			\includegraphics*[width=3.2in]{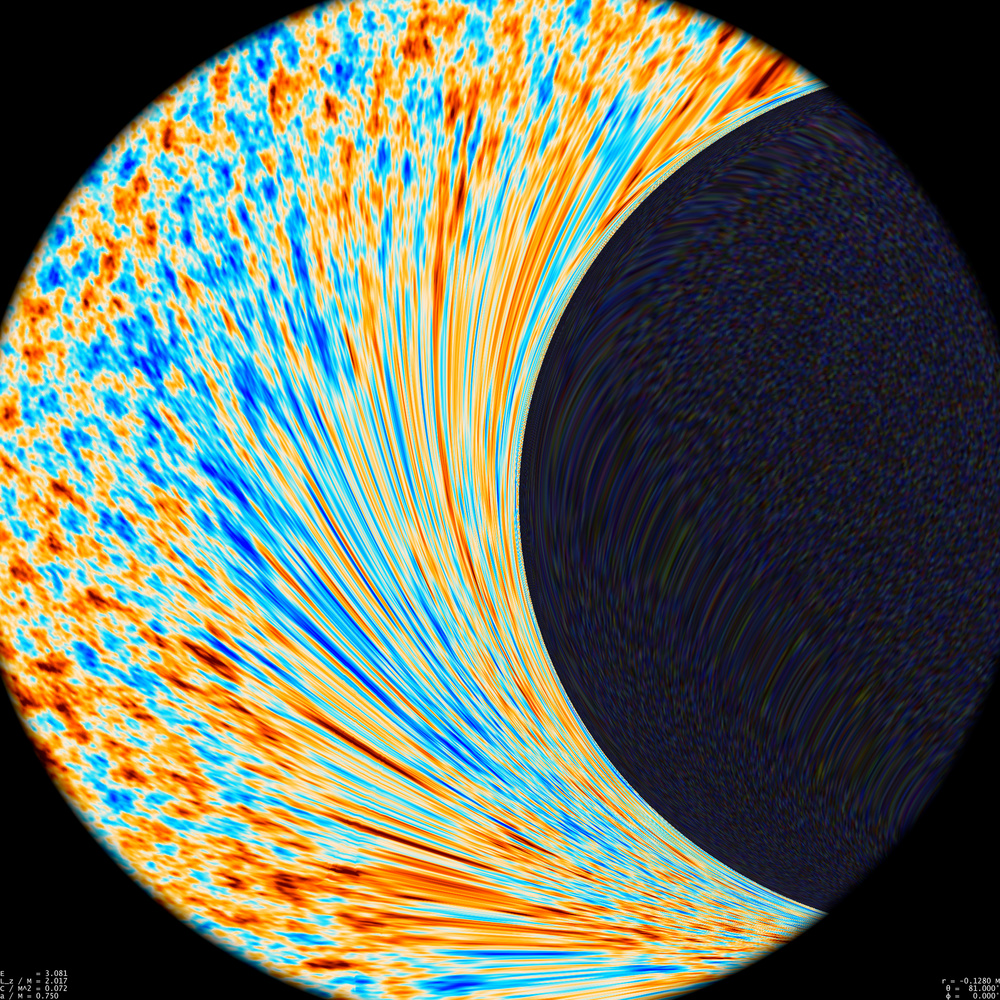}}
		\caption{Escaping toward the negative $r$ region along a
			quasi-static trajectory, at $- 0.0453 M$~(left) and $- 0.1810
			M$~(right).}
		\label{fig_seq7}
	\end{figure}

	Advancing into the negative $r$ region allows to see a larger and
	larger part of the negative $r$ region celestial
	sphere~(Fig.~\ref{fig_seq7}). We find that it is soon before $r \sim -0.09
	M$ that the negative $r$ region spread over more than half of the
	celestial sphere~(bottom left image).
	
\begin{figure}[ht]
		\centerline{
			\includegraphics*[width=4.8in]{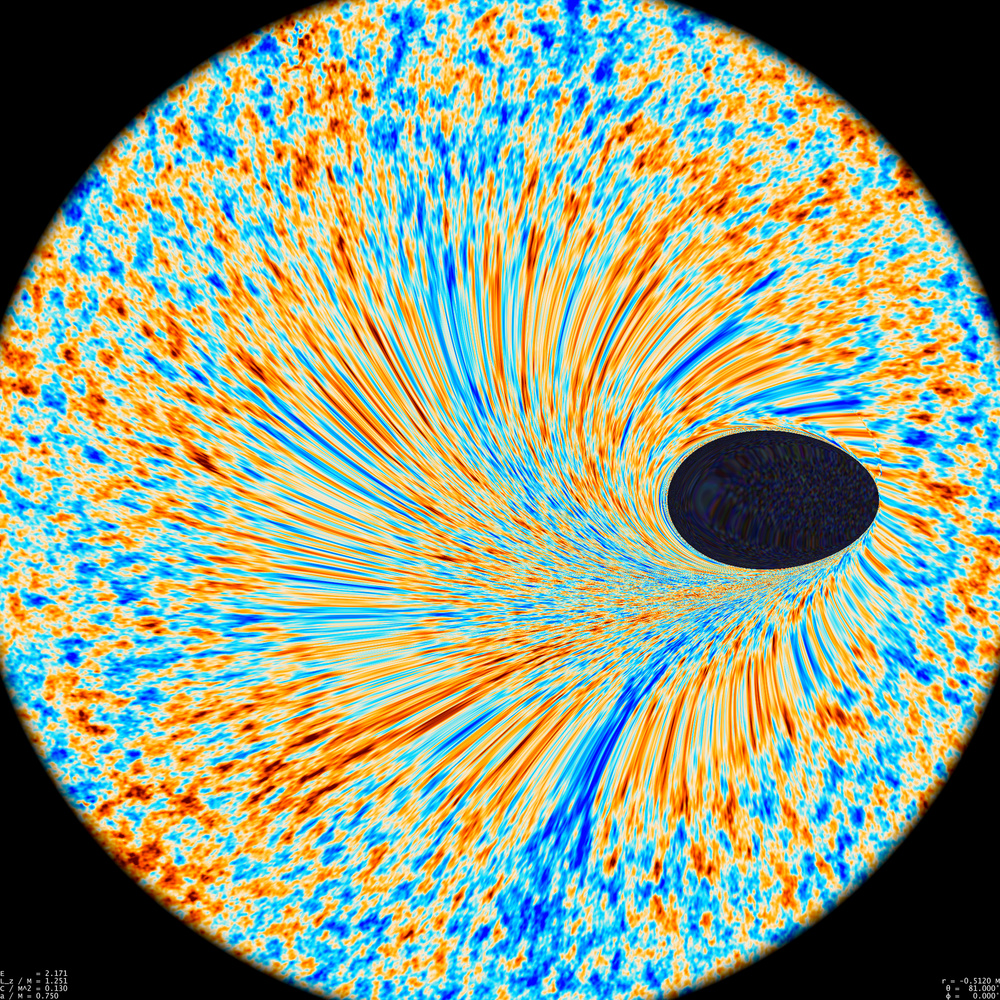}}
		\caption{A beautiful snapshot of the observer's region of origin
			seen from the negative $r$ region, $r = - 0.5120 M$.}
		\label{fig_seq8}
	\end{figure}

	The view at $r = - 0.5120 M$ is in our opinion the most aesthetic of
	this simulation~(Fig.~\ref{fig_seq8}). This image shows a nice
	swirling pattern around what remains visible from region~$1$. Such
	pattern is of course not unexpected since the metric deals with a
	rotating singularity, however, the apparent rotation of the swirl
	remains to be investigated as well. Another feature of interest is
	that we only see a limited part of region~$1$, the reason being the
	very same as the one which prevented from seeing the whole celestial
	sphere of region~$7$ from region~$5$.
	
\begin{figure}[ht]
		\centerline{
			\includegraphics*[width=3.2in]{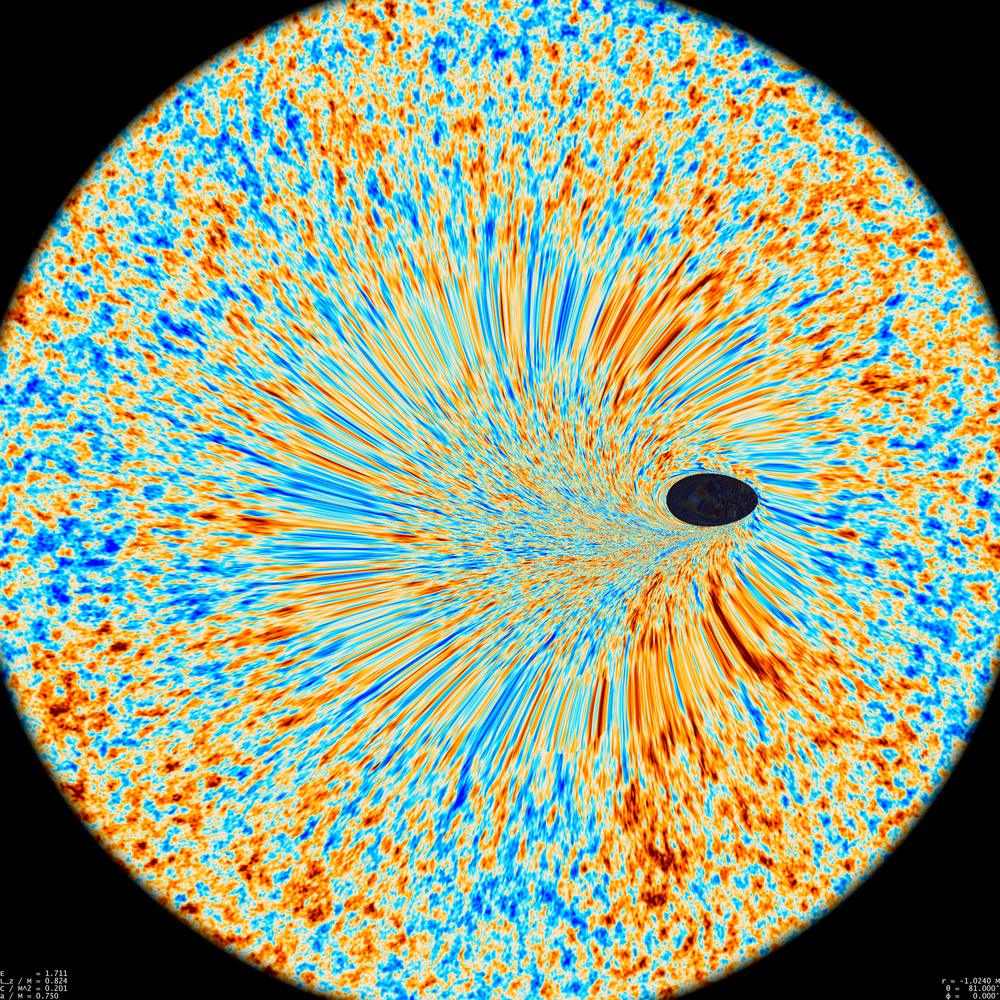}
			\includegraphics*[width=3.2in]{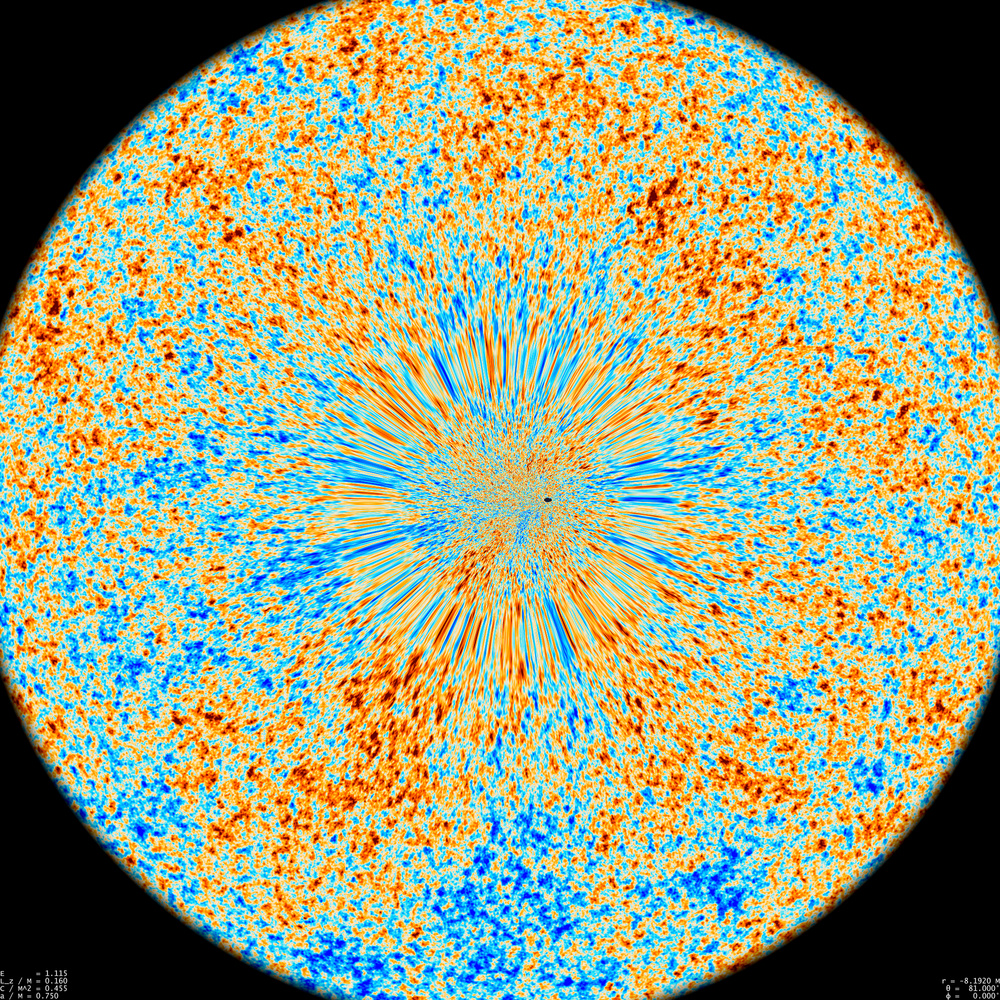}}
		\caption{Last steps of escape into the negative $r$ region, at $r =
			- 1.0240 M$ and $- 8.1920 M$.}
		\label{fig_seq9}
	\end{figure}

	Going even further within the negative $r$ region gives rise to
	phenomena which seem more easy to interpret~Fig.~\ref{fig_seq9}. The
	celestial sphere of region~$7$ shows multiple images along directions
	that are close to the singularity as well as a radial shear distortion
	which are characteristic of negative mass systems.
	
	\section{Conclusion}
	
	In this paper, we have studied from a visual point of view some aspects of the Kerr metric, focusing on its maximal analytic extension, but also on the very viciniry of a realistic black hole. Although several features of the metric can be known qualitatively without using raytracing techniques, the variety of phenomena that arise is much larger than what can be expected. This study is far from being complete. We did not study the maximal analytic extension of the extremal Kerr wormhole nor the naked singularity, which we will devote to in a future work.
	
	\appendix
	
	\section{Recovering the frame-dragging effect}
	\label{app_gravitom}

	We show here that with the correct approximation,
	Eqns.~(\ref{eq_xiy_final},\ref{eq_z_final}) lead to the well-known
	gravitomagnetic effect.
	The first step is to rewrite Eqns.~(\ref{eq_xiy_final},\ref{eq_z_final}) when
	neglecting $a$ as compared to $r$ since we consider a particle at~(very)
	large distance from the black hole. In particular, this amounts to make the
	substitution $4 r^2 / \Sigma - 3 \to 1$ and $4 r^2 / \Sigma - 1 \to 3$ and to neglect the
	$a \dot \varphi$ term as compared to $\dot r$. Moreover, we consider massive particles, so that
	$\kappa = 1$.
	Overall, this gives
	\begin{eqnarray}
	\label{eq_xiy_simp}
	\ddot x + i \ddot y
	& = & 4 i M a \frac{1}{r^3} \Ldot \left[ \dot x + i \dot y - \frac{x + i y}{r}
	\left\lbrace \dot r + \epsilon  \frac{\Ldot^2 - 1}{4 \Ldot }   \right
	\rbrace \right]  \\
	\nonumber & &
	- M \frac{x + i y}{r^3}
	\left[ 1 +  3  \frac{C - a^2 \Ldot^2}{r^2}  \right] , \\
	\label{eq_z_simp}
	\ddot z & = & - M \frac{z}{r^3} \left[ 1 + 3  \frac{C}{r^2} \right] .
	\end{eqnarray}
	The quantity $\Ldot$ is given in Eq.~(\ref{def_Ldot}). Its physical meaning is
	the frequency shift of ingoing or outgoing radial null rays seen by the
	observer. At lowest order, it is easy to check that it is, as expected, equal to $1 + \epsilon \dot r$. Therefore
	the term of the first equation above which is proportional to $\epsilon$ reduces to $\dot r / 2$. Finally, far
	from
	the black hole, and for a non relativistic observer ($\kappa = 1$, $E \simeq
	1$), the Carter constant can be approximated as
	\begin{equation}
	C \simeq \pi_\theta^2 + a^2 - 2 a E L_z + \frac{L_z^2}{\SN^2} 
	\simeq r^4 ( \dot \theta^2 + \SN^2 \dot \varphi^2) + a^2 - 2 a L_z .
	\end{equation} 
	
	The first two terms correspond to $L^2$, the observer's orbital angular
	momentum per unit
	of mass squared. Moreover, putting all the missing $c$'s, the $a^2$ term is in
	fact $a^2 c^2$~(still assuming that $a$ is a length, at most equal to the black
	hole coordinate radius). Consequently, $a^2 c^2$ is not larger than $G^2 \tilde M^2 /
	c^2$, where here $\tilde M$ is the black hole mass in true units. In comparison, if we
	consider for simplicity an observer in orbit around
	the black hole at some distance $r$, the angular momentum per unit of mass is
	given by $r G \tilde M$. The ratio
	between the two quantities is therefore of order $L^2 / (a^2 c^2) \sim r c^2 /
	G \tilde M \sim r / R_{\rm BH} \gg 1$. The $a^2$ term  can therefore  be dropped and
	we
	have
	\begin{eqnarray}
	\label{eq_xiy_simp2}
	\ddot x + i \ddot y
	& = & 4 i J \frac{1}{r^3} \left[ \dot x + i \dot y - \frac{3}{2}
	\frac{\dot r}{r} (x + i y) \right]  \\
	\nonumber & &
	- \frac{x + i y}{r^3}
	\left[ M +  3  \frac{M L^2}{r^2} - 6 \frac{J L_z}{r^2}  \right] , \\
	\label{eq_z_simp2}
	\ddot z & = & - \frac{z}{r^3} \left[ M  + 3  \frac{M L^2}{r^2} - 6 \frac{ J L_z}{r^2}
	\right] ,
	\end{eqnarray}
	Where $J = M a$ is the black hole spin. 
	We recognize in these two equations the Newtonian term, proportional to $- M x^i / r^3$ plus the
	Schwarzschild term, proportional to $- M L^2 x^i / r^5$.
	We need therefore to check that the first line of the first equation together
	with the $J L_z$ terms do corresponds to the usual gravitomagnetic term. 
	If we express the metric in Cartesian Kerr-Schild coordinates, the
	$g_{t\tilde\varphi}$ term transforms into
	\begin{equation}
	\label{gtphi_KScart}
	g_{t\varphi} \to g_{0i} = \frac{2 J}{r^3} \left( \begin{array}{c}
	- y \\ x \\ 0
	\end{array} \right) .
	\end{equation}
	The frame-dragging effect is recovered by computing the geodesic equation when
	considering only these terms and considering that the metric can be linearized
	with respect to this perturbation. For non relativistic trajectories, the only
	non-zero Christoffel symbols are $\Gamma^0_{ij}$ and $\Gamma^i_{0j},
	\Gamma^i_{j0}$. If one is interested in the evolution of the Euclidean velocity
	vector ${\pmb v}$, then the geodesic equation gives
	\begin{equation}
	\dot v^i = - 2 \Gamma^i_{0j} v^j ,
	\end{equation}
	where the summation runs on the three spatial indices. The Christoffel symbols
	are given by 
	\begin{equation}
	\Gamma^i_{0j} = - \Gamma_{i0j} = \frac{1}{2} \left(\partial_i g_{0j} -
	\partial_j g_{0i}\right) .
	\end{equation}
	If we consider the three components of $g_{0i}$ as those of a three-vector
	$\pmb g$, then from Eq.~(\ref{gtphi_KScart}) we have
	\begin{equation}
	\pmb g = \frac{2 \pmb J \wedge \pmb r}{r^3} ,
	\end{equation}
	together with
	\begin{equation}
	\dot {\pmb v} = \pmb v \wedge \left( - \pmb \nabla \wedge \pmb g \right) .
	\end{equation}
	The components of $- \pmb \nabla \wedge \pmb g$ are
	\begin{equation}
	- \pmb \nabla \wedge \pmb g = \frac{2 J}{r^5} \left( \begin{array}{c}
	-3 x z \\ - 3 y z \\ r^2 - 3 z^2
	\end{array}\right) .
	\end{equation}
	The components of the gravitational acceleration due to the frame-dragging are
	therefore
	\begin{equation}
	\pmb a = \frac{2 J}{r^5} \left(\begin{array}{c}
	\dot y r^2 - 3 z^2 \dot y + 3 y z \dot z \\
	- \dot x r^2 + 3 z^2 \dot x - 3 x z \dot z \\ 
	3 z (x\dot y - y \dot x)
	\end{array} \right) .
	\end{equation}
	The component along the $z$ axis can be rewritten using the fact that the
	constant of motion $L_z$ is equal to $x \dot y - y \dot x$, that is
	\begin{equation}
	a_z = \frac{6 z JL_z}{r^5} .
	\end{equation} 
	which is exactly the extra term of Eq.~(\ref{eq_z_simp2}). Regarding the $x$ and
	$y$ components, it is more convenient to use the complex combination $a_x + i
	a_y$:
	\begin{equation}
	a_x + i a_y = \frac{2 i J}{r^5} \left( (3 z^2 - r^2) (\dot x + i \dot y) - 3 z
	\dot z (x + i y) \right).
	\end{equation}
	In Eq.~(\ref{eq_xiy_simp2}) there are no $z$ nor $\dot z$, so that we now
	perform the substitution $z^2 \to r^2 - x^2 - y^2$ and $z \dot z \to r \dot r -
	x \dot x - y \dot y$, the last two terms of the last expression being re-expressed
	as 
	\begin{equation}
	x \dot x + y \dot y = (x - i y) (\dot x + i \dot y) - i (x \dot y - y \dot x) 
	= (x - i y) (\dot x + i \dot y) - i L_z .
	\end{equation}
	This leads to several simplifications whose net result is
	\begin{equation}
	a_x + i a_y = i \frac{4 J} {r^3} \left( \dot x + i \dot y - \frac{3}{2}
	\frac{\dot r}{r} (x + i y)\right) + \frac{6 J}{r^5} (x + i y) L_z ,
	\end{equation}
	which corresponds to the extra terms of Eq.~(\ref{eq_xiy_simp2}).
	
	\section{Symmetrizing the observer in the inter-horizon region}
	\label{app_lazy}
	
	In this appendix, we compute the Lorenz factor that allows to go from
	the four-velocity of an observer coming from region~$1$ with $E = 1,
	L_z = 0$ to its mirror analogue coming from region~$3$ with $E' = -1,
	L'_z = 0$. This of course can only happen in region~$2$~(and $12$
	also). We assume that both of them have $\dot \theta = 0$, so that
	they also have the same $\dot r$. The first observer's four-velocity
	has a norm of $1$, so that we can write, using the fact that $E =
	\pi_t$ and $L_z = - \pi_\varphi$,
	\begin{equation}
	g^{tt} E^2 - 2 g^{t\varphi} E L_z + g^{\varphi\varphi} L_z^2 + g_{rr}
	\dot r^2 = 1.
	\end{equation}
	With our choice of constants of motion, this means that
	\begin{equation}
	g_{rr} \dot r^2 = 1 - g^{tt} .
	\end{equation}
	The dot product between the two observers' four-velocity is then
	\begin{equation}
	\gamma = g^{tt} E E' - g^{t\varphi} (E L'_z + E' L_z) +
	g^{\varphi\varphi} L_z L'_z + g_{rr} \dot r^2 .
	\end{equation}
	This is of course the Lorentz factor of the boost that transforms one
	four-velocity into the other. Again, all the $L_z, L'_z$ terms cancel
	away and we have
	\begin{equation}
	\gamma = 1 - 2 g^{tt}
	= 1 + 2 \frac{1}{(- \Delta)}
	\left(r^2 + a^2 + \frac{2 M r a^2 \SN^2}{\Sigma} \right) .
	\end{equation}
	Given the observer's position we are considering~($\theta = 81^\circ$,
	$r = M$, the Lorentz factor is close to $9$, which corresponds to a
	fairly large boost indeed. Even going from one observer to the
	``lazy'', intermediate, observer with $E = 0$ gives a Lorentz factor
	of $\simeq 2.23$, that is, a relative velocity close to $0.9 c$. For
	other values of $r$, the Lorentz factor are even larger~(and diverge
	as $r$ gets close to any of the two horizons, see also~\cite{riazuelo19b} for the Reissner Nordstr\"om case).

\end{document}